\documentclass{revtex4}   

\usepackage{mathrsfs}
\usepackage{balance} 
\usepackage{graphicx}
\usepackage{amsmath}

\begin{document}

\title{Phase Separation in Soft Matter: Concept of Dynamic Asymmetry\\ \vspace{0.5cm}
{\it \small ---Lecture Notes for les Houches 2012 Summer School on
``Soft Interfaces''---} 
}

\author{Hajime Tanaka}
\affiliation{Institute of Industrial Science, University of Tokyo, 4-6-1 Komaba, Meguro-ku, Tokyo 153-8505, Japan}

\begin{abstract}
Phase separation is a fundamental phenomenon that produces spatially heterogeneous patterns in soft matter. In this Lecture Note we show that phase separation in these materials generally belongs to what we call ``viscoelastic phase separation'', where the morphology is determined by the mechanical balance of not only the thermodynamic force (interface tension) but also the viscoelastic force. The origin of the viscoelastic force is dynamic asymmetry between the components of a mixture, which can be caused by either a size disparity or a difference in the glass transition temperature between the components. We stress that such dynamic asymmetry generally exists in soft matter. The key is that dynamical asymmetry leads to a non-trivial coupling between the concentration, velocity, and stress fields. Viscoelastic phase separation can be explained by viscoelastic relaxation in pattern evolution and the resulting switching of the relevant order parameter, which are induced by the competition between the deformation rate of phase separation and the slowest mechanical relaxation rate of a system. We also discuss an intimate link of viscoelastic phase separation, where deformation fields are spontaneously generated by phase separation itself, to mechanical instability (or fracture) of glassy material, which is induced by externally imposed strain fields. We propose that all these phenomena can be understood as mechanically-driven inhomogeneization in a unified manner. 
\end{abstract}

\maketitle



\section{Introduction}

\begin{figure}[!h]
\begin{center}
\includegraphics[width=10cm]{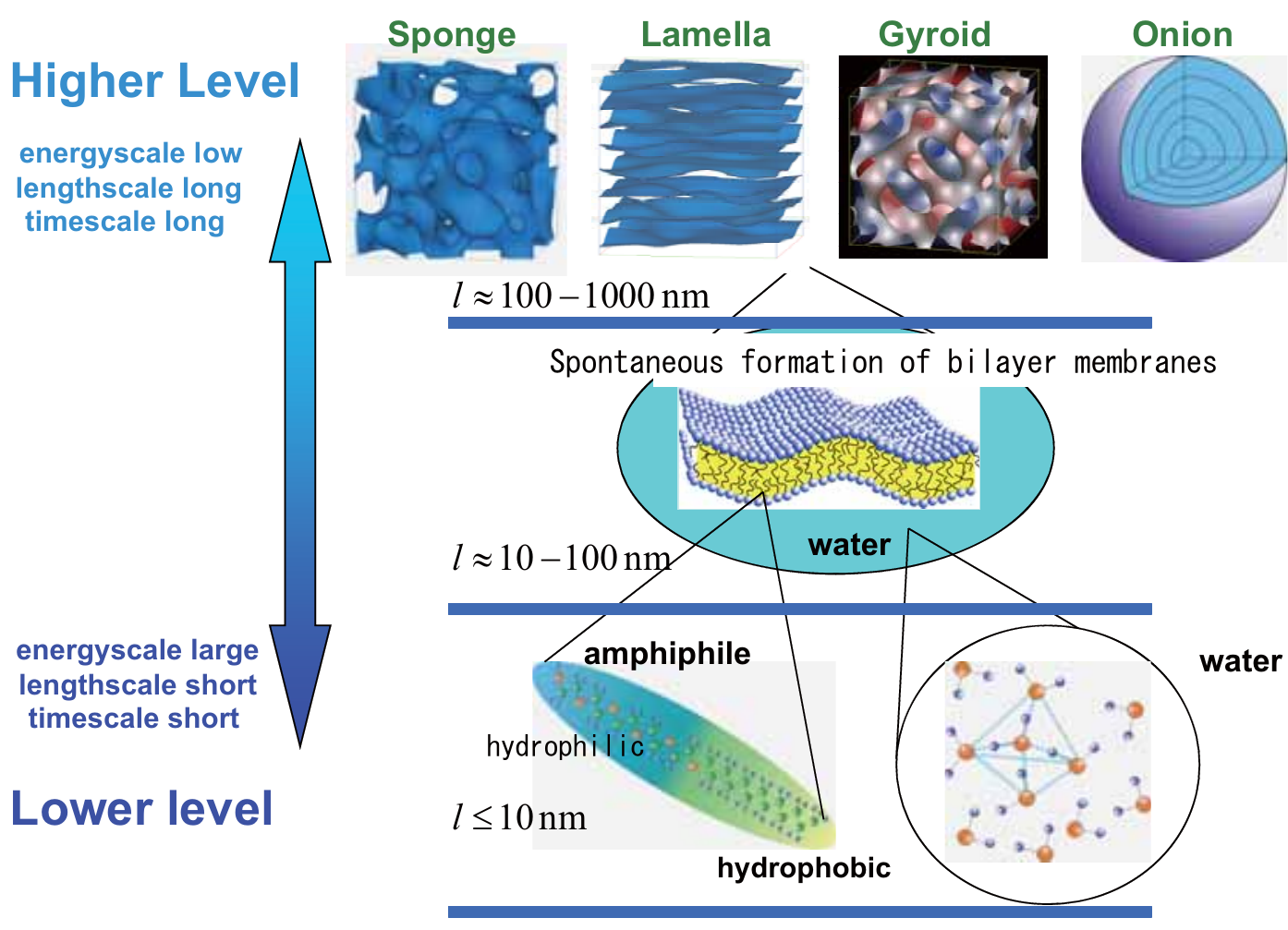}
\end{center}
\caption{
Schematic figure explaining spatio-temporally hierarchical structures in membrane systems. 
}
\label{fig:hierarchy}
\end{figure}

Soft matter is characterized by its spatio-temporally hierarchical structure, which is absent in ordinary 
classical fluids. This feature plays crucial roles in the dynamical behaviour of soft matter. 
As an example of typical hierarchical structures of soft matter, we show such a structure of an amphiphilic surfactant/water mixture 
in Fig. \ref{fig:hierarchy}. Amphiphilic molecules 
spontaneously form bilayer membranes, which further form higher order organizations such as sponge, lamella, gyroid, and onion structures in water. 
In this system, the motion of the low-level structure, i.e., hydrodynamic motion of water in this case, can be coupled with the motion of membranes 
in a dynamical manner: membrane motion causes flow and vice versa. This type of dynamical coupling between the different levels of the structure 
leads to intriguing dynamical behaviour of soft matter.  
We stress that the characteristic timescale of the high-level structures, e.g., membranes, is generally much longer than that of the low-level structure such as water, 
reflecting their size difference. This we call ``dynamic asymmetry'' between the components of a mixture. 
This can be easily understood by the following scaling arguments. 
Both softness and slowness of soft matter are consequences of the large size of a high-level structure: 
The elasticity $G$ is scaled as $G \sim k_BT_m/a^3$, where $a$ is the characteristic lengthscale of the high-level structure and 
$T_m$ is the characteristic ordering temperature of the system. 
On the other hand, the characteristic time $\tau_t$ scale as $\tau_t \sim a^2/D_a \sim 6\pi \eta a^3/k_BT$, with the diffusion constant $D_a=k_BT/6 \pi \eta a$, where $k_B$ is the Boltzmann constant, $T$ is the absolute temperature, and $\eta$ is the viscosity. 
Typically, the size disparity between the high- and low-level structure of soft matter is the order of $10^3 \sim 10^4$, 
which results in strong dynamic asymmetry between them. 
In this Lecture Note, we consider how such dynamic asymmetry affects phase separation dynamics and pattern evolution. 
Please refer to the Chapter by Prof. Mike Cates on phase separation of classical binary mixtures, which is the basis of 
our understanding of phase separation in soft matter.  

Phase-separation phenomena are commonly observed 
in various kinds of condensed matter including metals, semiconductors, 
simple liquids, soft materials such as polymers, surfactants, colloids, 
biological materials, and food materials. 
The phenomena play key roles in pattern evolution 
of immiscible multi-component mixtures of any materials. 
The resulting patterns are linked to optical, electrical, and mechanical 
properties of materials. 
Thus, phase-separation dynamics has been intensively studied from both fundamental  
and applications viewpoints~\cite{gunton,onuki}. 
For example, it was shown recently that a spatially heterogeneous pattern formed by protein phase separation causes a Bragg reflection of light, which is an origin of a colour of bird feathers~\cite{dufresne2009self}. We speculate that this phase separation 
may belong to viscoelastic phase separation, which we shall discuss below.

On the basis of the concept of dynamic universality of critical phenomena~\cite{hohenberg}, phase separation in various condensed matter systems 
were classified into a few groups. 
Phase separation in each group is described 
by a specific set of basic equations describing its dynamic process. 
For example, phase separation in solids is known as ``solid model (model B)'', 
whereas phase separation in fluids as ``fluid model (model H)''~\cite{gunton,hohenberg}. 
For the former the local concentration can be changed only by material 
diffusion, whereas for the latter by both diffusion and flow. 
The universal nature of critical phenomena in each model and the scaling concept based on the self-similar nature of domain growth 
have been established~\cite{gunton,onuki,siggia1979late}. 
In all classical theories of critical phenomena and phase separation, however, 
the same dynamics for the two components of a binary mixture, which we call ``dynamic symmetry''~\cite{tanaka1992} between the components, 
has been implicitly assumed. This assumption can always be justified very near a critical point, 
where the order parameter fluctuations are far slower than any other internal modes of a system (see Fig. \ref{fig:coupling}). 
For a mixture having strong dynamic asymmetry between the components, however, this is not necessarily the case far from a critical 
point, where most of practical phase separation takes place. 
As mentioned above, the presence of dynamic asymmetry means that there is also a large separation between the soft matter mode and the microscopic mode  
of a system. Furthermore, there is another gross variable of a system, the velocity field, whose relevance in dynamics 
comes from the momentum conservation law. Thus, dynamic asymmetry leads to complex couplings 
between the slow critical fluctuation mode, the slow soft matter mode, and the velocity field (see Fig. \ref{fig:coupling}).

In this article, we review the basic physics of viscoelastic phase separation~\cite{tanaka2000viscoelastic,tanaka2006simulation,tanaka2009formation} 
including fracture phase separation~\cite{koyama2009fracture}. 
We show that with an increase in the ratio of the deformation rate of phase separation to the slowest mechanical relaxation rate 
the type of phase separation changes from fluid phase separation, viscoelastic phase separation, to fracture phase separation. 
We point out that there is a physical analogy of this to 
the transition of the mechanical fracture behaviour of materials under shear from liquid-type, ductile, to brittle fracture.   
This allows us to discuss phase separation and shear-induced instability of disordered materials~\cite{furukawa2006violation,furukawa2009inhomogeneous} including soft matter~\cite{onuki1997phase,onuki1989elastic,helfand1989,doionuki,ji1995,milner1993,tanaka1999colloid}, 
on the same physical ground. Finally it should be noted that what we are going to describe in this article has not necessarily been 
firmly established and there still remain many open problems to be studied in the future. 
So this Lecture Note should be regarded as preliminary trials towards the understanding of complex interplay between thermodynamics, 
hydrodynamics, and mechanics, in nonequilibrium pattern formation dynamics in soft matter and glassy matter.

\section{Critical phenomena of dynamically asymmetric mixtures}

Before describing how slow modes of soft matter influence phase-separation behaviour, we consider 
how they affect dynamical critical phenomena. 
We first review some fundamental knowledges on dynamic critical phenomena, focusing on 
the mesoscopic spatial and temporal scales associated with critical fluctuations 
and then consider how the behaviour is modified by the presence of additional mesoscopic spatial and temporal 
scales of soft matter. 

\subsection{Dynamic critical phenomena}
\label{sec:dynamic}

Here we consider dynamic critical phenomena of a binary mixture, i.e., model H 
in the Hohenberg-Halperin classification \cite{hohenberg}. 
The basic equations of model H are given as follows: 
\begin{eqnarray}
\frac{\partial \phi}{\partial t} & = & -\vec{\nabla} \cdot (\phi \vec{v})+
L \vec{\nabla}^2 \frac{\delta (\beta \mathscr{H})}{\delta \phi}, \\
\rho_0 \frac{\partial \vec{v}}{\partial t} & =& - \phi \vec{\nabla} \frac{\delta (\beta \mathscr{H})}{\delta \phi} -\nabla p
+\eta \nabla^2 \vec{v},  \label{eq:momentum} \\
\vec{\nabla} \cdot \vec{v}&=&0, \label{eq:incompress}
\end{eqnarray}
where $\phi(\vec{r},t)$ is the composition deviation from the average value, $\vec{v}(\vec{r},t)$ is the velocity field, 
$\eta$ is the viscosity, and  $p$ is a part of the pressure, which is determined to satisfy the incompressible condition (\ref{eq:incompress}). 
Here we assume for simplicity that the Hamiltonian $\mathscr{H}(\phi)$ is 
given by the following Ginzburg-Landau form: 
\begin{eqnarray}
\beta \mathscr{H}(\phi) & = & \int d \vec{r} \ [\frac{r}{2} \phi^2+\frac{u}{4} \phi^4
+\frac{C}{2}(\nabla \phi)^2],  
\end{eqnarray}
where $\beta=1/(k_BT)$, $r=a(T-T_c)$ ($a$: a positive constant), and $u$ and $C$ are positive constants. 

Near the critical point $T_c$, the order parameter fluctuations become mesoscopic. 
This can be seen from the fact that the spatial correlation of the order parameter fluctuations can be obtained from the above Hamiltonian under 
a harmonic approximation as 
\begin{equation}
S(k)=\int d\vec{r} \ \langle \phi(r) \phi(0) \rangle \cong \frac{\chi_{\phi}}{1+k^2 \xi^2}, \label{eq:Sk}
\end{equation}
where $k$ is the wave number, 
$\xi=(C/|r|)^{1/2}$, and $\chi_{\phi}=1/|r|$ is the susceptibility. This functional form is 
known as the Ornstein-Zernike correlation function. 
Both the susceptibility $\chi_{\phi}$ and correlation length $\xi$ diverge towards $T_c$, respectively, as $\chi_{\phi}=1/|r| =\chi_0 \epsilon^{-\gamma}$ and 
$\xi =\xi_0 \epsilon^{-\nu}$, where $\chi_0$ is a positive constant and $\xi_0$ is the bare correlation length, 
$\epsilon=(T-T_c)/T_c$ is the reduced temperature. Here we note recent experimental \cite{royall2007bridging} and theoretical studies \cite{Campbell} 
suggest the more appropriate form for $\epsilon$ is $\epsilon=(T-T_c)/T$, which provides $\xi \rightarrow \xi_0$ in the limit of 
$T \rightarrow T_c$. 
The critical exponent for the susceptibility $\gamma$ 
is 1 for the mean-field approximation, but 1.24 for the 3D Ising universality after the renormalization. 
On the other hand, the critical exponent for the correlation length $\nu$ 
is 1/2 for the mean-field approximation, but 0.63 for the 3D Ising universality.

Reflecting the growth of the correlation length $\xi$, the dynamics of the order parameter fluctuations also slows down 
while approaching $T_c$, which is known as ``critical slowing down''. The lifetime of the fluctuations $\tau_\xi$ is estimated as 
$\tau_\xi=\xi^2/D_\xi=6 \pi \eta \xi^3/k_BT \sim \epsilon^{-\nu z}$, where $z$ is the standard dynamical 
critical exponent. Here $D_\xi$ is the diffusion constant of a mixture and given by $D_\xi=k_BT/(6\pi \eta \xi)$ (see below).

Here we briefly review how the diffusional transport can be described under a coupling to the velocity fields 
in the above framework of model H \cite{onuki}. 
The transport coefficients of fluids in the linear response regime can be expressed as the time integral of 
flux time correlation functions. By using the relevant flux $J_\phi(\vec{r},t)=\phi(\vec{r},t)\vec{v}(\vec{r},t)$, 
the renormalized transport coefficient $L_R(k)$ for $\phi$ is expressed as 
\begin{eqnarray}
L_R(k)=L_0+\int_0^\infty \int d\vec{r} \ \exp{(i \vec{k} \cdot \vec{r})} \langle \vec{J}_\phi (\vec{r},t) \cdot \vec{J}_\phi(\vec{0},t) \rangle, 
\end{eqnarray}
where $L_0$ is the (solid-type) bare transport coefficient and negligible compared to the second term.  
It can be calculated by decoupling the above four-point correlation function as 
\begin{eqnarray}
L_R(k)=L_0+\int_0^\infty \int d\vec{r} \ \exp{(i \vec{k} \cdot \vec{r})} \langle \vec{v}(\vec{r},t) \cdot \vec{v}(\vec{0},t) \rangle g(|\vec{r}|).  
\end{eqnarray}
Since the timescale of the order parameter fluctuations $\phi(\vec{r},t)$ is much slower than that of the momentum diffusion (or, the velocity field 
$v_x(\vec{r},t)$), we apply the adiabatic approximation in the above and assume $\langle \phi(\vec{r},t)\phi(\vec{0},t)\rangle=g(|\vec{r}|)$. 
Under the incompressible condition, the time correlation function of $\vec{v}_{\vec{k}}$ is expressed as 
\begin{equation}
\langle v_{ik}(t)v_{jk}(0)^\ast \rangle=\frac{k_BT}{\rho} \left(\delta_{ij}-\frac{1}{k^2}k_ik_j \right)\exp{\left( -\frac{\eta}{\rho}k^2t \right)}.  
\end{equation}
By using the Ornstein-Zernike form for the correlation function $S(k)$ (see Eq. (\ref{eq:Sk})), we obtain 
\begin{eqnarray}
L_R(k)=L_0+\left(\frac{k_BT}{6\pi \eta \xi} \right)\chi_{\phi} K(k \xi),
\end{eqnarray} 
where the scaling function $K(x)$ is called the Kawasaki function and $K(x) \cong 1$ for $x \ll 1$ whereas $K(x) \cong 3\pi/8x$ for $x \gg 1$ 
(see, e.g., Ref. \cite{onuki}). 

The order parameter decay rate $\Gamma_k$ is then given by 
\begin{eqnarray}
\Gamma_k \cong \frac{L_R(k) k^2}{S(k)} \cong \frac{k_BT}{6\pi \eta \xi}  K(k \xi)k^2 (1+k^2 \xi^2). \label{eq:Gamma_k}
\end{eqnarray}
This provides the interesting non-locality of the diffusional transport originating from the presence of the important 
mesoscopic lengthscale $\xi$. 

Next we consider the viscous transport in a critical binary mixture. 
The viscosity is expressed as 
\begin{equation}
\eta=\eta_0+\frac{1}{k_BT} \int_0^\infty dt \int d\vec{r} \langle \Pi_{xy}(\vec{r},t) \Pi_{xy}(\vec{0},0) \rangle, 
\end{equation}
where $\eta_0$ is the non-critical background viscosity of a mixture and 
$\Pi_{xy}=C(\partial \phi/\partial x)(\partial \phi/\partial y)$. By using a decoupling approximation, 
we can obtain the following expression: 
\begin{eqnarray}
\eta=\eta_0+\frac{C^2}{30 \pi^2 k_BT} \int_0^{\Lambda_0} dk \ \frac{k^6 S(k)^2}{\Gamma_{k}} \cong \eta_0+x_{\eta} \eta \log{(\Lambda_0 \xi)},  
\end{eqnarray} 
where $\Lambda_0$ is the short cut-off wavelength and $y_\eta=8/15\pi^2\cong 0.054$. Since $y_\eta \ll 1$, we can approximate the above as 
\begin{equation}
\eta \cong \eta_0 (\Lambda_0 \xi)^{y_\eta}. 
\end{equation}
The more exact calculation yields $x_\eta=\nu y_\eta \cong 0.042$, which indicates a very weak (almost logarithmic) divergence of the viscosity. 
The effects of dynamic asymmetry on this anomaly will be discussed below.

\subsection{Effects of dynamic asymmetry on dynamic critical phenomena} \label{sec:DA}
\begin{figure}[!h]
\begin{center}
\includegraphics[width=9cm]{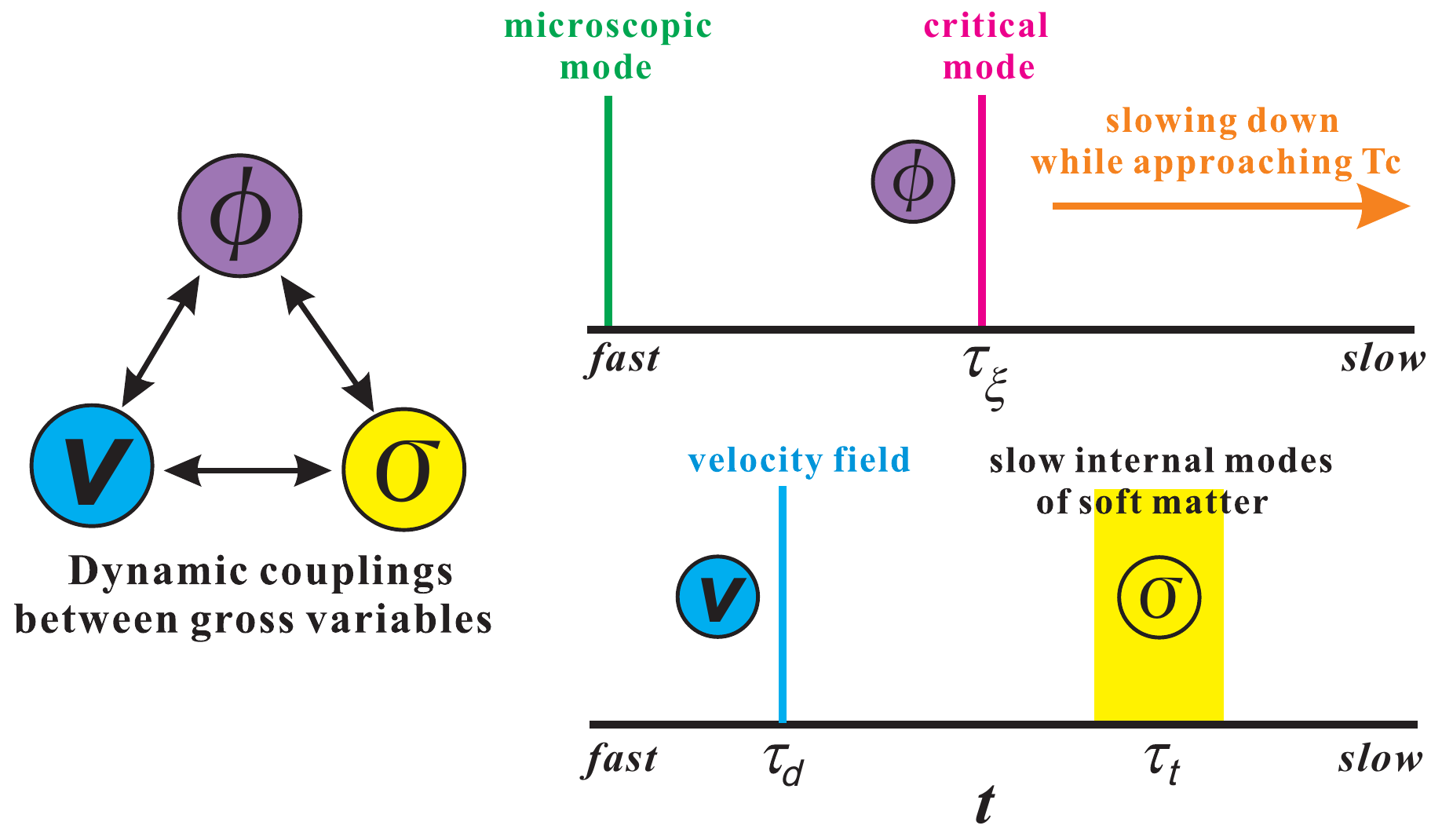}
\end{center}
\caption{
Schematic figure showing dynamical couplings among the three gross variables, 
the composition $\phi$, the velocity field $\vec{v}$, and the stress field $\mbox{\boldmath$\sigma$}$, 
The relation among these modes and the microscopic mode are also shown.  
When approaching the critical point $T_{\rm c}$, the order parameter fluctuation mode $\tau_\xi$ should eventually become the slowest mode 
in principle. In this limit, the relaxation of $\mbox{\boldmath$\sigma$}$ does not play any role and thus 
the dynamic universality should hold. However, this situation may not be practically realized 
for a system of strong dynamic asymmetry. In phase separation, we should also consider the characteristic time of 
deformation $\tau_d$. If the deformation rate $\tau_d$ is faster than the relaxation rate of the slow soft matter mode $\tau_t$, the viscoelastic 
effects have a drastic influence on phase separation. This figure is reproduced from Ref. \protect\cite{tanaka2012viscoelastic}. 
}
\label{fig:coupling}
\end{figure}

Now we consider how the dynamic asymmetry between the components of a mixture affects dynamic critical phenomena. 
Here we note that since the effects are purely dynamical, they do not affect the static critical phenomena. 
As schematically shown in Fig \ref{fig:coupling}, the effects of dynamic asymmetry originate from the fact 
that there is an additional slow mode coming from the large size of a component of a mixture, whose relaxation 
time we denote $\tau_t$. As described above, there is critical slowing down towards $T_c$ and thus the order parameter 
fluctuations become slower and slower while approaching $T_c$. The characteristic lifetime of the fluctuations $\tau_\xi$ always 
becomes the slowest mode near $T_c$. This clear separation between microscopic and critical length and time scale is the heart of the concept of dynamic universality \cite{hohenberg}. 
There is no exception for this limiting behaviour for $T \rightarrow T_c$, however, practically this limit of $\tau_\xi \gg \tau_t$ cannot be accessed 
experimentally in a system of strong dynamic asymmetry \cite{tanaka1994critical}. 
Then dynamic asymmetry affects dynamic critical phenomena significantly 
\cite{tanaka1993critical,tanaka1994critical,doionuki,furukawa2003,furukawa2003b,tanaka2002nonuniversal,kostko2002}. 
The relationship between key timescales are shown schematically in Fig. \ref{fig:coupling}.  

\begin{figure}[!h]
\begin{center}
\includegraphics[width=7cm]{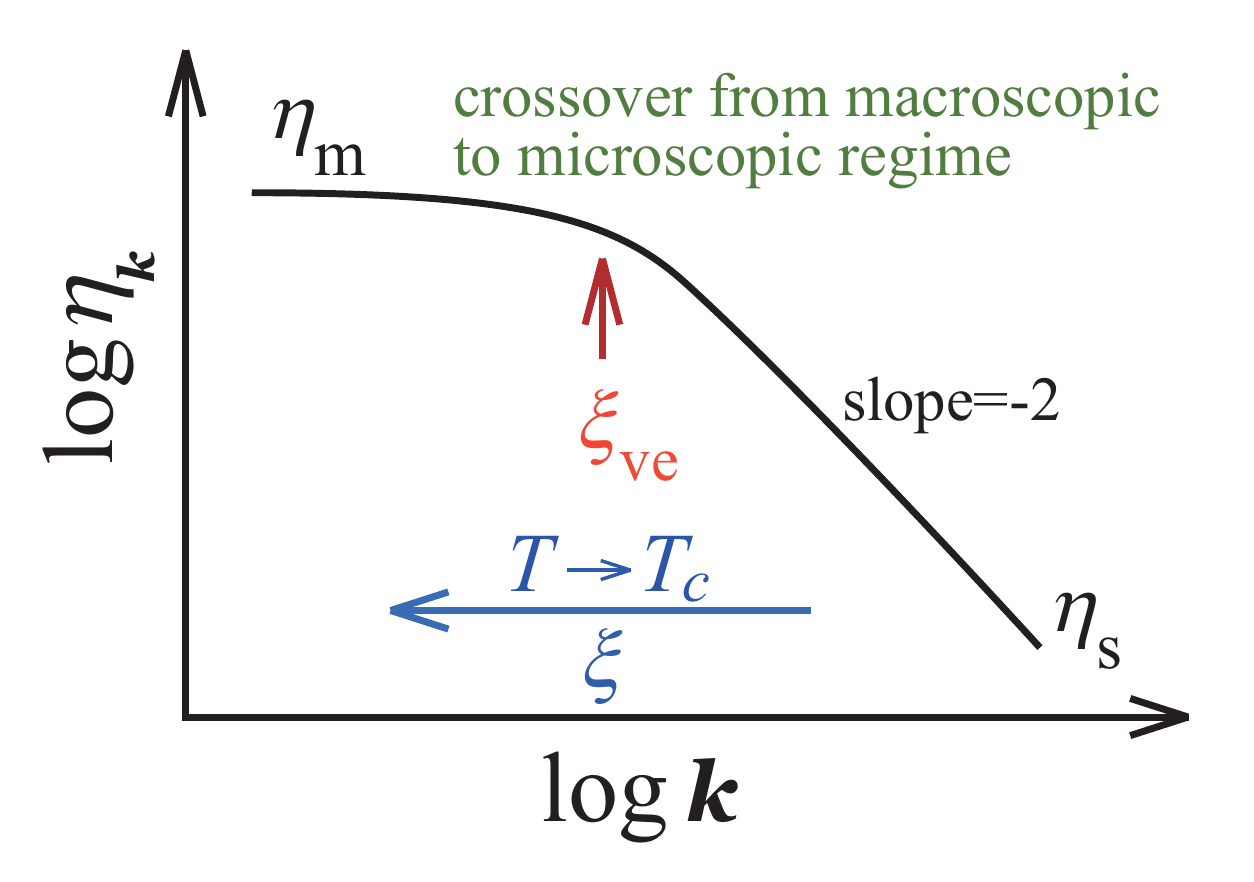}
\end{center}
\caption{
Schematic figure showing the wavenumber $k$-dependence of the viscosity. 
For example, the viscosity of a polymer solution at low $k$, or the macroscopic viscosity, is quite high due to entanglements, where as 
that at high $k$ is very low and comparable to the solvent viscosity $\eta_s$. This is natural, considering that polymer chains locally feels 
the viscosity of a solvent, but globally the macroscopic viscosity due to many-chain interactions. 
The crossover between these two regimes occurs at the mesoscopic lengthscale characterized by the viscoelastic length $\xi_{ve}$. 
We stress that this length has a purely dynamical origin. Ordinary critical phenomena and phase separation phenomena are recovered 
only when $\xi \gg \xi_{ve}$. 
}
\label{fig:eta}
\end{figure}

The slow dynamics characterized by the slow rheological relaxation time of soft matter $\tau_t$ affects 
the above decay rate of the order parameter fluctuations (see Eq. (\ref{eq:Gamma_k})) through the coupling between the two slow modes, 
critical and rheological modes \cite{doionuki,tanaka2002nonuniversal,kostko2002}. 
The dynamic asymmetry gives rise to an important new mesoscopic length $\xi_{ve}$, which is of purely rheological origin.  
We call this $\xi_{ve}$ the viscoelastic (or, magic) length \cite{brochard1977,brochard1983}, which is the characteristic length scale above which dynamics is 
dominated  by diffusion and below which by viscoelastic effects. This length is a rheological length scale
intrinsic to entangled polymer solutions \cite{brochard1977,brochard1983} and other dynamically asymmetric systems \cite{tanaka1999colloid}, 
and nothing to do with critical phenomena.  
We can also interpret this length scale as the length up to which the shear stress can transmit. 
The shear stress is dominated by the velocity fluctuations of the
distance over $\xi_{ve}$ and of the time $\tau_t$. 
Furukawa predicted that the dynamical asymmetry coupling 
between the velocity fluctuations and the viscoelastic stress affects the hydrodynamic relaxation process, resulting in a wavenumber-dependent shear viscosity: 
the viscosity of a polymer solution has the following wavenumber ($k$-) dependence associated with 
this length $\xi_{ve}$~\cite{furukawa2003,furukawa2003b}: 
\begin{equation}
\eta(\mbox{\boldmath$k$})=\eta_s+\frac{\eta_m}{1+\xi_{ve}^2 k^2}, 
\end{equation} 
where $\eta_m$ is the macroscopic viscosity and $\eta_s$ is the solvent viscosity. 
The viscoelastic length is approximately given by $\xi_{ve} \sim (\eta_m/\eta_s)^{1/2} \xi_b$, where $\xi_b$ is the blob size \cite{deGennes}. 
The $k$-dependence of $\eta$ is schematically shown in Fig. \ref{fig:eta}. 
We stress that the relation between the thermodynamic correlation length $\xi$ and the rheological viscoelastic length $\xi_{ve}$ 
depends upon how close to $T_c$ a system is, as illustrated in Fig. \ref{fig:eta}. So the viscosity felt by the composition fluctuations 
depends upon the relation between them. Thus, the relation between these lengthscales plays a crucial role in determining 
how phase separation proceeds in its early stage.  
This nonlocal nature of the viscous transport 
is a manifestation of the temporal hierarchical structure of dynamically asymmetric systems. 
Interestingly, the similar non-locality of the transport coefficient has recently been found by Furukawa and Tanaka for a supercooled liquid 
having dynamic heterogeneity over the lengthscale $\xi$ \cite{furukawa2009,furukawa2011,furukawa2012}.

\begin{figure}[!h]
\begin{center}
\includegraphics[width=8cm]{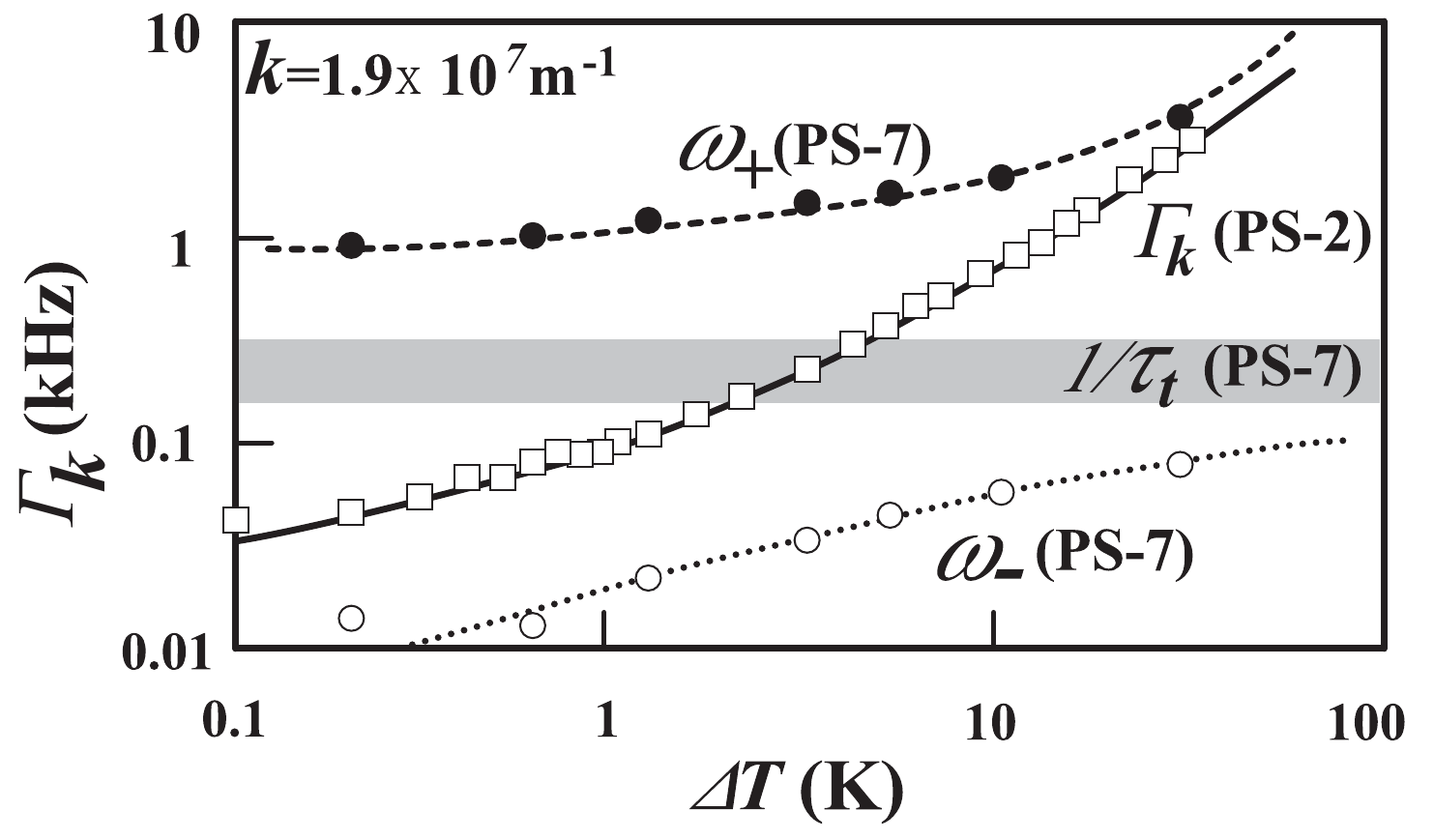}
\end{center}
\caption{
Temperature dependences of $\Gamma_k$ (open square) for PS-2 
and $\omega_+$ (filled circle) and $\omega_-$ (open circle) for PS-7. 
The viscoelastic relaxation rate $1/\tau_t$, which is 
estimated by the analysis, is also shown by the thick gray line. This figure is reproduced from Fig. 5 of Ref. \protect\cite{tanaka2002nonuniversal}. 
}
\label{fig:omega}
\end{figure}

Under the coupling between the viscoelastic and critical modes, the time correlation function of $\phi$, $h(t)$, 
should be given by \cite{doionuki}
\begin{eqnarray}
h(t)=f_+ \exp(-\omega_+ t)+f_- \exp(-\omega_- t), 
\end{eqnarray} 
where 
\begin{eqnarray}
\omega_\pm&=&\frac{1+k^2 \xi_{ve}^2+\tau_t \Gamma_k 
\pm \sqrt{(1+k^2\xi_{ve}^2+\tau_t \Gamma_k)^2-4 \tau_t \Gamma_k}}{2 \tau_t}, \\
f_{\pm}&=&\pm \frac{\omega_{\pm} \tau_t -(1+k^2 \xi_{ve}^2)}
{(\omega_+-\omega_-)\tau_t}
\end{eqnarray}
The temperature dependences of $\omega_+$ and $\omega_-$ 
for a critical solution of polystyrene (PS) in diethyl malonate (DEM), PS-7 (the molecular weight $M_w=3.84 \times 10^6$), and $\Gamma_k$ for PS-2 ($M_w=9.64 \times 10^4$) are shown in Fig. \ref{fig:omega}. 
Far above $T_c$ where $\tau_t \Gamma_k \gg 1$, 
the theory  \cite{brochard1977,brochard1983,doionuki} 
predicts that $h(t)$ should be approximated 
as 
\begin{equation}
h(t) \cong \exp(-\Gamma_k t)+\frac{\xi_{ve}^2 k^2}{\Gamma_k \tau_t} 
\exp(-t/\tau_t).
\end{equation} 
In this limit, $\omega_+=\Gamma_k$ and $\omega_-=1/\tau_t$. 
We also estimate $\xi_{ve}$ from the relation 
\begin{equation}
f_- \omega_+ +f_+ \omega_-=\frac{1+\xi_{ve}^2 k^2}{\tau_t} 
\end{equation}
as $\sim$0.1 $\mu$m for PS-7. 
It is also found that $\xi_{ve}$ only weakly depends on $T$, as expected from its rheological (non-critical) nature. 
The value of $\tau_t$ estimated from light scattering 
($\tau_t \sim 0.005$ s) is compatible with our rheological measurements \cite{tanaka1993critical}. 
Both $\tau_t$ and $\xi_{ve}$ are found to be 
larger for PS-8 ($M_w=8.42 \times 10^6$) than for PS-7, as expected. 
Very near $T_c$ where $\Gamma_k \rightarrow 0$, on the other hand, 
\begin{equation}
h(t)=\exp(-\frac{\Gamma_k}{1+\xi_{ve}^2 k^2} t). 
\end{equation}
This tells us that the critical mode can be free from the coupling to the viscoelastic mode 
only very near $T_c$ and at small $k$ \cite{tanaka1994critical}. 
Figure \ref{fig:omega} clearly demonstrates that the behaviour 
of the critical mode is strongly influenced by its dynamic coupling 
to the viscoelastic mode for PS-7, which supports the above scenario \cite{tanaka2002nonuniversal}. 

\begin{figure}[!h]
\begin{center}
\includegraphics[width=6.5cm]{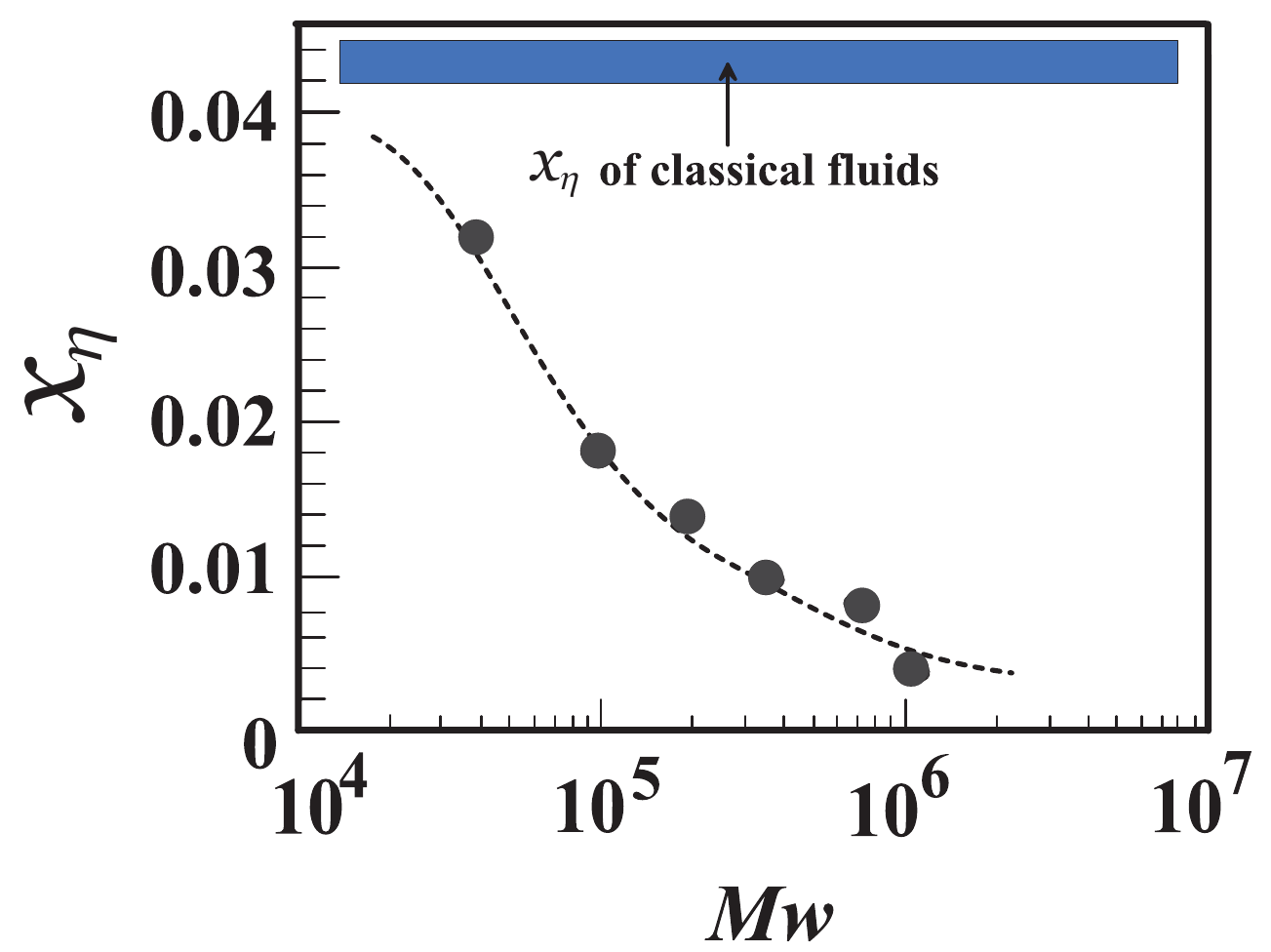}
\end{center}
\caption{
Molecular-weight ($M_w$-) dependence of the critical exponent $x_\eta$. 
Note that the universal value of $x_\eta$ for classical 
fluid mixtures is $\sim 0.042$. 
We can see that the presence of the slow rheological relaxation mode of a polymer solution 
significantly suppresses the critical anomaly of the viscosity. This figure is reproduced from Fig. 3 of Ref. \protect\cite{tanaka2002nonuniversal}. 
}
\label{fig:y}
\end{figure}

Next we consider the critical anomaly of the viscosity in polymer solutions \cite{tanaka2002nonuniversal}. 
We found that the viscosity anomaly is more strongly suppressed for a system 
of stronger dynamic asymmetry for polymer solutions. 
Figure~\ref{fig:y} plots the value of the exponent for the viscosity anomaly $x_\eta$ 
at the critical concentration against the molecular weight of polymers, $M_w$. 
This clearly demonstrates that the value of $x_\eta$ monotonically 
decreases with an increase in $M_w$. 
Since polymer solutions belong to the same dynamic universality class 
as classical fluids, this exponent $x_\eta$ should be equal to 
the universal value of $\sim 0.042$, irrespective of the 
molecular weight of the polymer, if we are close enough to $T_c$ (see above). 
Thus, the above results clearly indicate that 
the viscosity anomaly of critical polymer solutions 
(especially, of high molecular weight polymers) 
cannot be described by ``fluid model (model H)'' at least 
in the experimentally accessible temperature range. 
Note that $x_\eta$ approaches the universal value of classical fluids 
with a decrease in $M_w$, i.e, towards the limit of dynamic asymmetry.  

We ascribed this non-universal behaviour to 
the existence of an additional slow mode intrinsic to 
polymer solutions and its dynamic coupling to the critical mode, which is a direct consequence of dynamic asymmetry   
\cite{tanaka1993critical,tanaka1994critical,tanaka2002nonuniversal}. 
Note that the dynamic universality concept relies on 
the fact that there exists only one slow mode in a system, which is 
a critical mode. However, such a situation cannot be realized in practical experiments for a mixture of 
strong dynamic asymmetry (see Fig. \ref{fig:coupling}). 
For polymer solutions, the viscoelastic relaxation mode is dynamically coupled with 
the concentration fluctuation mode, which results in the viscoelastic suppression 
of the dynamic critical anomaly: The fluctuation modes whose 
characteristic length is shorter 
than the magic (viscoelastic) length $\xi_{ve}$ is significantly 
influenced by the viscoelastic effects. 
The theoretical account for this non-universal behaviour of the viscosity was 
provided by Furukawa \cite{furukawa2003,furukawa2003b}.

\section{Phenomenological theory of viscoelastic phase separation}

\subsection{Introduction}

Phase separation of soft matter such as polymer blends, polymer solutions, protein solutions, and emulsions had been believed 
to be the same as that of classical fluid mixtures (model H) \cite{gunton,hohenberg}. However, it was shown about two decades ago that 
it is not necessarily the case~\cite{tanaka1988,tanaka1992,tanaka1993,tanaka1994critical,tanaka1996,tanaka1992}. 
In normal phase separation observed in dynamically symmetric mixtures (model H), 
the phase separation morphology is determined by the balance between the thermodynamic (interfacial) force $\vec{F}_\phi$ and the viscous force $\vec{F}_v$
while satisfying the momentum conservation. 
In viscoelastic phase separation, on the other hand, the self-generated mechanical force $\vec{F}_\sigma$ also plays a crucial role in its 
pattern selection in addition to the thermodynamic $\vec{F}_\phi$ and viscous force $\vec{F}_v$. 
Thus we named this type of phase separation ``viscoelastic phase separation (VPS)''. 
In addition to the solid and fluid model, thus, we need the third model for phase separation in condensed matter, i.e., 
the ``viscoelastic model''~\cite{tanaka2000viscoelastic,tanaka1997general}. 
This model is actually a general model 
of phase separation of isotropic systems including the solid and fluid model as its special cases~\cite{tanaka1997general} (see Sec. \ref{sec:analogy}).  

Intuitively, viscoelastic phase separation can be explained as follows. 
When there is a large difference in the dynamics between the 
components of a mixture, phase separation tends to proceed 
in a speed between those of the fast and slow components. 
Then, the slow component cannot catch up with 
a deformation rate spontaneously generated by phase separation itself, $\tau_d$, and thus starts to behave as an 
elastic body, which switches on the elastic mode of phase separation. 
Thus, this phenomenon can be regarded as ``viscoelastic relaxation in pattern evolution'', which 
is the reason why we named it {\it viscoelastic phase separation}~\cite{tanaka1993}. 
Unlike ordinary mechanical relaxation experiments, the mechanical perturbation is characterized by the rate of deformation induced 
by phase separation, $\tau_d$, and the relaxation rate is that of the slowest mechanical relaxation, $\tau_t$, in a system (see Fig. \ref{fig:coupling}).  

Without dynamic asymmetry, the deformation rate is always slower than the relaxation rate. Thus, phase separation in such a mixture   
can always be described by the fluid model, no matter how slow the dynamics of the components is. 
This is because $\tau_\xi \gg \tau_t$ is always satisfied in the critical regime for dynamically symmetric systems 
and there is no $k$-dependence in $\eta$, i.e., $\xi_{ve}$ is a microscopic length.  
For example, this is the case for a mixture of two polymers having similar molecular weights and glass transition temperatures 
\cite{bates1989spinodal,hashimoto1988dynamics}. 
We emphasize that dynamic asymmetry, which is prerequisite to 
viscoelastic phase separation, usually exists in any materials, particularly, in soft matter. 
In this sense, we may say that ``normal phase separation (NPS)'' is a special case of viscoelastic phase separation.

\subsection{Two-fluid model of polymer solution and stress-diffusion coupling}

\label{sec:twofluid}
Shear effects on complex fluids have attracted much attention 
because of its unusual nature known as ``Reynolds effect'': For example, 
shear flow that intuitively helps the mixing of the components actually 
induces phase separation in polymer solutions~\cite{onuki1997phase,onuki}. 
This is caused by couplings between the shear velocity fields 
and the elastic internal degrees of freedom of polymers. 
To explain this counter-intuitive behaviour of polymer solutions under shear, 
there have been considerable theoretical efforts 
\cite{onuki1997phase,onuki1989elastic,helfand1989,doionuki,ji1995,milner1993}. 
Doi and Onuki~\cite{doionuki} established a basic set of coarse-grained equations describing 
critical polymeric mixtures, based on a two-fluid model whose 
original form was developed by de Gennes and Brochard~\cite{deGennes1976a,deGennes1976b,brochard1983} for polymer solutions 
and by Tanaka and Filmore~\cite{tanakafilmore} for chemical gels. 

Later we proposed that an additional inclusion of the strong concentration dependence of the bulk stress and/or the transport coefficient, 
which are not important in shear-induced instability, 
is necessary for describing viscoelastic phase separation of dynamically asymmetric mixtures, more specifically, 
the volume shrinking behaviour of the slow-component-rich phase~\cite{tanaka1997roles,tanaka1997general,tanaka1997}. 
We also argued its generality beyond polymer solutions to particle-like systems such as colloidal suspensions, emulsions, and protein solutions 
\cite{tanaka1999colloid}. 
That is, we showed that the internal degrees of polymer chains and entanglement effects peculiar to polymer 
systems are not necessary for viscoelastic phase separation to take place and strong dynamic asymmetry between the 
components of a mixture is the only necessary condition.  
A main difference between shear-induced phase separation and viscoelastic phase separation is that 
the velocity fields are induced by external shear fields 
in the former whereas they are self-induced by phase separation itself in the latter.

The dynamic equations for polymer solutions are given as follows (see Refs.~\cite{doionuki,tanaka1997general,doi2011onsager} for the derivation of these equations): 
\begin{eqnarray}
\frac{\partial \phi}{\partial t} & = & -\vec{\nabla} \cdot (\phi \vec{v})+
\vec{\nabla} \cdot \frac{\phi(1-\phi)^2}{\zeta}
\vec{\nabla} \cdot [\mbox{\boldmath$\Pi$}-
\mbox{\boldmath$\sigma$}]. \label{eq:conserve} \\
\vec{v}_p-\vec{v} &=& -\frac{(1-\phi)^2}{\zeta} \vec{\nabla} \cdot 
[\mbox{\boldmath$\Pi$}
-\mbox{\boldmath$\sigma$}]. \label{eq:relativev}\\
\rho_0 \frac{\partial \vec{v}}{\partial t} & =& - \vec{\nabla} \cdot 
[\mbox{\boldmath$\Pi$}-\mbox{\boldmath$\sigma$}] -\nabla p
+\eta_s \nabla^2 \vec{v}.  \label{eq:momentum2} \\
\vec{\nabla} \cdot \vec{v}&=&0. 
\end{eqnarray}
Here $\vec{v}_p(\vec{r},t)$ and $\vec{v}_s(\vec{r},t)$ 
are respectively the coarse-grained average velocities of polymer 
and solvent at point $\vec{r}$ and time $t$, 
and then the average velocity of a mixture  $\vec{v}$ is given by $\vec{v}=\phi \vec{v}_p +(1-\phi) \vec{v}_s$. 
$\phi(\vec{r},t)$ is the composition of polymer. 
$\mbox{\boldmath$\Pi$}$ is the osmotic stress tensor, which is 
related to the thermodynamic force $\vec{F}_\phi$ as 
\begin{equation}
\vec{F}_\phi= -\vec{\nabla} \cdot \mbox{\boldmath$\Pi$}=-\phi \nabla 
(\delta \mathscr{F}/\delta \phi), 
\end{equation}
where $\mbox{\boldmath$\sigma$}$ is the mechanical stress tensor, 
$\rho_0$ is the average density, $\eta_s$ 
is the solvent viscosity, and $\zeta$ is 
the friction constant per unit volume. 
Here $p$ is a part of the pressure, which is determined to satisfy the incompressible condition $\vec{\nabla}\cdot \vec{v}=0$. 
The free energy $\mathscr{F}(\phi)$ is 
given by the following Flory-Huggins-de Gennes form: 
\begin{eqnarray}
\mathscr{F}(\phi) & = & k_BT \int d \vec{r} \ [f(\phi)
+\frac{C(\phi)}{2}(\nabla \phi)^2],  \nonumber \\ 
f(\phi) & = & \frac{1}{N} \phi \ln \phi +
(1-\phi)\ln (1-\phi)+
\chi \phi(1-\phi), \nonumber 
\end{eqnarray}
where $N$ is the degrees of polymerization of polymer  
and $\chi$ is the interaction parameter between polymer and solvent. 
The terms containing the mechanical stress tensor cause couplings between the composition and the stress fields 
{\it via} the velocity fields. The above equation (\ref{eq:relativev}) clearly tell us that the relative velocity of polymers 
to the average velocity is determined not only by the thermodynamic osmotic force but also by the mechanical force. 
The role of mechanical force can be easily understood by considering a case of gel \cite{tanakafilmore}. 
To close these equations, we need a constitutive equation, which describes the time evolution 
of $\mbox{\boldmath$\sigma$}$. 

Here it is worth noting that in Eq. (\ref{eq:momentum2}) the inertia term is not relevant for the description of 
viscoelastic phase separation in ordinary situations since viscoelasticity suppresses the development of velocity fields. However, this 
is not necessarily the case for a shear problem and even a nonlinear velocity term plays an important role 
for high Reynolds number flow. We do not consider the nonlinearity since it is out of scope of this article. 

In the above, we consider a case of polymer solution, where only polymers can support viscoelastic stress, for simplicity. 
For a more general case, where viscoelastic stress is not supported only by one of the components, 
we need a more general set of equations \cite{tanaka1997general}. In such a case, the constitutive relation may also 
become more complex (see Sec. \ref{sec:division}).  

Finally, we mention a fundamental remaining problem of the two-fluid description. 
In the above derivation, the dissipation in a mixture is separated into the two contributions: 
One is viscous dissipation of the liquid component, and the other comes from the friction between the two components. 
This intuitively looks reasonable, however, the hydrodynamic couplings between the slow components are not considered in a systematic manner 
in the coarse-gaining procedure. This makes the validity of the above separation a bit obscure. 
Thus, we need theoretical justification for the treatment of dissipation, which remains a subject 
for future investigation.

\subsubsection{Kinetic equations describing the time evolution of stress tensor: Constitutive relation} 

We note that for mixtures composed of a large particle (or molecule) component (component 1) and a simple fluid (liquid or gas) (component 2)
the stress division becomes almost perfect ($\alpha_1 \cong 1$ and $\alpha_2 \cong 0$), reflecting the large size disparity 
and the resulting large difference in the friction constant. 
The mechanical stress of a system is selectively supported almost by the large component 1 alone. 
This is the case for polymer solutions~\cite{doionuki}, and suspensions of colloids, proteins, and emulsions \cite{tanaka1999colloid}. 
So the velocity relevant to the description of viscoelastic stress is the average velocity of component 1 ($\vec{v}_p$ for a polymer solution). 

As an example of this type of mixtures, here we consider how the mechanical stress, 
$\mbox{\boldmath$\sigma$}$, should be expressed in the case of polymer solution. 
In general, we should incorporate relevant constitutive equations into the above two-fluid model, depending upon the type of material. 
Doi and Onuki~\cite{doionuki} employed the upper-convective Maxwell equation as a constitutive relation 
describing its time evolution for polymer solution~\cite{doi,larson}: 
\begin{eqnarray}
\frac{D}{D t}\mbox{\boldmath $\sigma$}_S =\mbox{\boldmath $\sigma$}_S \cdot
 \vec{\nabla} \vec{v}_p+(\vec{\nabla} \vec{v}_p)^T \cdot 
\mbox{\boldmath $\sigma$}_S - \frac{1}{\tau_S(\phi)} 
\mbox{\boldmath $\sigma$}_S 
+ G_S(\phi)\{\vec{\nabla}\vec{v}_p+(\vec{\nabla}\vec{v}_p)^T\}, 
\label{eq:maxwell_S}
\end{eqnarray}
where $\frac{D}{D t}=\frac{\partial}{\partial t} 
+ \vec{v}_p \cdot \vec{\nabla}$ and 
$\tau_S$ and $G_S$ are the relaxation time and the modulus of the shear stress, 
respectively. Note that $(\vec{\nabla}\vec{v}_p)_{ij}=\partial_i v_{pj}$. 
To make the shear stress a traceless tensor, 
$\mbox{\boldmath $\sigma$}_S$ was defined as $\mbox{\boldmath $\sigma$}_S = 
\mbox{\boldmath $\sigma$}_S-\frac{1}{d}{\rm Tr}\mbox{\boldmath $\sigma$}_S 
\mbox{\boldmath $I$}$, 
where $\mbox{\boldmath $I$}$ is the unit tensor and $d$ is the space 
dimensionality. 

\begin{figure}
\begin{center}
\includegraphics[width=8cm]{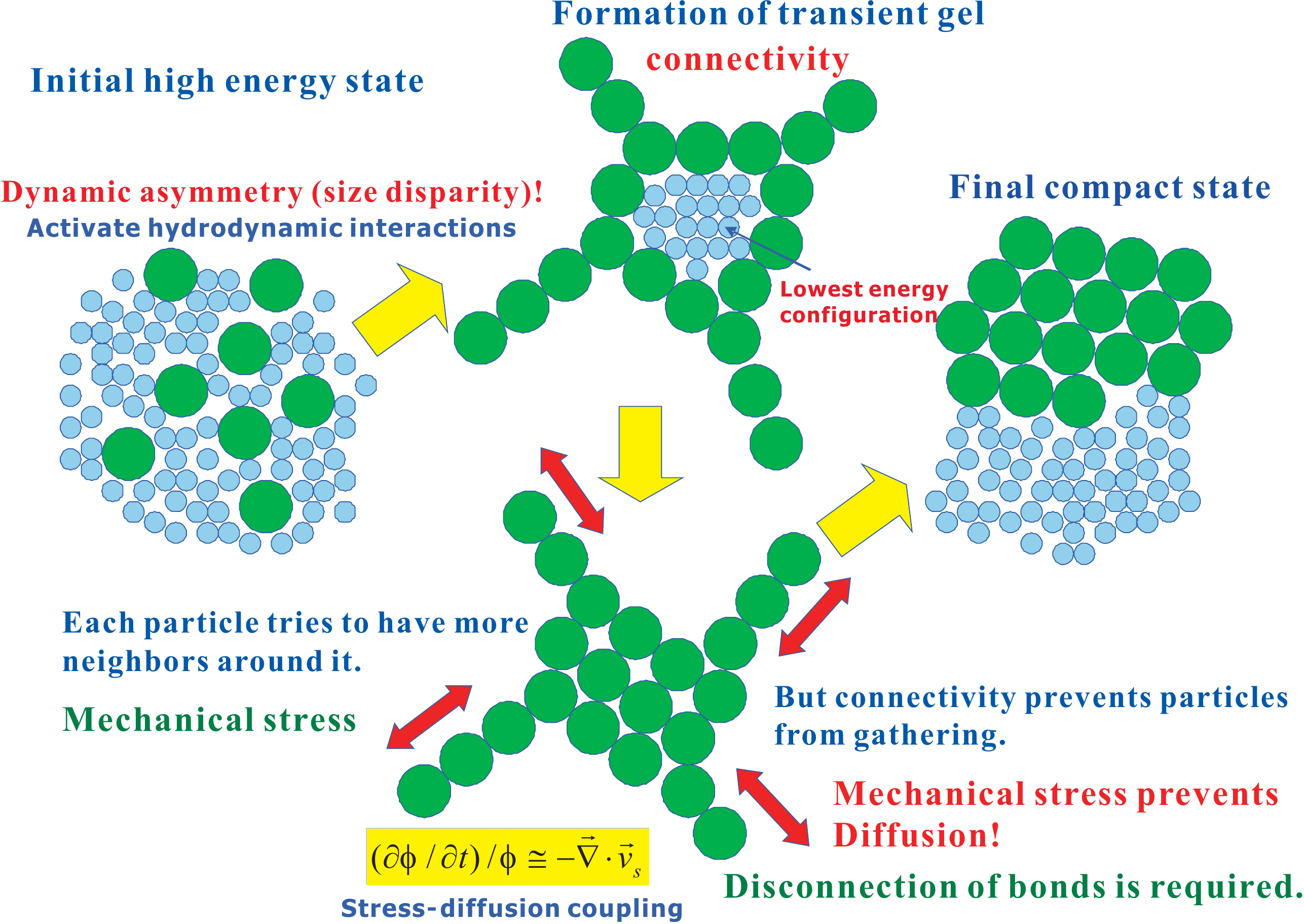}
\end{center}
\caption{Schematic figure explaining the physical mechanism of VPS in soft matter. Here, big green
particles represent slow components, such as polymers, colloids, and proteins, whereas small
blue particles represent solvent molecules. After initiation of phase separation, the small particles
can relax very quickly to the lowest-energy configuration. However, once the big particles are
connected to form a network structure with the help of hydrodynamic interactions, there is no
simple way to relax to the final lowest-energy configuration (the image furthest right). The
connectivity prevents big particles from lowering the contact energy by forming a more
compact structure. In other words, the diffusion is prevented by mechanical stress generated by
the connectivity: stress-diffusion coupling. This coarsening mechanism can be active even in the
absence of thermal noise (even at $T=0$), since it is of a purely mechanical nature \protect\cite{tanaka2007spontaneous}. 
This figure is reproduced from Fig. 1 of Ref. \protect\cite{tanaka2009formation}. 
}
\label{fig:TG}
\end{figure}

We proposed to introduce the bulk stress in addition to the shear stress to describe the volume shrinking behaviour of 
viscoelastic phase separation~\cite{tanaka1997roles,tanaka1997,tanaka1997general}.  
This stress expresses the connectivity of a transient gel formed in the early stage 
of viscoelastic phase separation and the resulting suppression of diffusion (see Fig. \ref{fig:TG}). 
Since the bulk stress is isotropic, it can be expressed by a scalar variable, namely, 
$\tilde{\sigma}=\frac{1}{d}{\rm Tr}\mbox{\boldmath $\sigma$}_B$: $\mbox{\boldmath $\sigma$}_B=\tilde{\sigma}\mbox{\boldmath$I$}$. 
Then, the bulk stress obeys the following equations: 
\begin{eqnarray}
\frac{D}{D t}\tilde{\sigma}=-\frac{1}{\tau_B(\phi)}\tilde{\sigma}
+G_B(\phi)\vec{\nabla}\cdot\vec{v}_p, \label{eq:bulk1}
\end{eqnarray}
where $\tau_B$ and $G_B$ are the relaxation time and the modulus of 
the bulk stress, respectively.

Here we discuss the rheological functions in the above constitutive equations. 
In the case of polymer solutions, 
$G_S(t)$ was estimated~\cite{doionuki,onuki1989elastic,helfand1989,milner1993}  
on the basis of rheological theories of polymer solution including the 
reptation theory~\cite{deGennes,doi} for good and $\theta$ solvents. 
The bulk stress related to $G_B(t)$ was not regarded to be important, since the longitudinal relaxation along a tube 
is much faster than the shear relaxation by reptation~\cite{doi}. This is true in good or $\theta$ solvents. 

However, elastic effects associated with the volume deformation may become important in polymer solutions under a poor solvent condition~\cite{tanaka1997roles,tanaka1997general}. 
It should be stressed that phase separation of polymer solutions always occurs 
in a poor-solvent condition. 
Thus, we cannot apply theories for polymers in good or $\theta$ solvents to our problem. 
In a poor solvent, there exists attractive interactions between polymer chains. Thus, we expect that there are 
temporal crosslinkings of energetic origin between polymer chains. 
The most natural model for polymer solutions under such a poor-solvent 
condition may be a transient gel model in which interpolymer 
attractive interactions produce temporal contact (crosslinking) points between polymer chains (see Fig. \ref{fig:TG}). 
Since such a connectivity is lost below a certain concentration $\phi^\ast$, there may be a steep concentration dependence 
of the bulk modulus on $\phi$. Thus, we mimic this situation by using the bulk modulus $G_B$, which is proportional to 
a step-like function $\Theta(x)$ 
($\Theta(x)=1$ for $x \geq 1$ and $\Theta(x)=0$ for $x<0$)~\cite{tanaka1997roles,tanaka1997general,tanaka1997}: 
$G_B(\phi)=G_B^0 \Theta(\phi-\phi^\ast)$, where $\phi^\ast$ is the threshold composition.  
We note that the introduction of a step-like $\phi$ dependence of the relaxation time $\tau_B$ has a similar effect. 

If we assume that the lifetime of temporal contacts between chains is 
$\tau_x$, we expect that the bulk relaxational 
modulus $G_B(t)$ has the relaxation time of the order of $\tau_x$. 
Then, the deformation described by $\vec{\nabla}\cdot \vec{v}_p$ 
that accompanies a change in the volume occupied by polymer chains 
causes bulk stress if the characteristic time of the deformation 
$\tau_d$ is shorter than $\tau_x$. 
However, since polymer dynamics in a poor solvent is far from being completely  
understood, we need further theoretical studies on this problem. 
We point out that this type of attractive interactions between 
molecules of the same component commonly exist in the unstable region 
of a mixture, which may generally result in formation of a transient gel 
in dynamically asymmetric mixtures.

We argued~\cite{tanaka1999colloid} that the same physics may be applied to particle-like systems such as colloidal suspensions, emulsions, and protein solutions 
on noting that under the action of attractive interactions particles tend to form a transient network with a help 
of hydrodynamic interactions~\cite{tanaka2000simulation,tanaka2006simulation,furukawa2010key}. 
The introduction of a steep $\phi$-dependence of $G_B$ allows us to include effects of transient gel formation and the resulting transient elasticity 
due to the gel-like connectivity~\cite{tanaka1997roles,tanaka1997general,tanaka1997}, as described above. 

Besides the above origin, there is a possibility that for particle suspensions 
the slow bulk stress relaxation may originate from hydrodynamic interactions 
under the incompressible condition: hydrodynamic squeezing effects~\cite{tanaka1999colloid}. What is the relative importance of the energetic and hydrodynamic 
origins in the bulk stress relaxation remains a problem for future investigation. 
This is related to the treatment of dissipation in the two-fluid description (see Sec. \ref{sec:twofluid}).

\subsection{A general rule of stress division between the components of a mixture} \label{sec:division}

The above perfect stress division only applies to a mixture of large size disparity. 
For example, in polymer blends~\cite{doionuki,onuki1994dynamic} and 
in a system where glass transition has 
a very different $T_{\rm g}$'s ~\cite{tanaka1997general}, mechanical stress is supported by both of the two components. 
In this case, the dynamical equations (\ref{eq:conserve})-(\ref{eq:bulk1}) must be generalized \cite{tanaka1997general}.  

Here we briefly discuss a general rule of the stress division in such a case. 
First, we introduce the rheologically relevant velocity $\vec{v}_r$, which appear 
in the constitutive relation. It is defined as   
$\vec{v}_r=\alpha_1 \vec{v}_1+\alpha_2 \vec{v}_2$, with 
$\alpha_1+\alpha_2=1$ \cite{onuki1994dynamic,tanaka1997general}. 
Here $v_k$ is the relative motion of component $k$ having the average velocity 
of $v_k$ to the mean-field rheological environment having 
the velocity of $\vec{v}_r$ and $\alpha_k$ is the stress division parameter. 
For simplicity, we neglect the transport and rotation of the stress tensor, which 
do not affect the pattern evolution so much since the transport and rotation are very slow in viscoelastic phase separation. 
In a linear-response regime, then, the most general expression 
of $\sigma_{ij}$ is formally written by introducing the 
time-dependent bulk and shear moduli in the theory of elasticity 
\cite{landau} as 
\begin{eqnarray}
\sigma_{ij}=\int^{t}_{-\infty} dt' [G_S(t-t')\kappa^{ij}_r(t')
+G_B(t-t') \vec{\nabla}\cdot \vec{v}_r (t') \delta_{ij}], \label{sigma}
\end{eqnarray}
where 
\begin{eqnarray}
\kappa_r^{ij}=\frac{\partial v_r^{j}}{\partial x_i} +
\frac{\partial v_r^{i}}{\partial x_j} -\frac{2}{d} 
(\nabla \cdot \vec{v}_r)\delta_{ij}. 
\end{eqnarray}
Here $\vec{v}_r$ is the velocity relevant to rheological 
deformation, and for polymer solutions $\vec{v}_r=\vec{v}_p$. 
$G_S(t)$ and $G_B(t)$ are material functions, which we call 
the shear and bulk relaxation modulus, respectively. 
It should be noted that the rheological relaxation functions 
$G_S(t)$ and $G_B(t)$ are functions of the local composition $\phi(\vec{r})$.  
We note that $G_B(t)$ is a purely mechanical 
modulus and different from the bulk 
osmotic modulus, $G_{os}=\phi^2 (\partial^2 f/\partial \phi^2)$. 
We have the relation $\eta= \int_{0}^{\infty} G(t) dt$, 
where $\eta$ is the viscosity of a material. 

The second term of the right-hand side of Eq. (\ref{sigma}) was introduced 
to incorporate the effect of volume change into 
the stress tensor~\cite{tanaka1997general,tanaka1997}.  
In a two-component mixture, the mode associated with 
$\vec{\nabla}\cdot \vec{v}_r$ can exist as far as $\vec{v}_r \neq \vec{v}$, 
even if the system is incompressible, $\vec{\nabla}\cdot \vec{v}=0$.  
As mentioned above, we proposed that this term plays a crucial role in viscoelastic phase separation~\cite{tanaka1997general,tanaka1997} (see also below), although 
it is not so important when we consider shear-induce demixing~\cite{doionuki,onuki1989elastic,helfand1989,milner1993}.  

Now we consider the stress division for the above general case. 
The local friction force is given by $\zeta_k(\vec{v}_r-\vec{v}_k)$, 
where $\zeta_k$ is the average local friction 
of component $k$ and the mean-field rheological environment 
at point $\vec{r}$, where the volume fraction of $k$ component 
is $\phi_k(\vec{r})$. 
Here $\zeta_k=\phi_k \zeta_k^{m}$ and 
$\zeta_k^{m}$ is proportional to the friction between an individual 
molecule or particle of the component $k$ and the mean-field rheological environment, 
which we call the generalized friction parameter. 
Because of the physical definition of the mean-field 
rheological environment, the two friction forces 
should be balanced. 
This fact guarantees that the rheological properties 
can be described only by $\vec{v}_r$.  
Thus, we have the following relation in general: 
\begin{eqnarray}
\zeta_1(\vec{v}_r-\vec{v}_1)+\zeta_2(\vec{v}_r-\vec{v}_2)=0. \label{balance}
\end{eqnarray} 
Then, the general expression of the stress division parameter $\alpha_k$ is obtained as  
\begin{eqnarray}
\alpha_k= \frac{\phi_k\zeta_k^{m}}{\phi_1\zeta_1^{m}+\phi_2\zeta_2^{m}}.  
\label{alphak}
\end{eqnarray}
The above relation is consistent with a simple 
physical picture that the friction is the only origin of the coupling 
between the motion of the component molecules and the rheological medium. 
We expect that this relation holds, irrespective of the 
microscopic details of rheological models, and, thus, 
we may apply it to a mixture of any material, the motion of both of whose 
components is described by a common mechanism. 
However, for theoretical estimation of friction coefficients, we need microscopic 
rheological theories, which are not available for a general case unfortunately.  
More importantly, as mentioned in Sec. \ref{sec:twofluid}, there is obscureness associated with the treatment of `nonlocal' hydrodynamic couplings 
in the coarse-graining procedure of the two-fluid model.

\subsection{Dynamic asymmetry in transport associated with glass transition}

Dynamic asymmetry affects not only the constitutive relation of a system, but also the kinetics of diffusion~\cite{tanaka1996,tanaka1997general}. 
This is particularly important in a mixture whose components having very different $T_{\rm g}$, since 
there is drastic slowing down of the dynamics towards $T_{\rm g}$, which gives rise to an extremely strong dependence 
of the diffusion coefficient $D$ on the composition $\phi$ near $T_{\rm g}$. 
Furthermore, formation of a transient gel is not expected for a mixture having little size disparity between its components. 
Formation of a transient gel or a gel should be specific to a mixture of large size disparity between its components. 
Thus, for a mixture whose components having very different $T_{\rm g}$, we do not expect a significant role 
of the bulk stress in suppressing the diffusion, unlike the case of a mixture of large size disparity (see above), 
since there may be no strong $\phi$-dependence of $G_B$. 
Even in this case, the strong $\phi$ dependence of $D$ can cause an effect similar to the bulk stress, as shown below. 

The $\phi$ dependence of $D$ near the glass transition point can be expressed by the following empirical Vogel-Fulcher-Tammann (VFT) relation: 
$D(\phi)=D_0 \exp(A\phi/(\phi_0-\phi))$, where $\phi_0$ is the VFT 
volume fraction and $A$ is the fragility index. 
Thus, we have to take into account this $\phi$-dependence of $D$, or the friction coefficient 
$\zeta$. Effects of a steep $\phi$-dependence of $D(\phi)$ were studied by numerical simulations~\cite{sappelt1997spinodal}. 
It should be noted that large bulk stress in the slow-component-rich phase (see above) 
and slow diffusion in the phase rich in the high $T_{\rm g}$ 
component play similar roles in phase separation: they both suppress 
rapid growth of the composition fluctuations and slow down 
the composition change in the more viscoelastic phase. 
Accordingly, a rate of the material transport between 
the two phases is limited or controlled by that in the slower phase. 
In this manner, a disparity in the diffusion coefficient $D$ 
between the two components of a mixture, i.e., a steep $\phi$-dependence of $D(\phi)$, has effects 
on phase separation, which are similar to those in the bulk relaxation modulus $G_B(\phi)$.

\subsection{Roles of the steep $\phi$ dependence of bulk stress and/or diffusion in viscoelastic phase separation}

Here we briefly discuss roles of the steep $\phi$ dependence of bulk stress and diffusion. The continuity equation 
\begin{equation}
\frac{\partial \phi}{\partial t}= -\vec{\nabla}\cdot (\phi \vec{v}_p)  
\end{equation}
tells us that it is $\vec{\nabla}\cdot \vec{v}_p$ that causes the composition change.  
The bulk stress caused by the deformation type of $\vec{\nabla}\cdot \vec{v}_p$, thus, suppresses 
growth of composition fluctuations if $\tau_d$ is shorter than $\tau_x$. 
In this way, the bulk stress is directly coupled with the composition change 
and the volume shrinking ~\cite{tanaka1997roles,tanaka1997general}. Note that the volume change of the 
polymer-rich (slow-component-rich) phase is directly associated with the deformation described 
by $\vec{\nabla}\cdot \vec{v}_p$. 
So the volume shrinking behaviour peculiar to viscoelastic phase separation is a consequence of (i) the slow bulk 
stress relaxation due to the connectivity of a transient gel formed by the large component and/or hydrodynamic squeezing effects 
or (ii) a steep composition dependence of the diffusion constant $D$.

\subsection{Beyond the linear constitutive relations}

As will be discussed later, viscoelastic phase separation accompanies mechanical fracture of the slow-component-rich phase 
and thus nonlinearity of rheology may also play an important role in the pattern evolution. 
In the above, we assume a simple Maxwell-type constitutive relation. However, we may need more complicated constitutive relations to describe the rheology of materials, since viscoelastic phase separation accompanies mechanical fracture 
of a transient gel, which intrinsically involves non-linear phenomenon. Two key deformation types in viscoelastic phase separation 
are elongation (uniaxial stretching) and extension 
(biaxial extension). The former is important in network-forming viscoelastic phase separation whereas the latter 
is in cellular one. 
The importance of strain hardening in the formation of cellular patterns has been recognized for both synthetic polymers \cite{spitael2004strain} 
and food polymers (e.g., breads) \cite{dobraszczyk2004physics,dobraszczyk1994strain}. 
Qualitatively, strain hardening makes the more viscoelastic phase mechanically more resistive to fracture, or makes the morphological selection due to mechanical force balance 
more robust (see Eq. (\ref{eq:balance})). 
Furthermore, the inclusion of the yielding behaviour to the constitutive relation is also of crucial importance in describing 
ergodic-to-nonergodic transitions such as gelation during phase separation (see Sec. \ref{sec:arrest}).  
It is highly desirable to incorporate these features into the constitutive relation for characterizing these nonlinear effects on 
a quantitative level.

\section{Dynamics of viscoelastic phase separation and the resulting pattern evolution}

\subsection{State diagram} 

\begin{figure}[!h]
\begin{center}
\includegraphics[width=8cm]{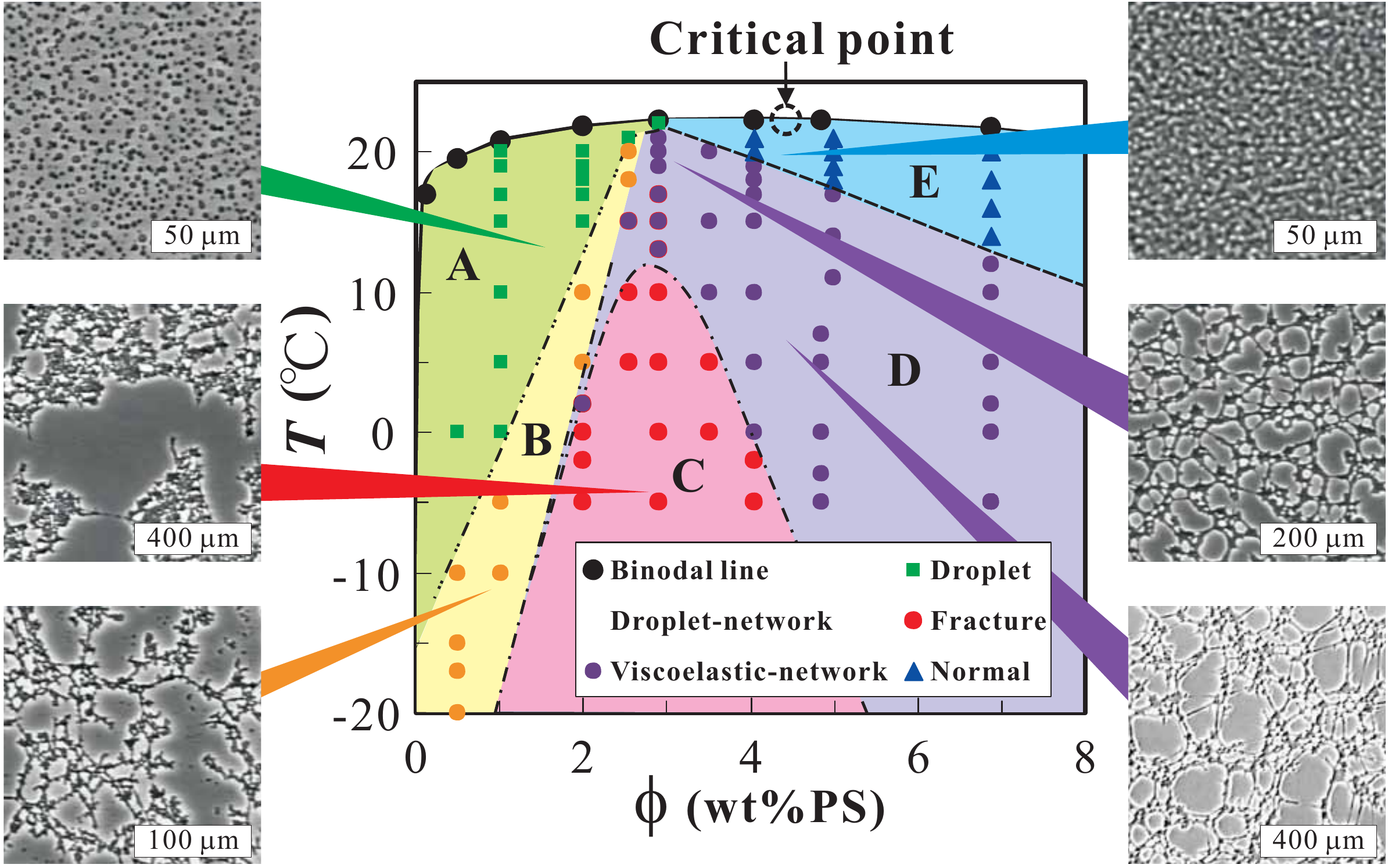}
\end{center}
\caption{State diagram and typical 
phase-separation patterns observed in a polystyrene (PS)/diethyl malonate mixture ($M_w=7.04 \times 10^5$).  
The gap between the two cover glasses was 5 $\mu$m. 
The boundaries of region C in the phase diagram are dependent upon 
the thickness of the cell, which is characteristic of elastic instability. 
This figure is reproduced from Fig. 1 od Ref. \protect\cite{koyama2009fracture}. 
} 
\label{fig:state}
\end{figure}

In normal phase separation of a dynamically symmetric mixture, there are only two types of pattern evolution in the unstable region: 
droplet spinodal decomposition for an asymmetric composition and bicontinuous spinodal decomposition for a symmetric composition. 
Thus, the morphological selection depends solely on $\phi$. 
Contrary to this, the morphological selection in a dynamically asymmetric mixture depends on not only $\phi$ but also the relation between 
$\tau_d$ and $\tau_t$. 
Here we show what kind of phase separation proceeds in a polymer solution, depending upon the quench condition ($\phi$, $T_f$), 
where $\phi$ is the composition and $T_f$ is the final temperature after a quench from a one-phase region. 
The classification based on microscopy observation of pattern evolution is summarized in Fig. \ref{fig:state}. 
In region A, droplet-type phase separation takes place. The polymer-rich phase 
appears as droplets. In region B, polymer-rich droplets are initially formed, 
but they aggregate and form a percolated network. After the percolation, 
the morphology is determined by the mechanical force balance in the network structure. 
In region C, phase separation proceeds by mechanical fracture (fracture phase separation). 
In region D, typical viscoelastic phase separation is observed. 
After the formation of a transient gel, the minority polymer-rich phase 
forms a network structure. In this region, phase inversion is observed: 
The initially majority phase eventually becomes the minority phase with time due to its volume shrinking. 
In region E, normal phase separation without accompanying phase inversion 
takes place. The solvent-rich phase appears as droplets here. 
Later we explain how phase separation proceeds in regions A-D.

\begin{figure}[!h]
\begin{center}
\includegraphics[width=10.cm]{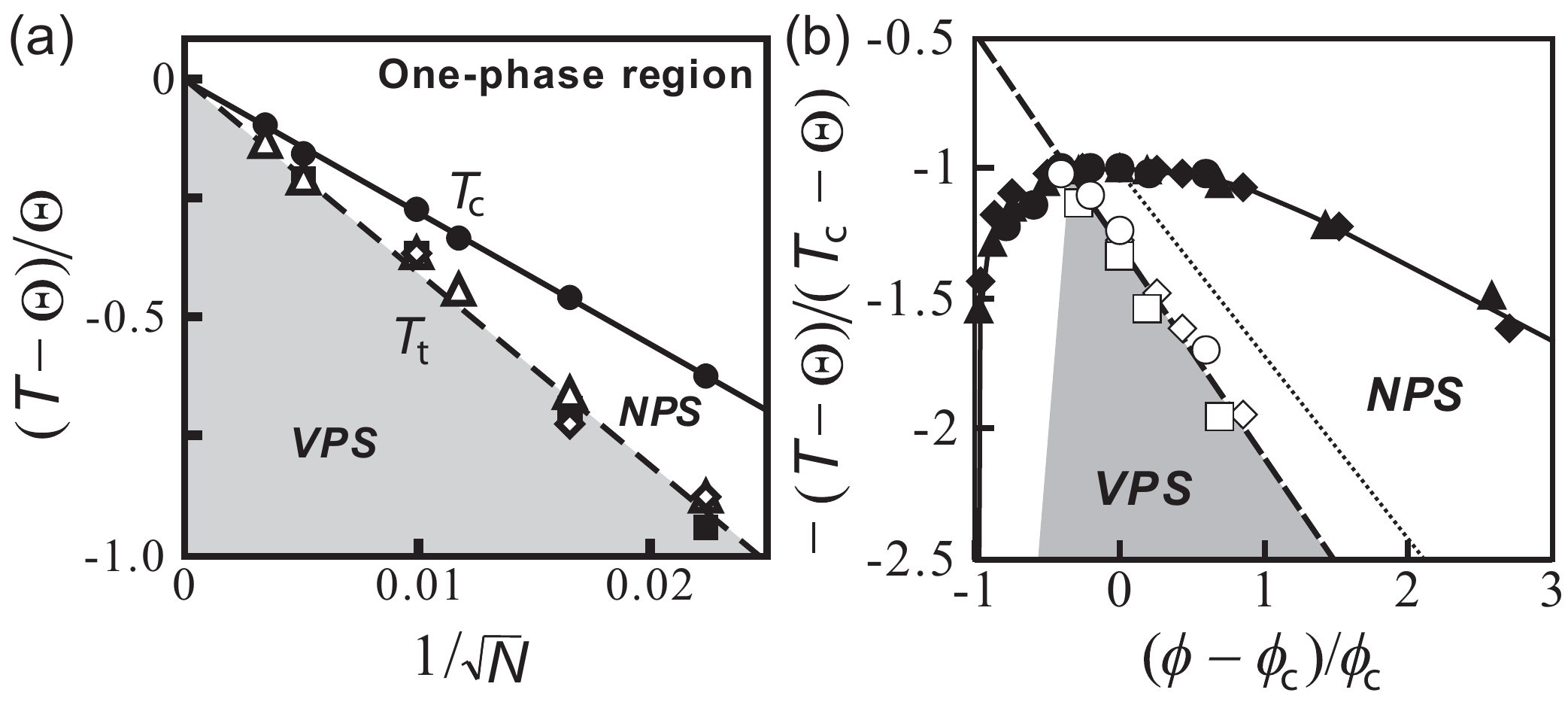}
\end{center}
\caption{
(a) $N$ dependence of $T_c$ and $T_t$ in critical solutions of polystyrene (PS) and 
diethyl malonate (DEM).
Filled circles: $T_c$. $T_t$, which separates regions of viscoelastic (VPS) and normal phase separation (NPS), were 
determined by morphological observation (triangles), the temporal 
change of $q_p(t)$ and $S(q_p(t))$ (see Fig. \ref{fig:Sk}) (diamonds), 
and the incubation time (filled squares). 
Solid and dashed lines are eye guides for $T_c$ and $T_t$, respectively. 
(b) The scaled static and dynamic phase diagrams for PS/DEM mixtures  
of three different degrees of polymerization $N$ of PS. 
The solid line is the binodal line, the dotted line is the symmetric line, 
and the dashed line is the $T_t$ line. 
Filled and open symbols are experimentally 
measured binodal temperatures and $T_t$'s, respectively. 
diamonds: PS-2 ($N=3.41 \times 10^3$); squares: PS-3 ($N=6.78 \times 10^3$); circles: PS-5 ($N=3.69 \times 10^4$).  
In the left-hand side of the VPS region, PS-rich droplets are 
observed: moving droplet phase (see below). 
Panels (a) and (b) are reproduced from Figs. 5 and 6 of Ref. \protect\cite{koyama2007generic}, respectively. 
}
\label{fig:Ndep}
\end{figure}

Next we consider how the critical temperature $T_c$ and the temperature $T_t$ (the border of region D and E (the dashed line)
in Fig. \ref{fig:state}), below which viscoelastic phase separation 
accompanying transient gelation is observed, 
depend upon the degree of dynamic asymmetry, which is controlled by the degree of polymerization $N$, for a critical polymer solution ($\phi=\phi_c$). 
In Fig. \ref{fig:Ndep}(a), we plot $T_c$ and $T_t$ as 
a function of $1/\sqrt{N}$ for polystyrene (PS)/diethyl malonate (DEM) mixtures. 
According to the standard theory of polymer solutions \cite{deGennes}, 
$T_c$ should be on the straight line in this plot, which is indeed confirmed. 
Interestingly, we found that $T_t$ is also 
on the straight line and furthermore the $T_c$ and $T_t$ lines 
meet at the $\Theta$ temperature in the limit of $N \rightarrow \infty$:
$\Theta-T_k=a_k/\sqrt{N}$, 
where $k$=$c$ or $t$ and $a_k$ is a constant. 
Thus, the $\Theta$ temperature, where binary interactions 
apparently disappear, is determined as $\Theta=32.0$ $^\circ$C. 
In the limit of $N \rightarrow \infty$, the polymer has an infinite molecular weight, 
and thus it can be regarded as a gel according to the definition 
based on the concept of percolation. The fact that the $T_c$ and $T_t$ lines meet at the $\Theta$ 
temperature in the limit of $N \rightarrow \infty$ implies a connection of a transient gel and a special gel at $N=\infty$, which 
looks physically very natural. 
This further suggests a general link between phase separation and transient gelation, both of which are 
driven by attractive interactions between like species. 

The fact that $T_c$ is on the straight line passing through the $\Theta$ temperature at $N \rightarrow \infty$ 
is a consequence of that the equilibrium phase diagrams of polymer solutions of various $N$ can be 
mapped on the master curve after scaling $\phi$ by $\phi_c \cong 1/\sqrt{N}$ 
and $T$ by $(\Theta - T_c) / \Theta \cong 2/\sqrt{N}$. 
Thus, the above result suggests an interesting possibility that the dynamic phase 
diagrams concerning the transient gelation can also be scaled together with the equilibrium phase diagrams. 

To check this possibility, we plot the dynamic phase diagrams for the three 
solutions of different $N$'s (PS-2, PS-3, PS-5) in the scaled form (see Fig. \ref{fig:Ndep}(b)). 
We use $- (T - \Theta)/(T_c - \Theta)$ as a reduced temperature. 
We find  that the dynamic phase diagrams are indeed scaled on the master one, implying the universal 
nature of this dynamic phase diagram including NPS and VPS. 
Note that $T_t$ is a decreasing function of the composition separating VPS from NPS, $\phi_t$: 
\begin{eqnarray}  
T_t \cong -0.8 (\Theta-T_c) \phi_t/\phi_c+0.5 (T_c+\Theta). 
\end{eqnarray} 
Thus, VPS occurs below the binodal line around $\phi\cong 0.7 \phi_c$. 
This finding may have an impact on our understanding of transient gelation in polymer solutions. 
We speculate that $T_t(\phi,N)$ may be determined by the dynamic 
crossover between the characteristic deformation rate and the 
characteristic rheological relaxation rate as a function of $\phi$ and $N$ 
\cite{tanaka1994critical,tanaka2000viscoelastic,araki20013D}. 
This remains a problem for future investigation.

\subsection{The early stage of viscoelastic phase separation}
\label{sec:early}

First we consider viscoelastic effects on the early stage of phase separation. 
We note that this theory applies to any regions in Fig. \ref{fig:state}, including normal and viscoelastic phase separation, 
as far as phase separation takes place in the unstable region 
(spinodal decomposition). 
For simplicity, here we do not consider a difference in the relaxation time 
between shear and bulk stress and assume $\tau_B=\tau_S=\tau$. 
Using the relation $\mbox{\boldmath$\nabla$} \cdot \mbox{\boldmath$v$}_p=
-\frac{1}{\phi}\frac{\partial \phi}{\partial t}$, we obtain 
the linearised equation for $Z_k=[\mbox{\boldmath$\nabla$}\cdot 
\mbox{\boldmath$\nabla$}\cdot 
\mbox{\boldmath$\sigma$}_p]_k$: 
\begin{eqnarray}
\frac{\partial Z_k}{\partial t} 
\cong -\frac{Z_k}{\tau} +\frac{2G}{\phi}k^2 
\frac{\partial \phi_k}{\partial t}, \nonumber 
\end{eqnarray}
where $G=G_B+\frac{4}{3}G_S$. 
Here $\phi_k$ is the Fourier component of the 
deviation from the initial composition $\phi_0$,  
and it obeys, to linear order,~\cite{doionuki,onukitaniguchi} 
\begin{eqnarray}
\frac{\partial \phi_k(t)}{\partial t} \cong -\Gamma_k \phi_k(t)
-\frac{2LGk^2}{\phi^2}\int^{t}_0 dt' e^{-\frac{t-t'}{\tau}} 
\frac{\partial \phi_k(t')}{\partial t'}. \label{phi}
\end{eqnarray}
Here we use the Ginzburg-Landau-type free energy 
$f(\phi)=k_BT[\frac{r_0}{2}(\phi-\phi_c)^2+\frac{u}{4}(\phi-\phi_c)^4]$. 
This form of the free energy is reasonable as far as 
we concern only a shallow quench near a critical point. 
Then $\Gamma_k=Lk^2(r_\phi+Ck^2)$ (see Sec. \ref{sec:dynamic}), 
where $L=\phi^2(1-\phi)^2/\zeta(\phi)$, 
is the decay rate in the absence of viscoelastic couplings.  
$r_\phi=r_0+3u(\phi_0-\phi_c)^2$, where $r_0=a (T-T_c)$ ($a$: 
a positive constant). 
The correlation length is given by $\xi=[\frac{C}{|r_\phi|}]^{1/2}$ in the mean-field theory. 
For a case when the time scale of $\phi_k$ change 
is slower than $\tau$, we can set $\frac{\partial \phi_k(t')}{\partial t'}=
\frac{\partial \phi_k(t)}{\partial t}$ in Eq. (\ref{phi}) and, thus, 
the growth rate of $\phi_k$ is given by 
\begin{equation}
A(k)=L|r_\phi| k^2(1-\xi^2k^2)/(1+\xi_{ve}^2k^2), \label{Aq}
\end{equation}
where $\xi_{ve}=(2 \eta L/\phi^2)^{1/2}$ is the so-called 
viscoelastic length~\cite{deGennes1976a,deGennes1976b,brochard1977} (see Sec. \ref{sec:DA}). 
Without viscoelastic couplings, the relation $A(k)=L |r_\phi| k^2(1-\xi^2k^2)$ should hold as the Cahn's linear 
theory~\cite{gunton} predicts. It was shown~\cite{takenaka2006viscoelastic} that the early stage of phase separation of a polymer solution 
is well explained by the above Onuki-Taniguchi theory~\cite{onukitaniguchi}. 

\begin{figure}
\begin{center}
\includegraphics[width=8cm]{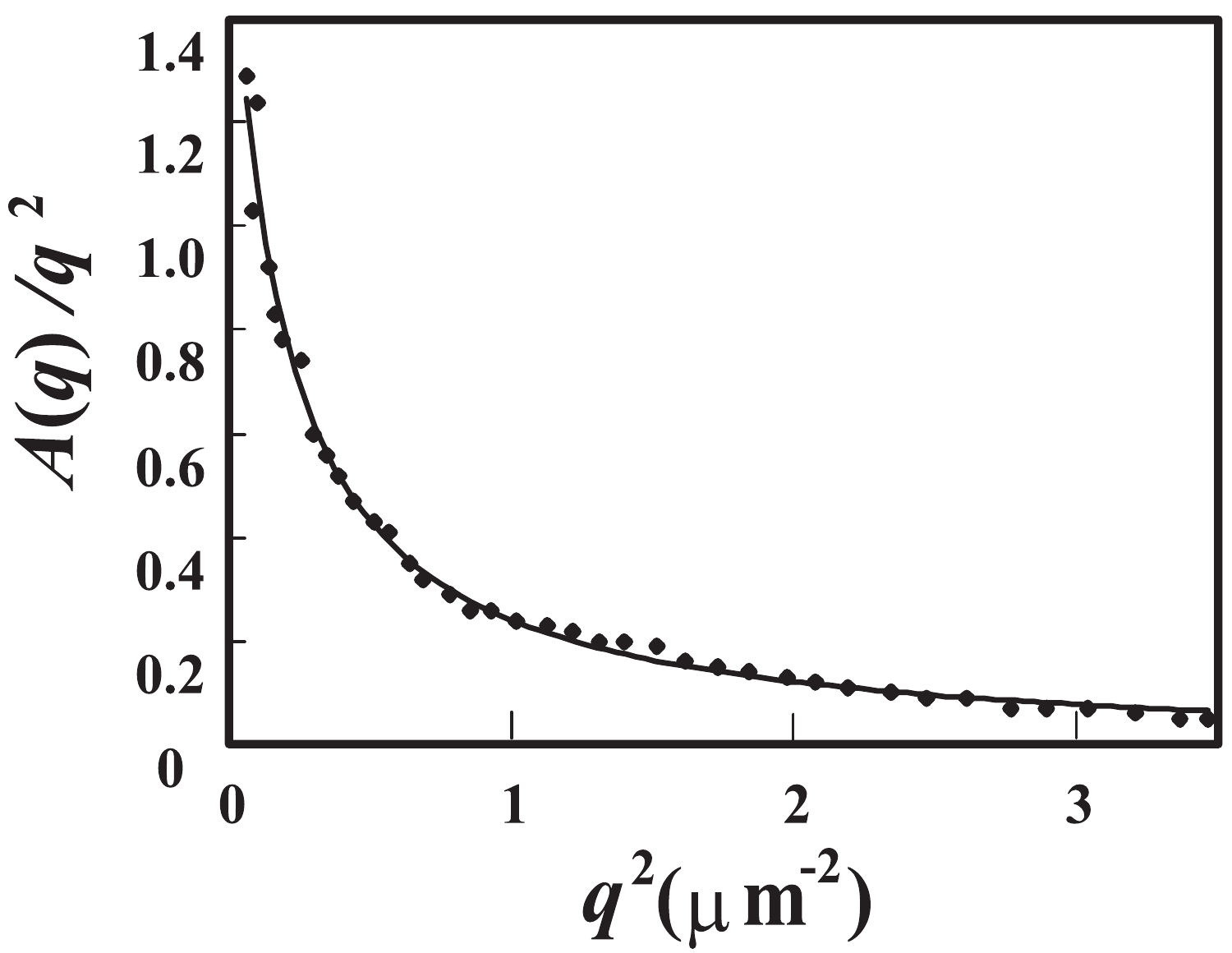}
\end{center}
\caption{Fitting of Eq. (\ref{Aq}) to the experimentally observed $A(q)$ for a colloidal suspension undergoing spinodal decomposition \protect\cite{verhaegh1996}. The solid curve 
represents a theoretical curve. This figure is reproduced from Fig. 3 of Ref. \protect\cite{tanaka1999colloid}.  
}
\label{fig:Aq}
\end{figure}

We emphasize that the early stage of phase separation in dynamically asymmetric mixtures including soft matter 
should be analysed by this theory. Applications of the Cahn's theory without considering viscoelastic effects 
may not be appropriate in many cases since $\xi_{ve}$ can easily become mesoscopic in dynamically asymmetric mixtures. 
In relation to this, we note that the above relation 
[Eq. (\ref{Aq})] also well explains 
the unusual $q$-dependence of $A(k)$ experimentally observed 
in colloid phase separation~\cite{tanaka1999colloid} (see Fig. \ref{fig:Aq}). 
This suggests the relevance of the viscoelastic model to phase separation not only in polymer solutions, 
but also in colloidal suspensions, emulsions, and protein solutions, 
which further indicates the importance of viscoelastic effects in any dynamically asymmetric mixtures 
including oxides and metals~\cite{tanaka2000viscoelastic}.

\subsection{Network-forming viscoelastic phase separation}
\label{sec:general}

\begin{figure}[!h]
\begin{center}
\includegraphics[width=11cm]{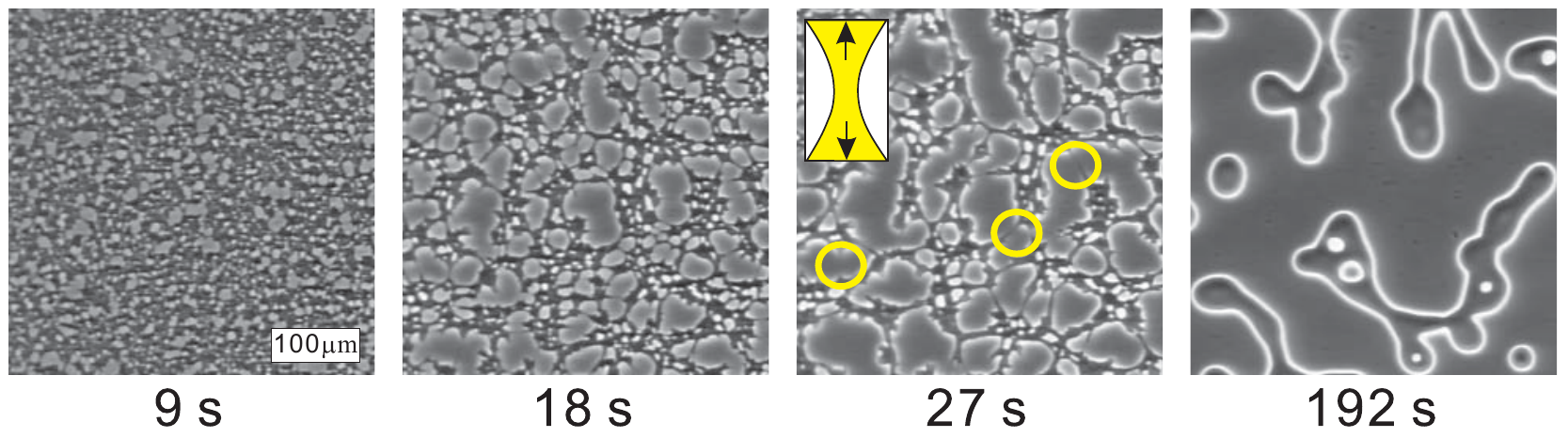}
\end{center}
\caption{Network-forming viscoelastic phase separation 
of a polystyrene (PS)/diethyl malonate mixture ($M_w=7.04 \times 10^5$). 
It is observed after a quench to 
15 $^\circ$C for a mixture (2.91 wt\% PS). 
The gap between the two cover glasses 
was 5 $\mu$m. 
In the yellow circles, we can see that 
the polymer-rich phase is elongated 
under the stretching force generated by its volume shrinking 
(see also the small schematic figure). 
This resembles a typical liquid-type or ductile fracture of material 
under elongational deformation. 
This figure is reproduced from Fig. 2 of Ref. \protect\cite{koyama2009fracture}. 
}
\label{fig:NPS}
\end{figure}

Here we describe the process of pattern evolution in typical viscoelastic phase separation, 
which are observed in region D of the state diagram in Fig. \ref{fig:state}. 
The phase separation in this region is characterized by (i) the formation of a network structure 
of the minority phase (see Fig. \ref{fig:NPS}), which is contrary to the well-accepted belief of normal phase separation 
that the minority phase always forms droplets to minimize the interfacial energy, and (ii) phase inversion .  

In normal phase separation, the late stage phase separation is discussed on the basis of the scaling concept, 
which relies on the fact that the volume of each phase is conserved in the late stage and thus there is only one characteristic length scale, i.e., the domain size, in a system: {\it self-similarity of pattern evolution}. 
For viscoelastic phase separation, however, such a scaling concept is not valid because of the volume shrinking 
of the slow-component-rich phase during phase separation. 
The absence of self-similarity is obvious in Fig. \ref{fig:NPS}. 
Because of this difficulty, there has been no analytical 
theory of domain coarsening so far. 
In the following, thus, we describe pattern evolution in phase separation on a qualitative level. 

\begin{figure}
\begin{center}
\includegraphics[width=10cm]{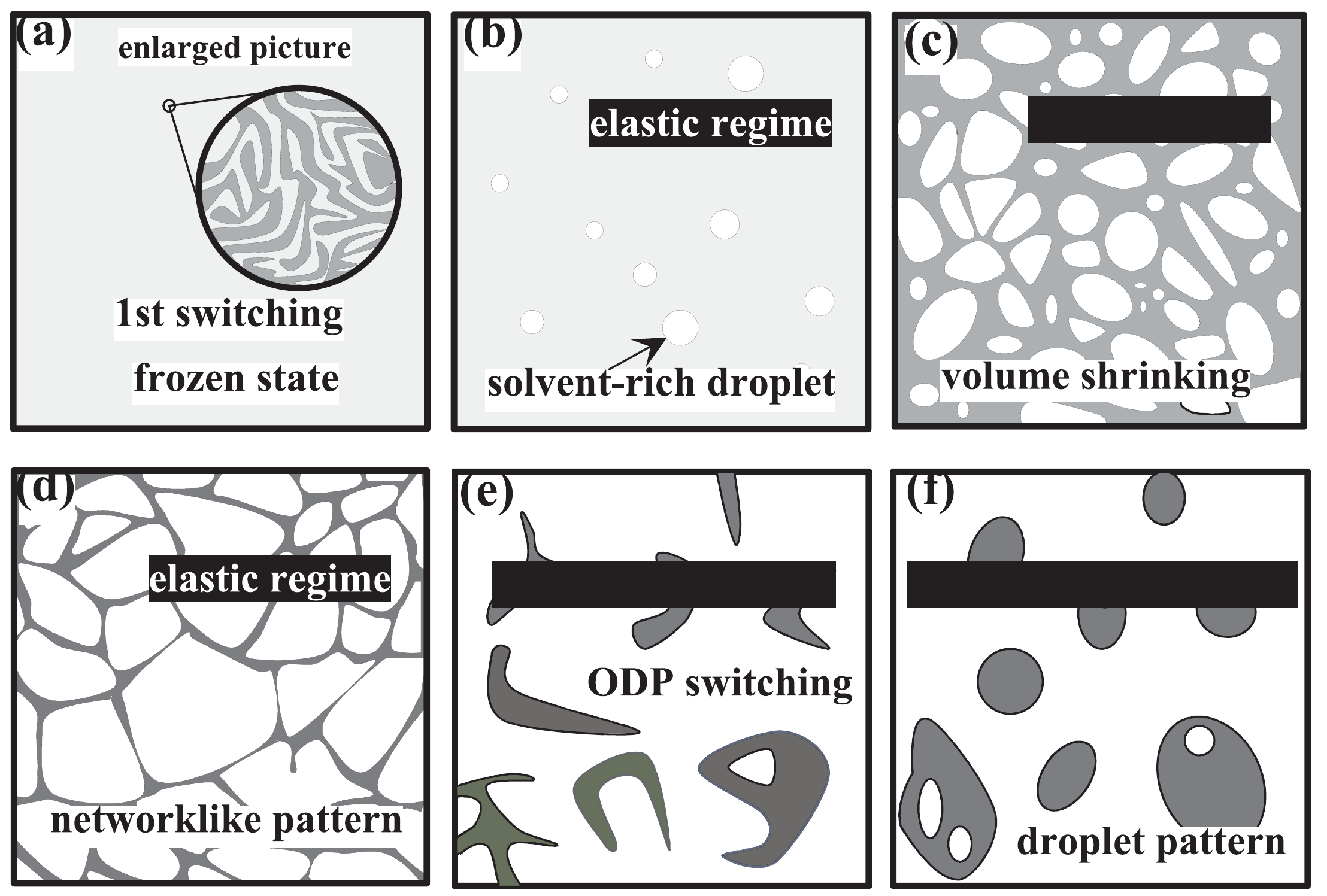}
\end{center}
\caption{Schematic figure showing the characteristic features of the pattern evolution during viscoelastic phase 
separation of mixtures having nearly critical composition. This figure is reproduced from Fig. 9 of Ref. \protect\cite{tanaka2006simulation}. }
\label{fig:schematicVPS}
\end{figure}

The process of viscoelastic phase separation is schematically shown in Fig. \ref{fig:schematicVPS}. 
Just after the temperature quench, a mixture first becomes cloudy due to the growth of composition fluctuations 
(see Sec. \ref{sec:early}), then 
after some incubation time, small solvent holes 
start to appear (see Fig. \ref{fig:schematicVPS}(b)). We call this incubation period the ``frozen period'', which 
is the stage of transient gel formation. 
Then the number and the size of solvent holes  
increase with time. In this process, the slow-component-rich matrix phase shrinks by expelling the fast liquid component to solvent holes, which leads to the growth of holes made of the fast-component-rich 
phase. In this way, the matrix phase keeps shrinking its volume and becomes network-like or sponge-like, 
(see Figs. \ref{fig:schematicVPS}(c) and (d)). 
In this volume-shrinking process, the bulk mechanical stress plays a crucial role~\cite{tanaka1997roles,tanaka1997}. 
Thin parts of a network-like structure are elongated and eventually broken and disconnected. 
This results in the release of mechanical stress and allows local mechanical relaxation. 
In this network-forming process, 
the pattern is dominated by the mechanical shear force balance condition and thus the shear stress plays a major role~\cite{tanaka1997}. 
In the final stage, a network-like structure tends to relax to a structure of rounded shape and the domain shape starts to be 
dominated by the interface tension as in usual fluid-fluid 
phase separation (see Figs. \ref{fig:schematicVPS}(e) and (f)). Domains finally become spherical. 
When the slow-component-rich phase is the minority phase, thus, there is a phase inversion during phase separation. 
This phase inversion is a characteristic feature of viscoelastic phase separation. 

If the concentration of the slow-component-rich phase reaches the glass transition composition and the yield stress of the slow-component-rich phase exceeds the mechanical stress generated, a structure is dynamically arrested and becomes stable. 
This may be regarded as the general scenario for formation of colloidal gels (see below) 
\cite{tanaka1999colloid,cates2004theory,lu2008gelation,PaddyNM}

\begin{figure}[!h]
\begin{center}
\includegraphics[width=8cm]{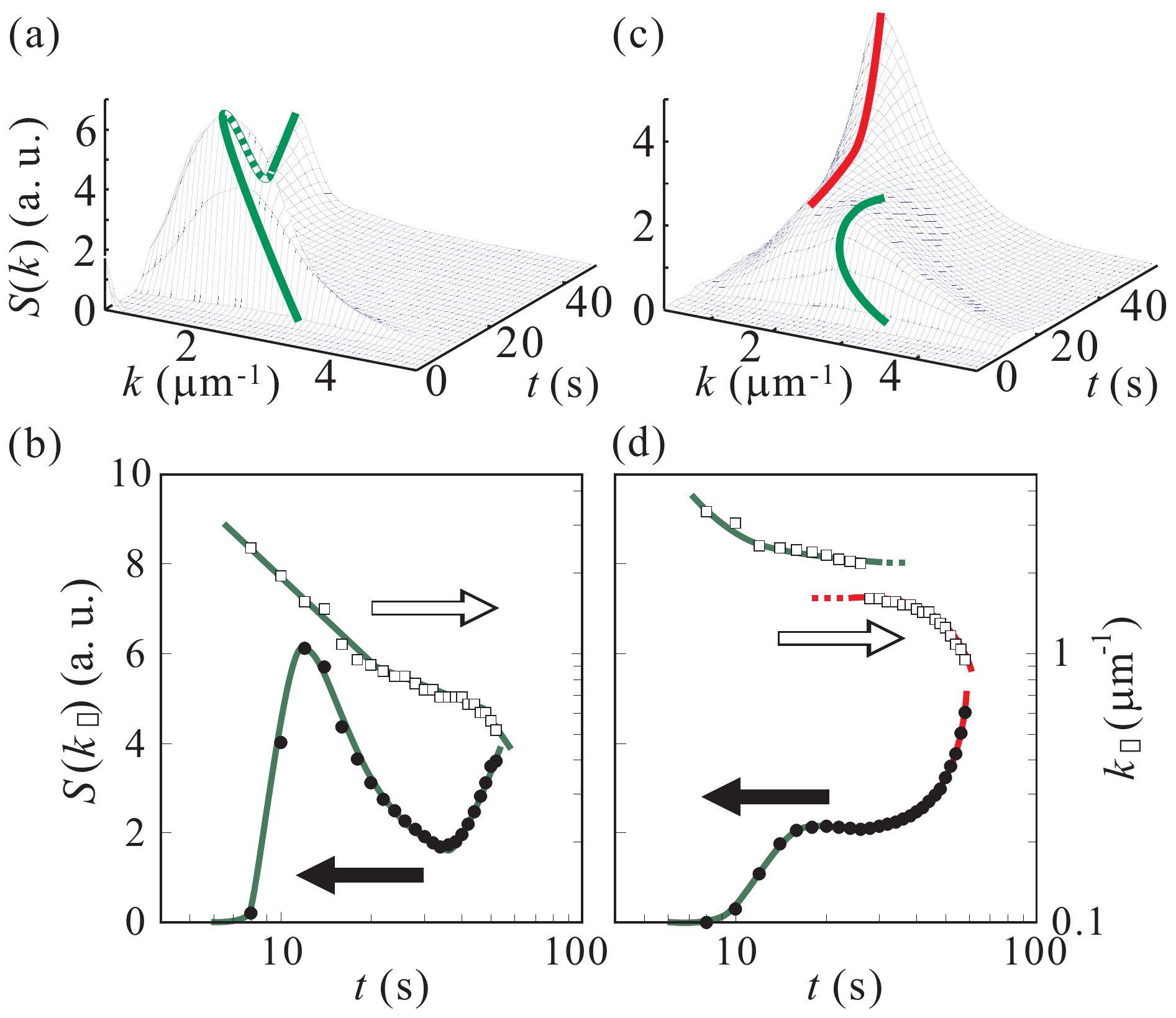}
\end{center}
\caption{(colour online) 
Temporal change of $S(k)$, $k_p(t)$ (squares), and 
$S(k_p(t))$ (circles) in the critical 
solution of PS/DEM mixture ($M_w =1.09 \times 10^6$, $N=1.05 \times 10^{4}$, $\phi_c=4.11$ \%, $T_c=23.5$ $^\circ$C). 
(a),(b): $\Delta T = 1.5$ K (shallow quench). 
(c),(d): $\Delta T = 5.5$ K (deep quench). 
Solid lines are eye guides. $S(k)$ is calculated by digital image analysis (DIA) \protect\cite{tanaka1986application}. 
This figure is reproduced from Fig. 3 of Ref. \protect\cite{koyama2007generic}. 
}
\label{fig:Sk}
\end{figure}

Now we describe the temporal evolution of the structure factor $S(k)$ in PS/DEM mixtures. 
For a shallow quench (Figs. \ref{fig:Sk}(a) and (b)), 
fluid phase separation affected by wetting to walls was observed. 
In this case, $S(k_p(t))$, where $k_p$ is the peak wavenumber, initially
increases, then decreases and increases again, which is
characteristic of phase separation of a thin film of a binary
mixture under wetting effects \cite{tanaka2001interplay}. The first increase
reflects initial bicontinuous phase separation in bulk. Then the
decrease reflects the following process. The DEM-rich
phase is more wettable to the glass walls. Thus it is transported towards the glass walls by a hydrodynamic pumping
mechanism \cite{tanaka2001interplay} such that a bicontinuous phase-separated
structure transforms into a three-layer structure. Accordingly, a phase-separation structure tends to disappear
in the observation plane. The final increase reflects the
growth of the remaining DEM-rich droplets bridging the wetting layers. 
During this process, $k_p(t)$ monotonically decreases, since
the domains keep growing via a hydrodynamic mechanism \cite{tanaka2001interplay}. 

For a deep quench (Figs. \ref{fig:Sk}(c) and (d)), on the
other hand, we observed viscoelastic phase separation. In this case, $S(k_p(t))$ increases initially, then has a plateau,
and finally increases again. Dynamic arrest of the coarsening by transient gelation of the PS-rich phase leads to
the plateau of both $S(k_p)$ and $k_p$. Finally, shrinking of the
PS-rich transient gel increases the concentration difference
between the two phases, which results in the increase in
$S(k_p)$. After the formation of a transient gel, a large-scale
structure due to ``elastic instability'' is superimposed
onto the initial phase-separation structure and becomes
more and more dominant with time,
which leads to a double-peaked shape of $S(k)$. This
results in the crossover of $k_p(t)$ from one branch to
another (Figs. \ref{fig:Sk}(c) and (d)). Thus, the border between viscoelastic and normal phase separation 
(between regions E and D in Fig. \ref{fig:state}) can be determined as
the temperature where $S(k_p(t))$ and $k_p(t)$ change their
behaviour between the above two types. 
We note that this two-step phase separation has not been captured by numerical simulations 
on the basis of the viscoelastic model yet. This might be due to a too steep $\phi$ dependence of the bulk modulus and/or 
the setting of $\phi^\ast$ to the average composition of a mixture in our simulations. 

According to the common sense of normal phase separation, after the formation 
of a sharp interface between the coexisting phase (namely, 
in the so-called late stage) the concentration of each phase 
almost reaches the final equilibrium one and, thus, there should be no change 
in the volume and concentration of each phase~\cite{gunton,onuki,siggia1979late}. 
Thus, the sharp front formation of the interface before the saturation of the order parameter to the equilibrium values  
is a very interesting feature of viscoelastic phase separation. 
We believe that it is a consequence of the steep $\phi$ dependence of $D(\phi)$ or the bulk stress, 
which makes the diffusion in a dilute region much faster than that in a concentrated region. 
We pointed out~\cite{tanaka2000viscoelastic} that 
the volume decrease of the more viscoelastic phase 
with time after the formation of a sharp interface 
is essentially the same as the volume shrinking 
of gels during volume phase transition~\cite{matsuo1988kinetics,matsuo1992patterns,sekimoto1989sponge}. 
The physical reason of this similarity to gel will be discussed later. 

The scaling law established in normal phase separation is a direct consequence of the 
conservation of the volumes of the two phases after the formation of a sharp interface 
and the resulting self-similar growth of domains. 
The volume shrinking of the slow-component-rich phase inevitably leads to the absence 
of self-similarity during viscoelastic phase separation and thus the absence of an extended scaling regime. 
Nevertheless, we observe a transient scaling law (the characteristic domain size $R \sim t^{1/2}$) in the intermediate coarsening stage 
for a few systems~\cite{tanaka2005network,tanaka2005protein,tanaka2007spontaneous}, although its physical 
mechanism remains elusive. 

In sum, the whole pattern evolution process can be clearly divided into three 
regimes: the initial, intermediate, and late stages. 
The crossovers between these regimes can be explained 
by viscoelastic relaxation in pattern evolution and the 
resulting switching of the primary order parameter, as will be described below (see \ref{sec:switching}). 

\begin{figure}[!h]
\begin{center}
\includegraphics[width=8cm]{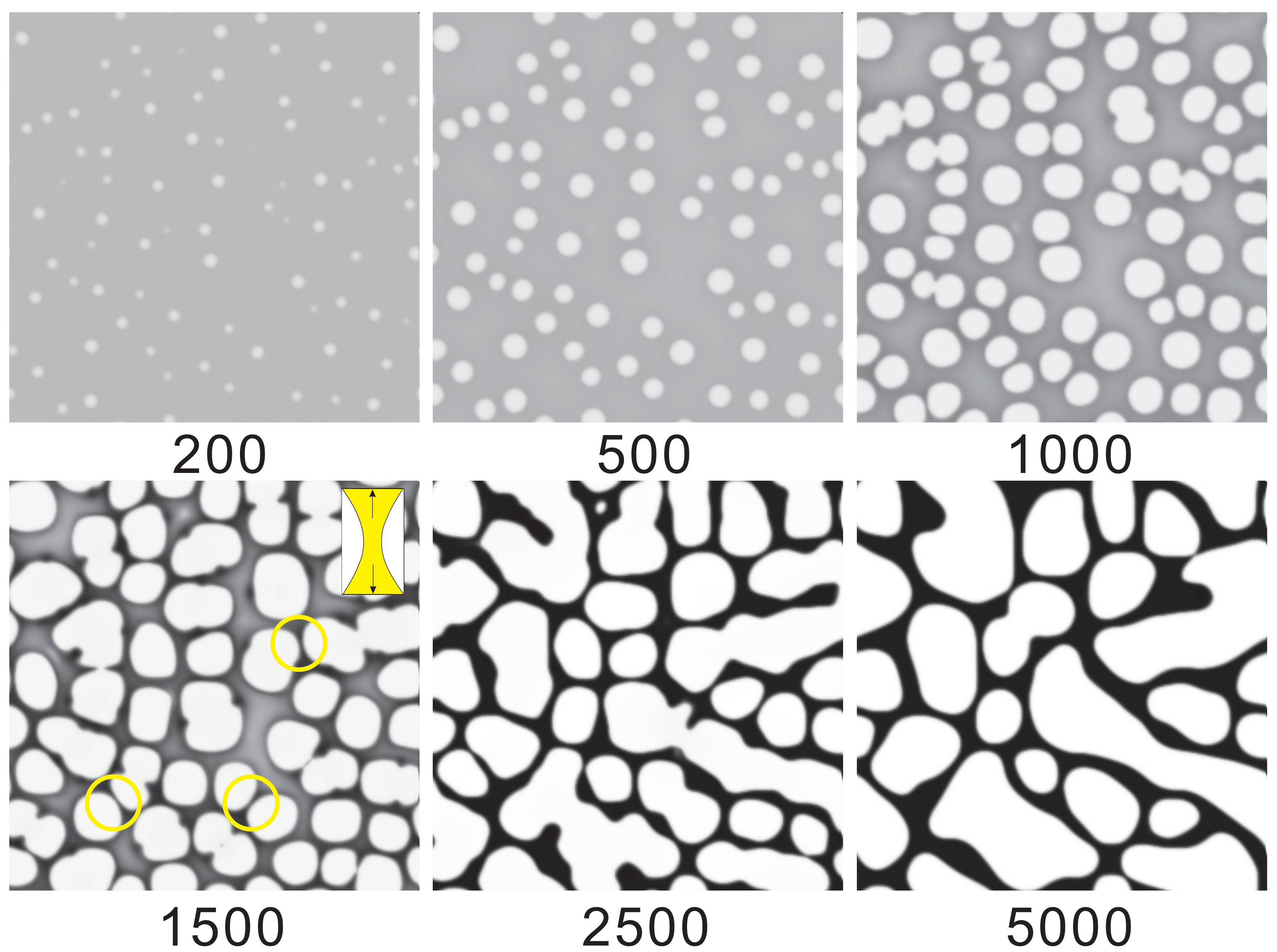}
\end{center}
\caption{2D pattern evolution of viscoelastic phase separation 
simulated by our model.  
Here $\phi_0=0.35$, $G_B=5 \theta(\phi-0.35)$, 
$G_S=0.5  \phi^2$, and $\tau_B=\tau_S=50 \phi^2$. 
This should be compared with the experimental results in Fig. \ref{fig:NPS}, for which the same annotation was used.  
} 
\label{fig:visco}
\end{figure}

\begin{figure}[!h]
\begin{center}
\includegraphics[width=8cm]{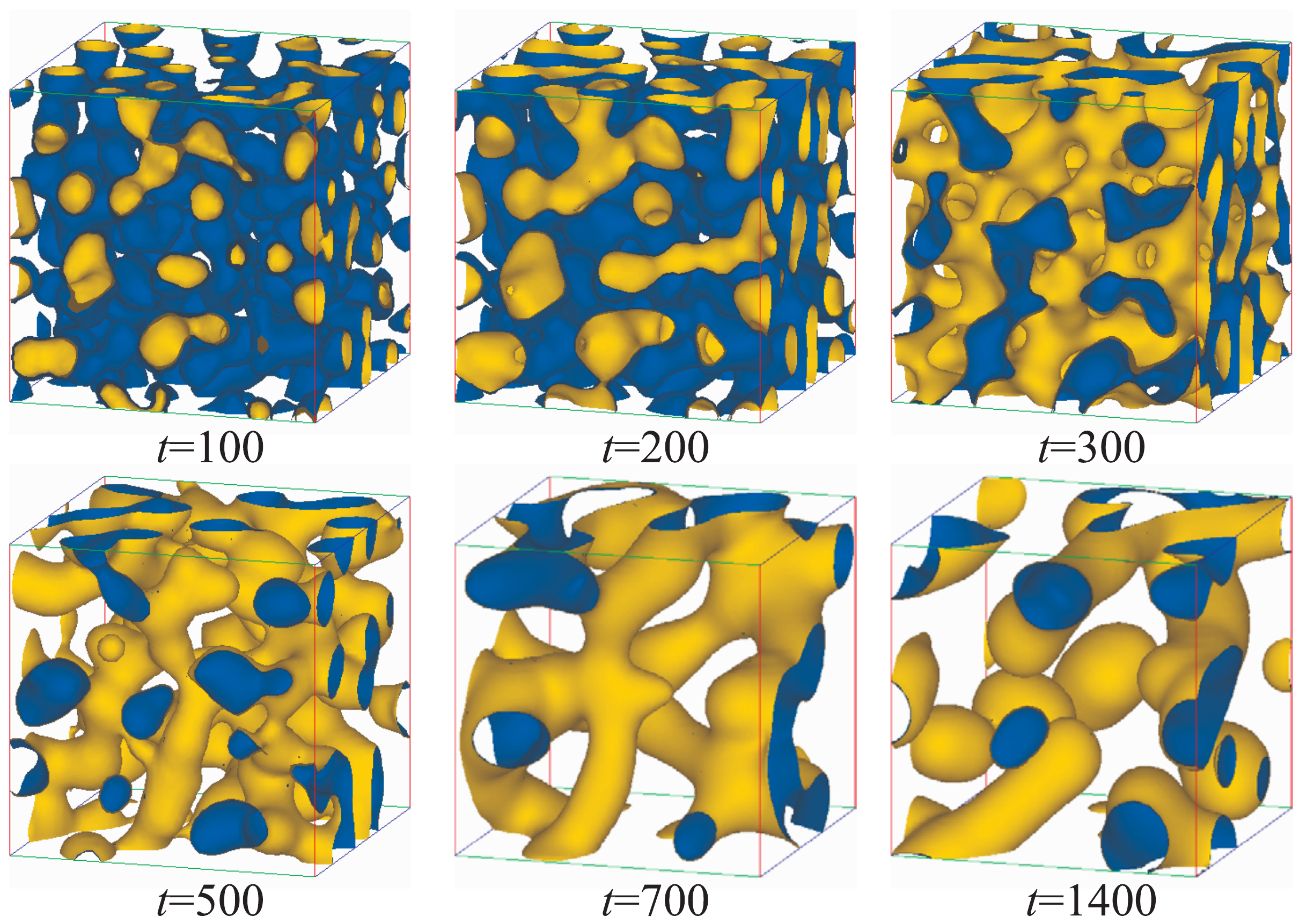}
\end{center}
\caption{An example of 3D pattern evolution of 
viscoelastic phase separation simulated by our model. We can see a well-developed network structure at $t=700$.
This figure is reproduced from Fig. 23 of Ref. \protect\cite{tanaka2006simulation}. 
} 
\label{fig:3DVPS}
\end{figure}

Besides the early stage, we do not have any reliable analytical predictions and thus 
numerical simulations based on the viscoelastic model play a crucial role in its understanding 
\cite{taniguchi1996network,tanaka1997,bhattacharya1998,araki20013D,araki2005simple,tanaka2006simulation,tanaka2007spontaneous,zhang2001kinetics}. 
We showed that a steep composition dependence of the bulk modulus or the diffusion constant is the key to volume shrinking and the resulting phase inversion and 
a rather smooth $\phi^2$-dependence of the shear modulus is responsible for the formation of a network-like structure 
\cite{tanaka1997,araki20013D,tanaka2006simulation}. A typical numerical simulation result of the process of viscoelastic phase separation is shown in Fig. \ref{fig:visco}. An example of 3D pattern evolution of 
viscoelastic phase separation obtained by numerical simulations is also shown in Fig. \ref{fig:3DVPS} (see Refs. \cite{araki20013D,tanaka2006simulation}). 

\begin{figure}[!h]
\begin{center}
\includegraphics[width=7cm]{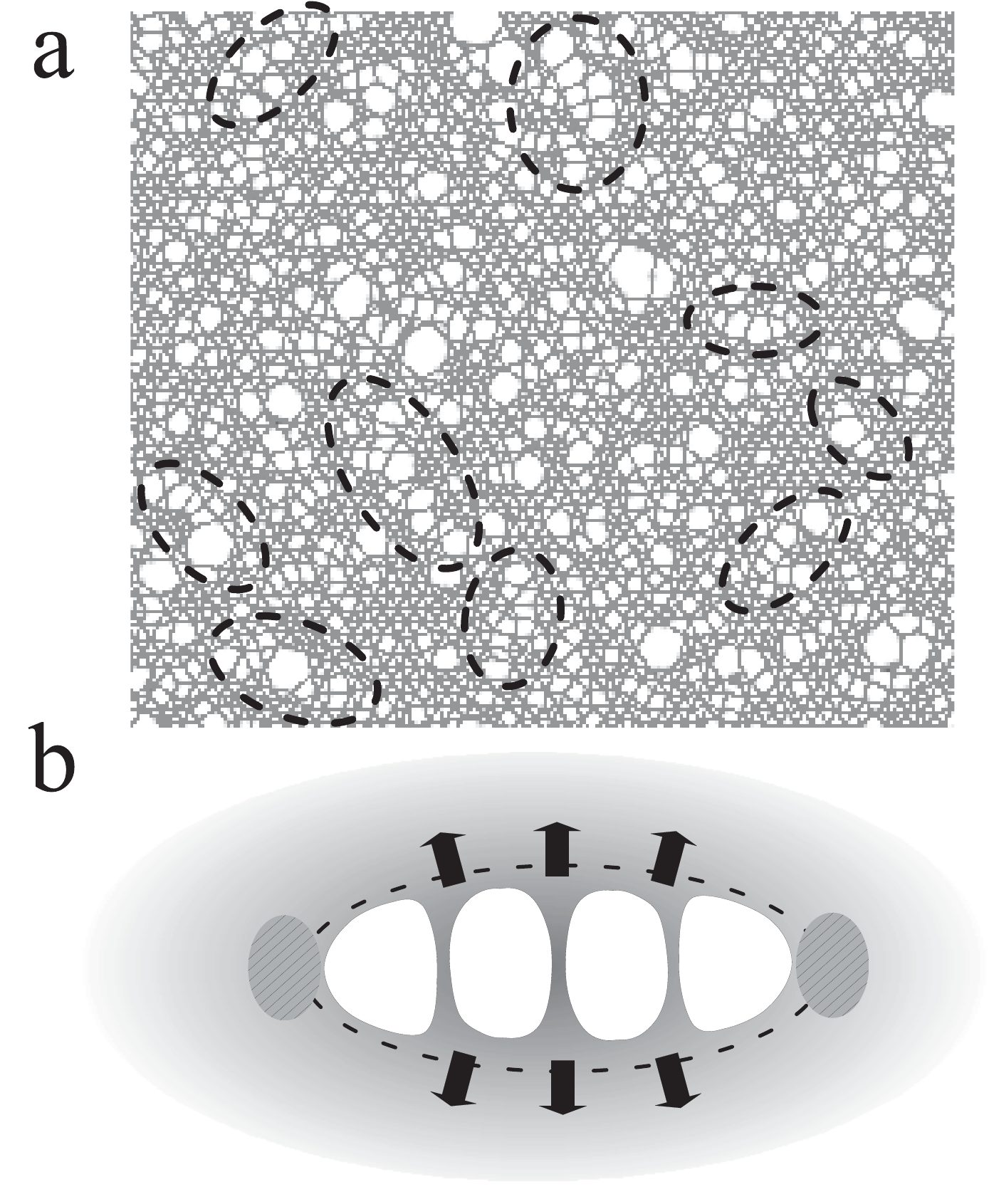}
\end{center}
\caption{(a) Arrays of nucleated droplets observed in the simulation
of a network of disconnectable strings. (b) Schematic picture
of an array of nucleated droplets and the elastic force field in the shrinking
viscoelastic matrix phase. 
This force field helps nucleation of solvent holes at the edges of the ellipsoidal region (the hatched region). 
This figure is reproduced from Fig. 11 of Ref. \protect\cite{araki2005simple}.}
\label{fig:spring}
\end{figure}

Finally we discuss the spatial correlation in the formation of solvent holes in the slow-component-rich phase. 
One-dimensional arrays of droplets are often formed in viscoelastic phase separation 
(see Fig. \ref{fig:spring}(a), which was simulated by our disconnectable spring model \cite{araki2005simple}. A similar
pattern is also often observed in experiments.
This indicates the presence of spatial elastic coupling in the hole (or, solvent droplet) formation. Figure \ref{fig:spring}(b) schematically shows such an
array of nucleated solvent-rich droplets and the mechanical forces acting on the shrinking transient gel. The formation of a droplet
array can be explained as follows. 
First it is energetically more favourable to have two solvent droplets nearby rather than independently since the former situation 
costs less elastic energy than the latter one. 
Then, if two droplets are nucleated close to each other in this way, the deformation
field around them, which is induced by the shrinking of the more viscoelastic matrix phase, becomes
anisotropic. The stress is more concentrated at the edges of the array of droplets. This helps formation of solvent-rich holes there 
(see the hatched regions in Fig. \ref{fig:spring}(b)), and thus leads to a further
increase in the number of droplets in the array. The overall domain shape of the array is approximated by an elongated
ellipsoidal droplet (see the ellipsoids of the broken curve in Fig. \ref{fig:spring}(a)), 
if we neglect the thin bridges between droplets. This elastic coupling between emerging droplets 
is a characteristic of mechanically-dominated pattern evolution.

\subsubsection{Mechanical nature of the coarsening of a network structure}

\begin{figure}[!h]
\begin{center}
\includegraphics[width=11cm]{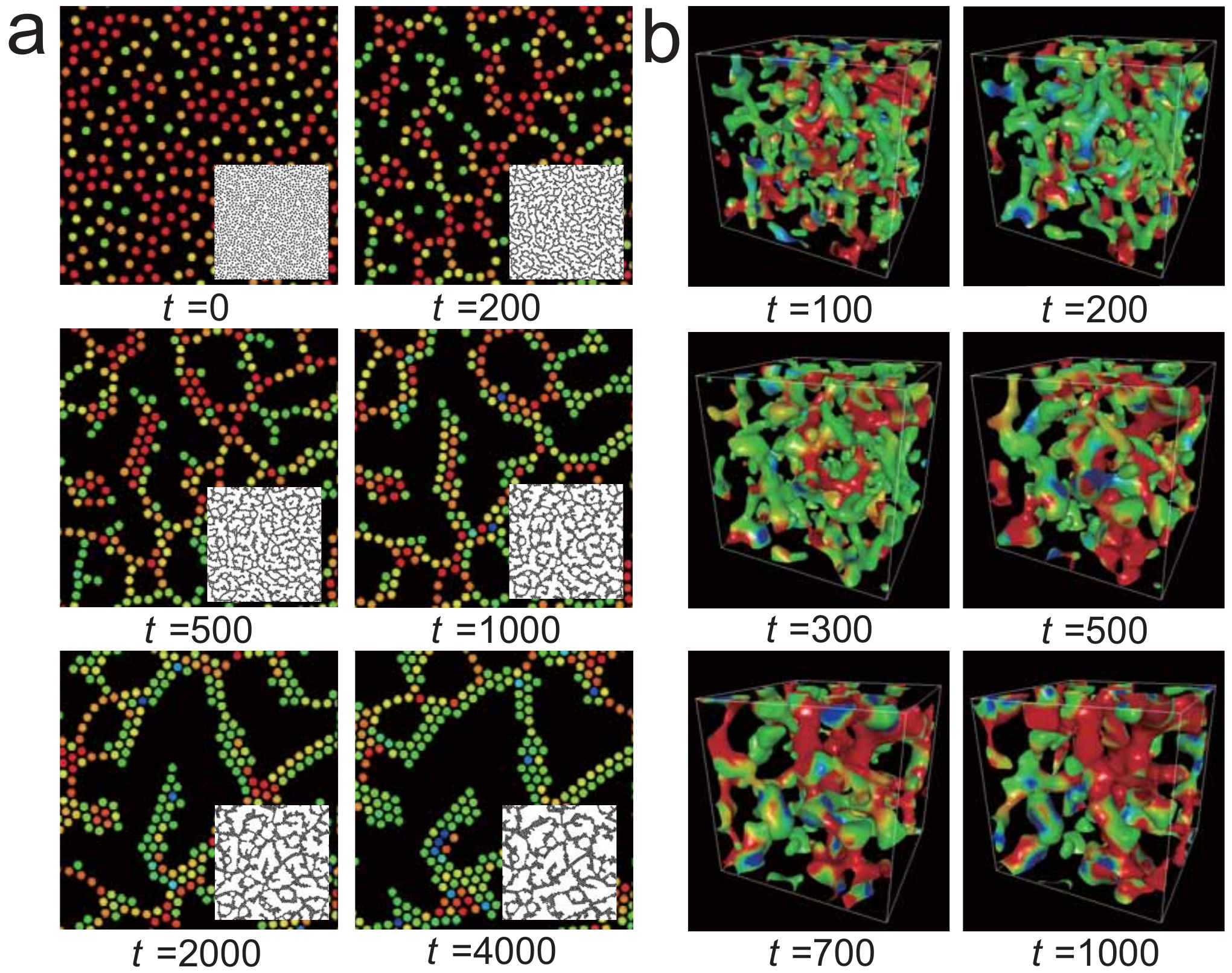}
\end{center}
\caption{
Phase separation processes of colloidal suspensions interacting 
with the Asakura-Oosawa potential, whose range is characterized 
by $R=d_p/D_p$, where $D_p$ is the particle diameter and $d_p$ is the range of the potential. 
(a) 2D pattern evolution (the volume fraction is 0.248 and $R=0.7$). 
(b) 3D (coarse-grained) pattern evolution 
(the volume fraction is 0.100 and $R=0.6$). 
We coarse-grained structures by replacing a particle by a Gaussian field and extracting the interface by applying 
a black\&white operation to the field.  
The details of the simulations  
are described in Ref. \protect\cite{tanaka2007spontaneous}. 
In both (a) and (b), red particles are stretched and in a high energy state, whereas blue particles are in a low energy state. }
\label{fig:colloidsimu}
\end{figure}

In normal phase separation, the late-stage coarsening is driven by the thermodynamic force 
coming from the interface energy. 
In viscoelatic phase separation, on the other had, it is driven mainly by the mechanical force 
in the volume shrinking stage, whereas by the thermodynamic force in the final stage where 
the order parameter reaches its equilibrium values and the mechanical force decays almost completely.    
We revealed that the mechanical stress accumulated in a network structure leads to its coarsening 
by repeating the following sequence: stress concentration on a weak part of the network, its break up and 
the resulting stress relaxation, and structural 
rearrangements towards a structure of lower interface energy~\cite{tanaka1997,araki20013D,tanaka2006simulation}. 
Such examples can be seen in 
2D and 3D colloid simulations, as shown in Fig. \ref{fig:colloidsimu}. 
We stress that this process can proceed without any thermal activation. Actually, the simulations in Fig. \ref{fig:colloidsimu} 
were performed without any thermal noises, namely, at $T=0$. This indicates that the coarsening of 
network-type viscoelastic phase separation can proceed purely mechanically: mechanically driven coarsening. 
This is markedly different from a conventional picture based on the activation-type coarsening process. 
It has been widely believed that no coarsening proceeds if the activation barrier exceeds 10 $k_{\rm B}T$, but this argument 
cannot be applied to a network structure formed by viscoelastic phase separation.  
Stress concentration can provide a way to overcome such a barrier. 
We emphasize that mechanically driven coarsening cannot be characterized by the strength of attractive interactions measured by 
the thermal energy $k_{\rm B}T$ alone.

\subsubsection{Universality of viscoelastic phase separation to dynamically asymmetric mixtures} 

\begin{figure}[!h]
\begin{center}
\includegraphics[width=7cm]{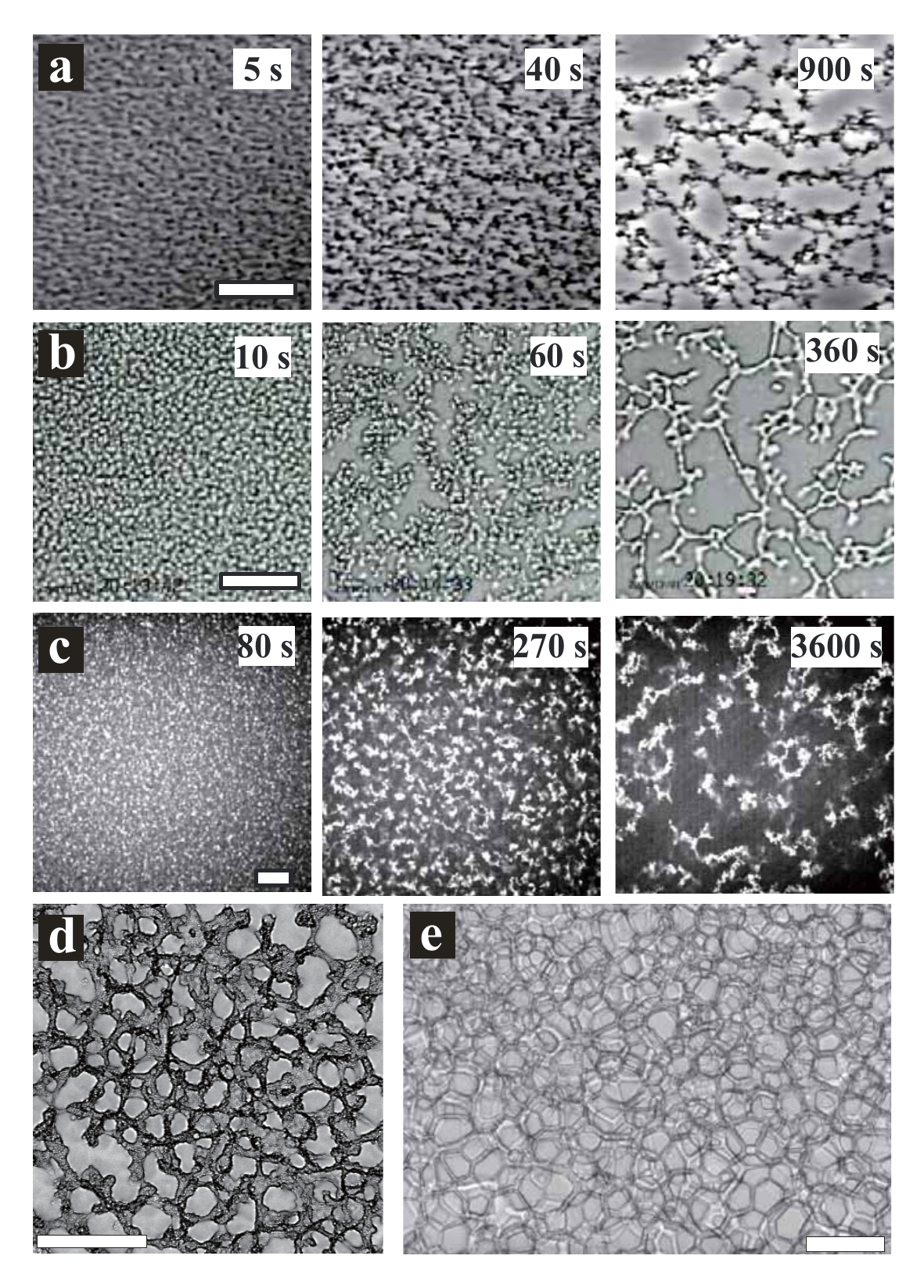}
\end{center}
\caption{
Comparison of pattern evolution observed in (a) a polymer solution, (b) a protein solution, (c) a colloidal suspension, and (d) a surfactant solution. The scale bar corresponds to 50 $\mu$m for (a)-(c). (a) is observed in a mixture of polystyrene (its molecular weight being 706,000) and diethyl malonate (0.5 wt\% polystyrene) at 20 $^\circ$C with phase-contrast microscopy. (b) is observed in a protein solution ($\phi$=200 mg/ml, salt concentration 7.5 wt\%) at 37.0 $^\circ$C with conventional optical microscopy. (c) is observed in a colloidal suspension ($\phi$=0.25 v\%,  salt concentration 15 wt\%) with confocal microscopy. (d) Network pattern observed for a surfactant (C$_{10}$E$_3$)/water mixture (19.9 wt\% C$_{10}$E$_3$) at 42.7 $^\circ$C  (5 s after the stop of heating with  the rate of 16.0 K/min). The scale bar corresponds to 500 $\mu$m. 
 The patterns observed and their temporal changes in (a)-(d) are striking similar to each other. This strongly suggests that viscoelastic phase separation may be universal for any dynamically asymmetric mixtures.
(e) A 3D cellular structure observed for a surfactant (C$_{12}$E$_5$)/water mixture of 19.9 wt\% C$_{12}$E$_5$ at 74.0 $^\circ$C. The sample thickness was 140 $\mu$m. The scale bar corresponds to 200 $\mu$m. This figure is reproduced from Fig. 2 of Ref. \protect\cite{tanaka2009formation}. 
} 
\label{fig:univ}
\end{figure}

Finally, we emphasize that pattern evolution in viscoelastic phase separation is essentially the same between the two types of dynamically asymmetric mixtures~\cite{tanaka2000viscoelastic,tanaka2006simulation,tanaka2009formation}: one is a system like polymer solutions 
\cite{tanaka1992,tanaka1993,tanaka1994critical,koyama2007generic}, colloidal suspensions~\cite{tanaka2005network}, and protein solutions ~\cite{tanaka2005protein}, where the strong dynamic asymmetry comes from a large difference in the molecular size and topology between the components, and the other is a system whose components have a large difference 
in the glass transition temperature~\cite{tanaka1996}.  Here we show viscoelastic phase separation observed in 
four different systems, polymer solutions (a), 
colloidal suspensions (b), and protein solutions (c), and surfactant systems (d) and (e) in Fig. \ref{fig:univ}. 
We can see the universal nature of the pattern evolution in any dynamically asymmetric mixtures, (a)-(d). 
A pattern in (e) is different from the others because of an interplay between viscoelastic phase separation and 
smectic ordering (see Sec. \ref{sec:smectic}). 

\subsection{Viscoelastic droplet phase separation: Moving droplet phase}

Here we mention another type of viscoelastic phase separation which takes place in a mixture with a low volume fraction 
of the slow component. 
This type of phase separation is observed in region A of the state diagram in Fig. \ref{fig:state}. 
At such a low volume fraction, the slow-component-rich phase 
immediately forms droplets while shrinking their size by expelling a solvent. 
This shrinking process is finished rather quickly because of the small sizes of droplets in the time scale of $R^2/D$ 
($R$: the droplet size). 
Thus, the volume fraction inside droplets may rapidly reach a strongly entangled jammed state. 
Such situations are realized also if the final volume fraction is higher 
than the glass transition volume fraction. 
If the collision timescale is shorter than the structural relaxation time, droplets behave as elastic glassy balls.  
This may also be expressed as follows. If the contact time during droplet collision due to Brownian motion 
is shorter than the material transport between droplets, droplets do not coalesce and stay without growing 
for a fairly long time. 
We referred to this interesting metastable state due to the elastic or glassy nature of droplets 
as ``moving droplet phase (MDP)''~\cite{tanaka1988,tanaka1992,tanaka1994critical,tanaka2000viscoelastic}. 

\begin{figure}[!h]
\begin{center}
\includegraphics[width=10cm]{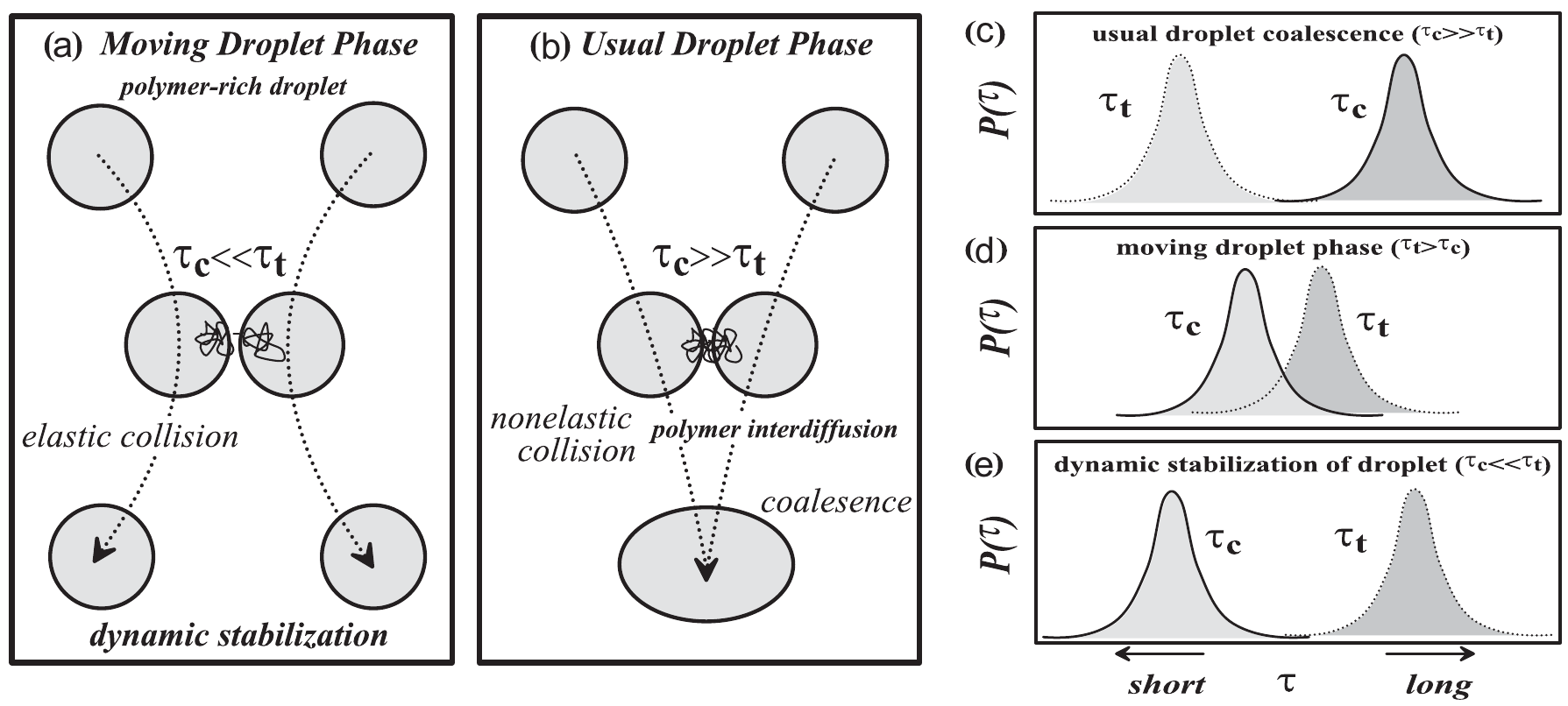}
\end{center}
\caption{
Schematic figures showing the elementary process of 
droplet collision and the resulting coalescence. (a) MDP with dynamic 
stabilization of droplet and (b) usual droplet coarsening. 
In case (a), polymers cannot inter-diffuse between droplets 
during $\tau_{\rm c}$ and the collision may be elastic. 
In case (b), on the other hand, polymers can inter-diffuse 
between droplets within $\tau_{\rm c}$ and droplets can coalesce 
with each other by Brownian coagulation mechanism. 
We also show schematic figures showing 
the relation between $P(\tau_{\rm t})$ and $P(\tau_{\rm c})$ for 
droplet phase separation. 
(c) Usual droplet phase separation; (d) MDP with coarsening; 
(e) MDP without coarsening. Droplets could be dynamically 
stabilized in this case. Panels (a)-(c) are reproduced from Figs. 18 and 19 in Ref. \protect\cite{tanaka2000viscoelastic}. 
}
\label{fig:mdp}
\end{figure}

Here we consider the droplet stabilization mechanism using polymer solutions as an example. 
The two important time scales characterizing the situation 
may be the characteristic time of the collision between two droplets (or 
the contact time) $\tau_{\rm c}$ and the characteristic 
rheological time of the polymer-rich phase $\tau_{\rm t}$. 
Viscoelastic effects should play a role 
when $\tau_{\rm c}$ is shorter than or comparable to $\tau_{\rm t}$. 
Brownian motion of a droplet with mass $m$ 
is characterized by a randomly varying thermal velocity 
of magnitude $\langle v \rangle \sim (k_{\rm B}T/m)^{1/2}$ and duration 
$\tau_{\rm r} \sim mD_R/k_{\rm B}T$ ($D_R$: the diffusion constant 
of a droplet with radius $R$).  
Thus $\tau_{\rm c}$ should satisfy the relation 
$r_i/\langle v \rangle <\tau_{\rm c}<r_i^2/D_R$, 
where $r_i$ is the range of interaction. 
On the other hand, $\tau_{\rm t} \sim \eta_{\rm s} b^3N^3\phi^{\alpha}
/k_{\rm B}T$, where $b$ is the segment size. 
For the typical values of the parameters, 
$\tau_{\rm t}$ could be longer than $\tau_{\rm c}$ for a large $N$ or 
for a deep quench, in particular, if attractive interactions between chains are 
taken into account. 
This means that a droplet may behave as an {\it elastic 
body} on the collision time scale for $\tau_{\rm t}>\tau_{\rm c}$. 
For $\tau_{\rm t}<\tau_{\rm c}$, on the other hand, droplets 
can coalesce with each other. 
This viscoelastic effect is probably responsible for 
the slow coarsening and the unusual dependence 
of the coarsening rate on the quench depth. 
Since $\tau_{\rm t}$ is strongly dependent on $N$ and $\phi$ of 
a droplet phase, it is natural that this phase exists 
only for a polymer solution having a large $N$ under 
a deep quench condition. 
Figures \ref{fig:mdp}(a) and (b) schematically show the elementally process 
of droplet collision and the resulting coalescence for MDP and 
that for usual liquid-like droplet phase, respectively. 
To consider the coarsening behaviour in more detail, 
we need the information on the distribution functions of 
$\tau_{\rm c}$ and $\tau_{\rm t}$, which are expressed by 
$P(\tau_{\rm t})$ and $P(\tau_{\rm c})$, respectively. 
The typical situations are schematically shown in Figs. \ref{fig:mdp}(c)-(e).  
The coarsening rate may be determined by the relation between 
$P(\tau_{\rm t})$ and $P(\tau_{\rm c})$. 
For $\tau_{\rm t}/\tau_{\rm c} \ll 1$, the coarsening process is similar to 
usual binary liquid mixtures 
and described by the Brownian 
coagulation mechanism \cite{gunton,siggia1979late}. 
With an increase in $\tau_{\rm t}/\tau_{\rm c}$, the coarsening rate should 
become slower and finally MDP might be kinetically stabilized for 
$\tau_{\rm t} \gg \tau_{\rm c}$. The stabilization of MDP may be almost 
complete for the case when $\tau_{\rm t} \gg \tau_{\rm c}$ and 
$P(\tau_{\rm t})P(\tau_{\rm c})\sim 0$.

This phenomenon may be used to make rather monodisperse particles whose size is the order of sub microns to microns. 
The monodisperse nature is a direct consequence of the formation of droplets due to the growth of concentration fluctuations 
with a characteristic wavenumber and little coarsening after that. 
So this phenomenon may provide us with a new very simple and low cost method to make particles with a desired size, 
which may be useful in both soft materials and foods industries. 
For example, we speculate that the formation of elastic particle gels of proteins \cite{van2010elastic} 
may share a common mechanism with the moving droplet phase. Recently, it was also shown 
that even random nonionic amphiphilic copolymers can form stable aggregates, a mesoglobular phase between
individual collapsed single-chain globules and macroscopic precipitation \cite{wu2004viscoelastic}. 

Here we note that if the droplet concentration becomes too high, droplets are no more stable and tend to aggregate 
to form networks due to inter-droplet attractions~\cite{tanaka2005network}. After the formation of network, the behaviour is similar to 
viscoelastic phase separation. This may be regarded as two-step viscoelastic phase separation: 
the formation of elastic gel balls followed by network formation.  
Such behaviour is observed in region B. This suggests that viscoelastic phase separation 
is also observed in colloidal suspensions \cite{tanaka1999colloid}.

\subsection{Pattern evolution in fracture phase separation}

Here we show a special case of viscoelastic phase separation, where phase separation proceeds accompanying 
brittle mechanical fracture of a mixture. This type of phase separation is observed in region C of the state diagram 
in Fig. \ref{fig:state}. 

\subsubsection{Physical mechanism}

\begin{figure}[!h]
\begin{center}
\includegraphics[width=11cm]{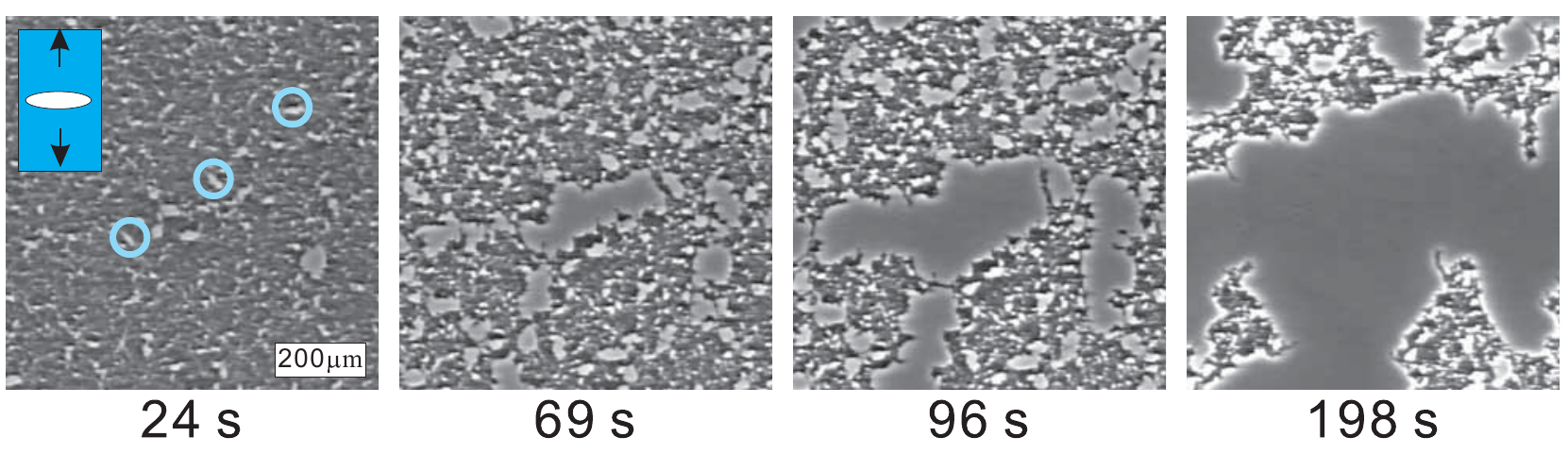}
\end{center}
\caption{Fracture phase separation observed in a PS/DEM mixture ($M_w=7.04 \times 10^5$; $\phi=$2.91 wt\% PS) after a quench to 0 $^\circ$C. 
The gap between the two cover glasses was 5 $\mu$m. 
In the blue circles we can see that the polymer-rich phase is fractured 
under the stretching force generated by its volume shrinking 
(see also the small schematic figure). This resembles a typical brittle fracture of material 
under elongational deformation. This figure is reproduced from Fig. 2 of Ref. \protect\cite{koyama2009fracture}. }
\label{fig:FPS}
\end{figure}

\begin{figure}[!h]
\begin{center}
\includegraphics[width=12cm]{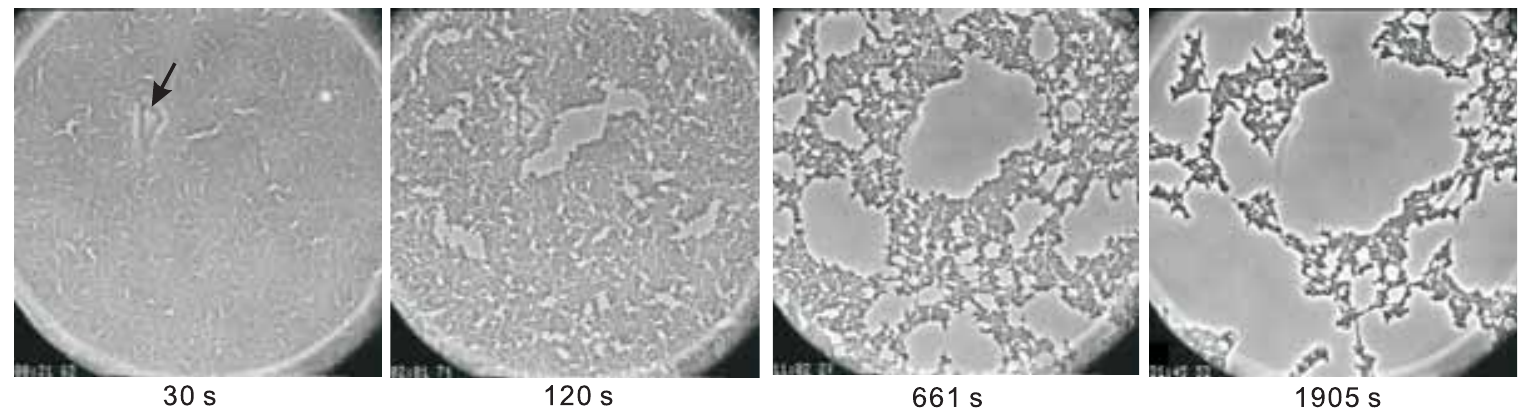}
\end{center}
\caption{Fracture phase separation observed 
at a low magnification for 
a PS/DEM mixture ($M_w=3.84 \times 10^6$, 4 wt\% PS) after a quench to 22 $^\circ$C. 
Crack formation and crack growth are clearly observed. 
The temporal increase in the contrast means 
the increase in $\phi$ in the polymer-rich phase. 
Coherent with this, we can also see a significant volume shrinking of the polymer-rich transient gel. 
The sample thickness is 5 $\mu$m. 
The diameter of the circular window is 1 mm.
The object indicated by the arrow (30 s) is a crack 
in the glass window of the hot stage. 
Since it is located outside the sample, it has nothing to do with the phenomena. 
This figure is reproduced from Fig. 3 of Ref. \protect\cite{koyama2009fracture}.  
} 
\label{fig:macroFPS}
\end{figure}

\begin{figure}[!h]
\begin{center}
\includegraphics[width=8cm]{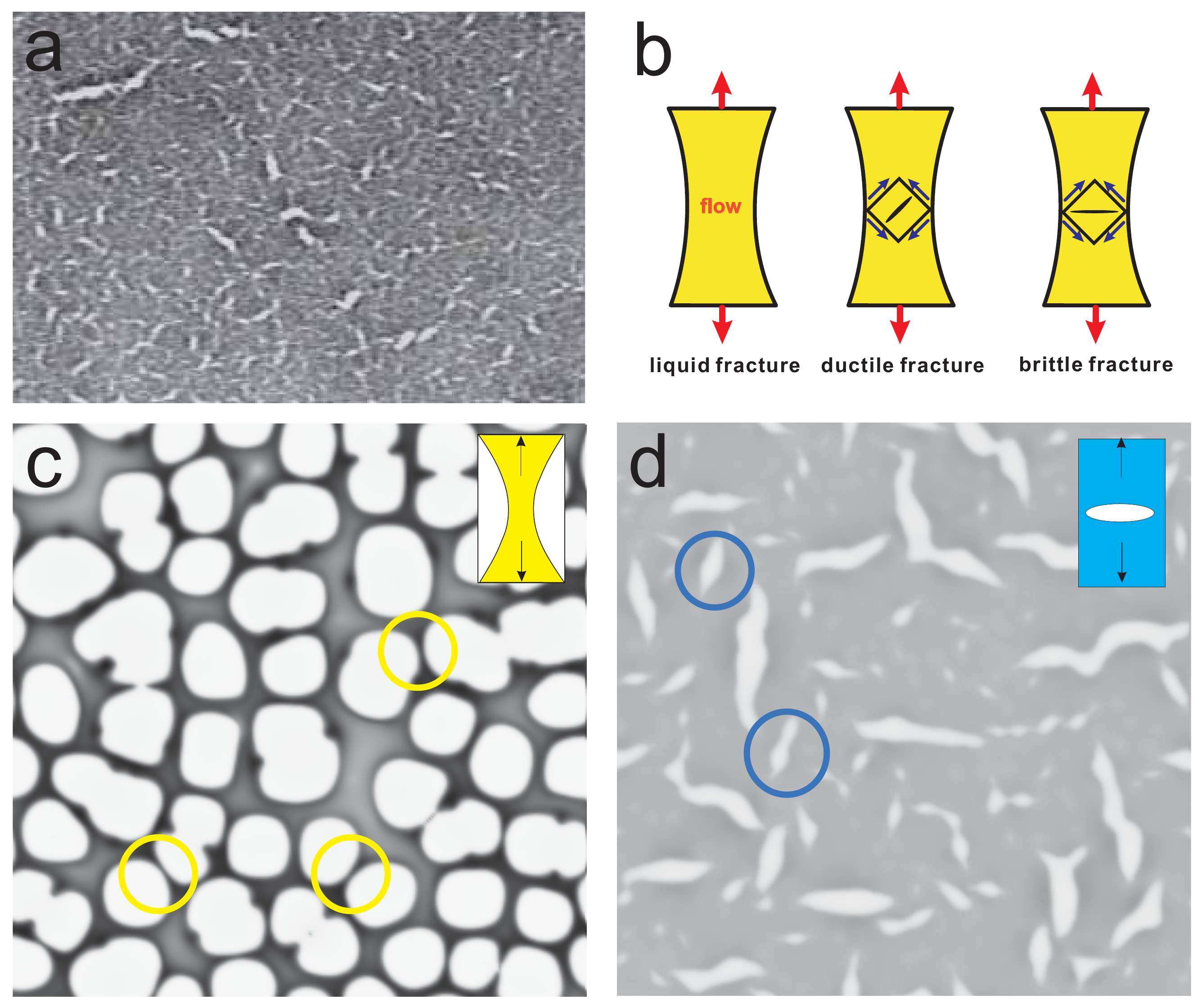}
\end{center}
\caption{
(a) Crack formation in the initial stage of fracture phase separation. 
Fracture phase separation observed for 
a polystyrene(PS)/diethyl malonate (DEM) mixture (4 wt\% PS) after a quench to 22 $^\circ$C. 
Crack formation is clearly observed. Cracks are solvent-rich domains. 
The sample thickness is 5 $\mu$m. 
The width of the image corresponds to 0.5 mm.
(b) Schematic figure showing liquid-type, ductile and brittle fracture of a material under elongational deformation. 
For ductile fracture a crack is formed along 45$^\circ$ from a stretching direction, whereas for brittle fracture 
it is formed perpendicular to a stretching direction. Brittle fracture is also characterized by crack formation just after the linear 
Hookian regime. On the other hand, liquid and ductile fracture occur after large nonlinear deformation. 
Viscoelastic phase separation accompanies liquid-type or ductile fracture for self-generated shear deformation, whereas fracture phase separation 
accompanies brittle fracture. 
Viscoelastic phase separation (c) and fracture phase separation (d)
simulated on the basis of the viscoelastic model.  
We can see typical patterns of liquid and solid fracture in (c) and (d), respectively.   
} 
\label{fig:VPS_FPS}
\end{figure}

The above-described mechanical nature of viscoelastic phase separation implies its close analogy to the mechanical response 
of materials. 
Indeed, we recently found novel phase-separation behaviour accompanying mechanical 
fracture (``fracture phase separation'') in polymer solutions~\cite{koyama2009fracture} (see Figs. \ref{fig:FPS} 
and \ref{fig:macroFPS}). 
Surprisingly, mechanical fracture becomes the dominant coarsening process in this  
phase separation. 
This type of phase separation is observed when the deformation rate of phase separation becomes much faster than 
the slowest mechanical relaxation time of a system. 
In this sense, the transition from viscoelastic to fracture phase separation 
corresponds to the ``liquid-ductile-brittle transition'' in fracture 
of materials under shear deformation~\cite{furukawa2009inhomogeneous} (see Fig. \ref{fig:VPS_FPS}(b)). 
The only difference between fracture phase separation and 
material fracture is whether 
the deformation is induced internally by phase separation 
itself or externally by loading. 

We argue that fracture phase separation is the process of mechanical fracture of a transient gel 
against {\it self-generated shear deformation}, which is caused by volume shrinking of the slow-component-rich phase. 
For slow shear deformation, a transient gel behaves as viscoelastic 
matter and exhibits liquid fracture behaviour for shear deformation: {\it viscoelastic phase separation}. 
A network is stretched continuously under stress, elongated along the stretching direction, 
and eventually breaks up. This process resembles the process of liquid fracture of a material under a stretching force (see Fig. \ref{fig:VPS_FPS}(c)) 
\cite{Spaepen,Argon,Carsson1996,Falk-Langer}. 
For fast shear deformation, a transient gel should behave solid-like  
and exhibit brittle (or ductile) fracture behaviour: {\it fracture phase separation}  (see Fig. \ref{fig:VPS_FPS}(d)) .  
At this moment, it is not so clear whether crack formation in fracture phase separation belongs 
to ductile or brittle fracture, since we are not able to visualize the deformation field in the coarse of phase separation. 
We speculate, however, that cracks are formed perpendicular to the stretching direction (see Fig. \ref{fig:VPS_FPS}(b)), 
which is characteristic of brittle fracture.  
This fracture behaviour is a manifestation of solid-like (or, elastic) behaviour 
\cite{Carsson1996,Falk-Langer} of a transient gel.

The physical mechanism of this mechanical instability is basically the 
same as shear-induced fracture of a viscoelastic matter: 
self-amplification of density fluctuations under shear~\cite{furukawa2006violation,furukawa2009inhomogeneous}. 
In our view, a steep composition dependence of the bulk stress 
leads to instability of the interaction network 
for the volume deformation of type $\vec{\nabla} \cdot \vec{v}_p<0$, 
whereas that of the shear stress leads to its instability for 
shear-type deformation, which should be the origin of fracture-like 
behaviour. In fracture phase separation, elastic couplings between cracks also play a crucial role in pattern formation. 
We studied this problem by using a simple spring model~\cite{araki2005simple} (see, e.g., Fig. \ref{fig:spring}), 
but further detailed studies are necessary to elucidate roles of spatio-temporal elastic coupling. 

\begin{figure}[!h]
\begin{center}
\includegraphics[width=8cm]{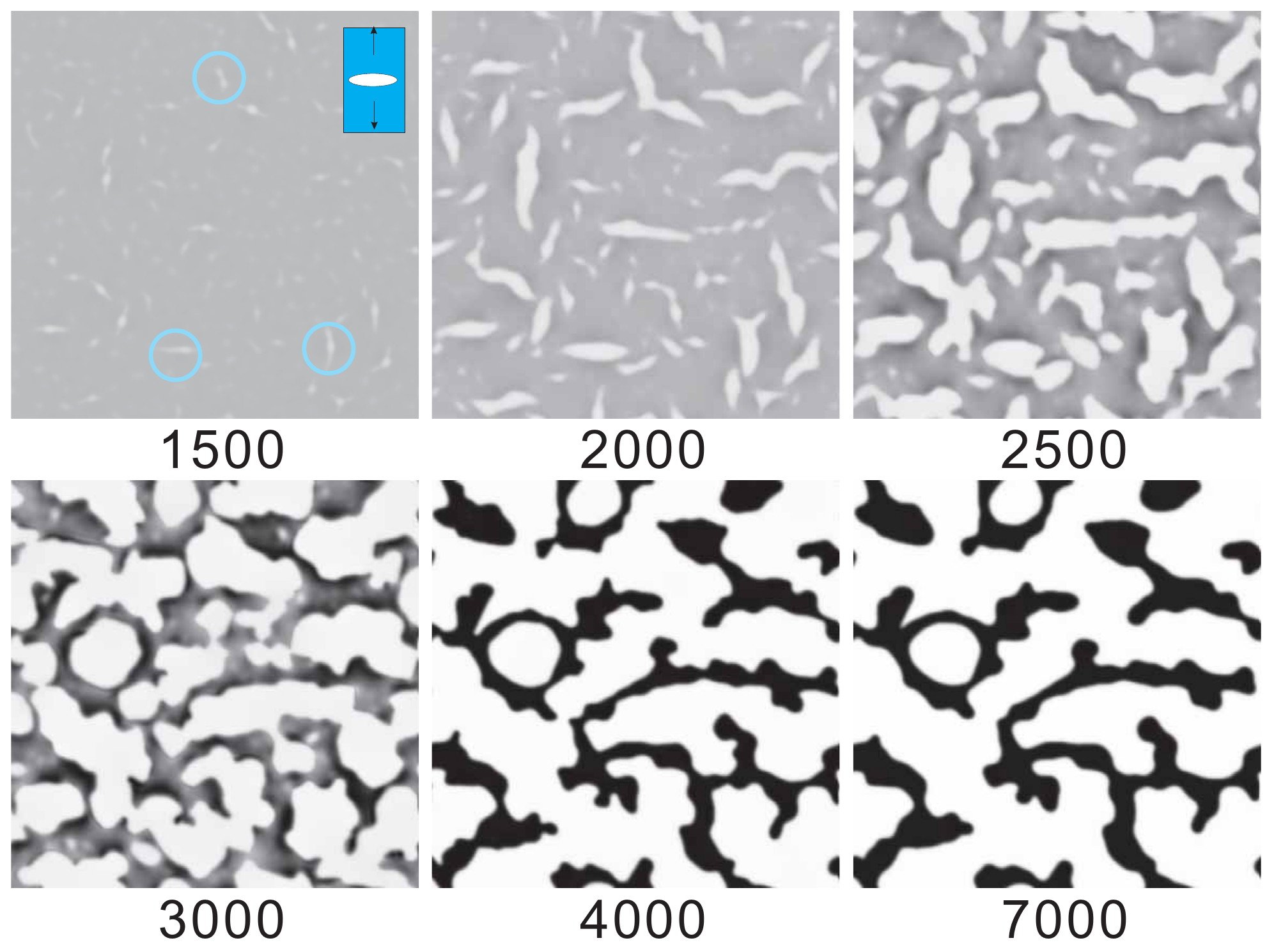}
\end{center}
\caption{Pattern evolution kinetics of fracture phase separation simulated by our model.  
Here $\phi_0=0.35$, $G_B=5 \theta(\phi-0.35)$, 
$G_S=5 \theta(\phi-0.33)$, and $\tau_B=\tau_S=50 \phi^2$. 
The anisotropic solvent holes appear due to the strongly nonlinear 
shear modulus. This captures the characteristic features of 
fracture phase separation observed experimentally. 
This should be compared with the experimental results in Fig. \ref{fig:FPS} 
(for which the same annotation was used).  
} 
\label{fig:FPSsim}
\end{figure}

In fracture phase separation, 
the breakup of bonds is required not only 
for volume deformation, but also for shear deformation of the network. 
To represent such a strongly nonlinear behaviour, we introduce 
a steep (actually, step-like) composition dependence also 
for the shear modulus~\cite{koyama2009fracture}: 
$G_S(\phi)=G_S^0 \Theta(\phi-\phi_0^S)$, where $\phi_0^S$ 
is the threshold polymer composition for the shear modulus. 
The simulated pattern evolution in this way is shown in Fig. \ref{fig:FPSsim}, 
which captures the characteristic features of fracture phase separation observed in experiments 
(see Fig. \ref{fig:FPS}).   
$\phi_0^S$ may be material specific, reflecting its constitutive relation. 
We speculate $\phi_0^S <\phi_0^B$ since 
the instability occurs for volume deformation 
before it occurs for shear deformation. This is because 
only volume deformation can induce 
a composition change and shear deformation cannot. 
We confirmed that the introduction of a step-like $\phi$ dependence 
for the relaxation time $\tau_S$ has a similar effect \cite{koyama2009fracture}. 

Finally we stress that fracture phase separation also provides a mechanism for the formation of 
shrinkage crack patterns, which are widely observed in both nature (tectonic plates, dried mud layers, 
and cracks on rocks) and materials (cracks in concretes, coatings and grazes on a ceramic mug 
and crack formation in dried foods). 
Thus, we expect that the viscoelastic model may be the basis of the understanding of this wide class of  
pattern formation linked to mechanical instability.

\subsection{Arrest of viscoelastic phase separation}
\label{sec:arrest}

\subsubsection{Gelation as dynamically arrested viscoelastic phase separation}
\label{sec:gelation}

Gels and glasses are important nonergodic states of condensed matter, both of which are dynamically arrested 
nonequilibrium states \cite{zaccarelli2007colloidal}. Unlike crystals, their static elasticity does not come from translational order. 
These states are particularly important in soft matter and foods. 
Here we briefly consider the nature of gel and the mechanism of its formation.  

\begin{figure}[!h]
\begin{center}
\includegraphics[width=8cm]{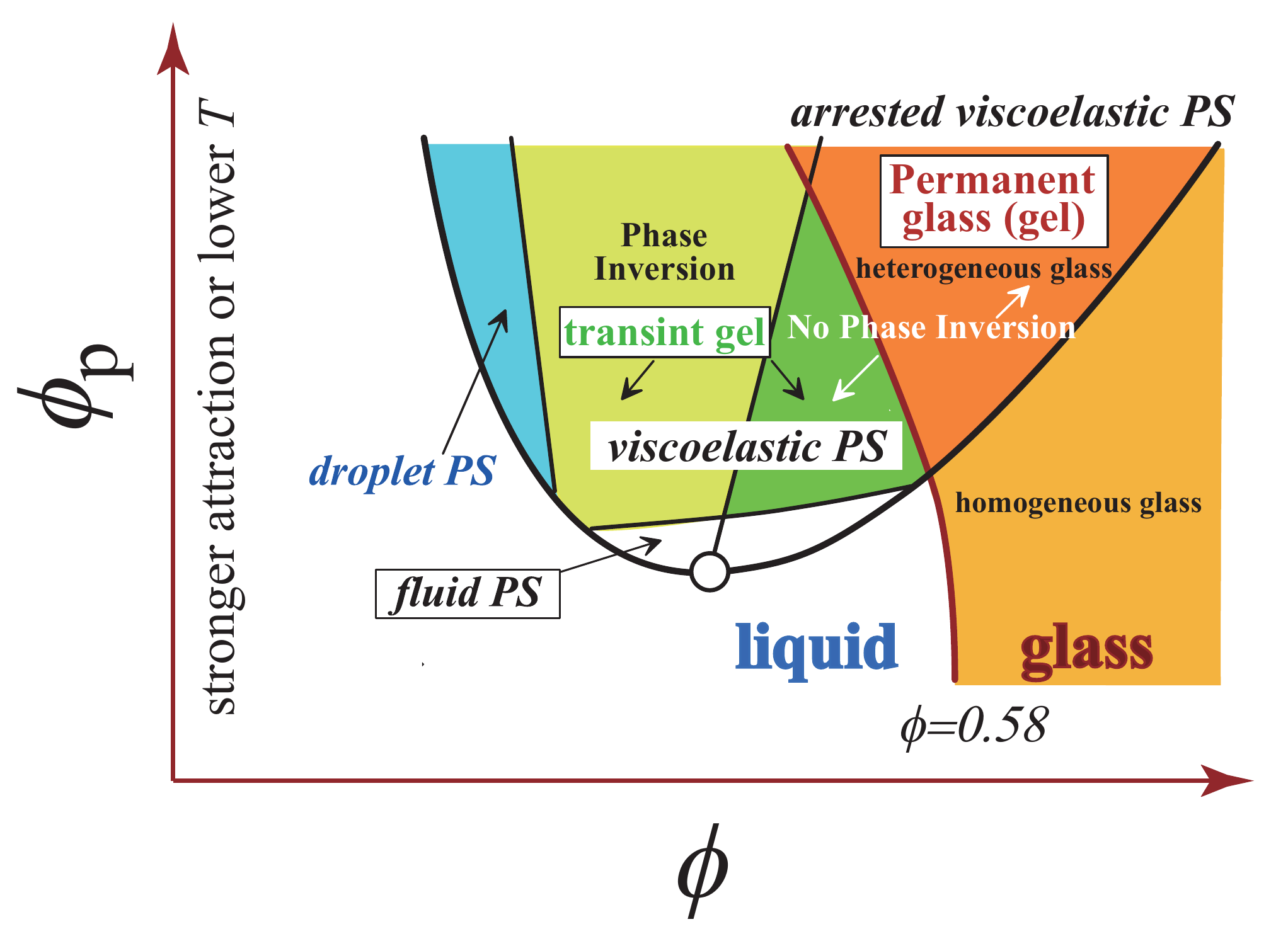}
\end{center}
\caption{Schematic state diagram for colloidal suspensions, emulsions, and protein solutions. 
Whether phase inversion takes place or not is determined by the static symmetry line on which the two separated phases
occupy the same volume. In the left-hand side of this line, a network pattern is formed, whereas in the right-hand side 
a sponge-like structure is formed.  Whether viscoelastic phase separation is arrested or not is determined by the 
glass-transition line. 
The timing when viscoelastic phase separation is arrested by glass transition, or the degree of coarsening of a phase-separated structure, 
is dependent on the quenching condition (the composition, the effective temperature, and the quench speed).  
This figure is reproduced from Fig. 7 of Ref. \protect\cite{tanaka2012viscoelastic}.     
} 
\label{fig:gel}
\end{figure}

A schematic state diagram for colloidal suspensions shown in Fig. \ref{fig:gel} indicates that a transient gel is a consequence of 
viscoelastic phase separation and a permanent gel is that of viscoelastic phase separation dynamically arrested by 
glass transition~\cite{tanaka1999colloid}. Recently, by combining careful experiments and simulations, Lu et al.~\cite{lu2008gelation} 
showed evidence that colloidal gelation is spinodal decomposition dynamically arrested by glass transition. 
Here it is worth pointing out that spinodal decomposition may not be the necessary condition, 
but phase separation including nucleation-growth type is enough to cause gelation if the slow-component-rich phase 
is the majority phase~\cite{tanaka1999colloid}. 
In this scenario, there is an intimate relation between gels and glasses, since the source of dynamic arrest for these two 
nonergodic states is the same. However, there are many distinct differences in both structures and dynamics between them (see, e.g., Ref.~\cite{tanaka2004nonergodic}). 
Locally the dynamic arrest is a consequence of glass transition. However, since gelation is a consequence of 
phase separation, it intrinsically has macroscopic spatial heterogeneity. 
This is always the case if a gel is formed by ordinary attractive interactions between particles. 
Furthermore, the glassy state forming a gel network is far from equilibrium because of 
a rapid density increase induced by phase separation. This is particularly the case of spinodal decomposition. 
Local structural analysis provided such evidence \cite{PaddyNM}.  
Furthermore, a gel formed by viscoelastic phase separation is inevitably 
under the influence of self-generated mechanical stress. 
For the gel to be (quasi-)stable, this stress should be below its yield stress. This feature is absent in a glass. 

In some cases, however, gelation involves specific interactions such as strong hydrogen or covalent bonding and microcrystallite 
formation (e.g., gelatin gels and agarose gels). We note that the mechanism of gelation in these cases is different from the above scenario, reflecting 
the difference in the mechanism of local dynamic arrest. 
For example, in gelatin and agarose crosslinking points are formed by microcrystallites of polymers. 
The difference in the physical interactions stabilizing a gel network leads to the difference in the 
stability and yield stress. Upon viscoelastic phase separation, mechanical stress is always generated in the polymer-rich phase 
but the formation of crosslinkings leads to the increase in the yield stress, which results in the stabilization of the gel 
under the mechanical stress. We emphasize that this mechanical stress is induced by many body effects (the sum of attractions 
between many molecules) and thus can well exceed the interaction strength per bond, which is often measured in the unit of $k_{\rm B}T$.  
Thus, even for strong attractions ($\gg k_{\rm B}T$), coarsening can proceed upon phase separation accompanying gelation 
(see the discussion in Sec. \ref{sec:general}) if there is a strong driving force for volume shrinking, 
although stronger bonds of course tend to increase the yield stress and make a gel more stable. 
The level of coarsening can also crucially depend upon at which stage gelation takes place upon phase separation and 
what is the level of the yield stress (see the state diagram 
and the caption of Fig. \ref{fig:gel}).

\subsubsection{Ageing of gels and glasses}

The scenario that gelation is viscoelastic phase separation dynamically arrested by 
glass transition immediately tells us a crucial difference in the ageing mechanism 
between gelation and vitrification. 
In viscoelastic phase separation, the coarsening is driven by elastic stress associated with volume shrinking 
and interfacial tension. These features are absent in the ageing of glass transition at least in colloidal suspensions 
due to the conservation of the composition. 
In ordinary glass transition, which takes place under a condition of constant pressure, the volume of a sample decreases 
(or, the density increases) during ageing since the ageing accompanies the densification due to attractive interactions. 
We can say that the ageing of gels proceeds under the momentum conservation while satisfying Eq. (\ref{eq:balance})  
\cite{tanaka2007spontaneous}. The intrinsic macroscopic heterogeneity of gels, which comes from its link to 
viscoelastic phase separation, leads to strong inhomogeneity of particle mobility, i.e., faster dynamics near the interface~\cite{ohtsuka2008local}. 

Whether viscoelastic phase separation is dynamically arrested or not may be determined by whether 
the connectivity of the slow-component-rich phase remains when the system reaches a nonergodic state or not. 
Once the volume shrinking is completed, the driving force for domain coarsening becomes only the interfacial tension. 
If the yield stress of a gel is higher than the force exerted by this interfacial tension, the system 
is basically frozen and only exhibits slow ageing towards lower free-energy configuration.

\subsubsection{Arrest of viscoelastic phase separation by smectic order} \label{sec:smectic}

So far we have considered only phase separation of a mixture of isotropic disordered materials. 
However, there is a possibility that viscoelastic phase separation accompanies other ordering phenomena. 
The most obvious such example is freezing by crystallization. 
Here we show an interesting example in which viscoelastic phase separation is arrested  
by smectic ordering~\cite{iwashita2006self}.  We studied phase separation of an ordered phase (lamella) of a lyotropic liquid crystal 
(tri-ethyleneglycol mono n-decyl ether (C$_{10}$E$_3$)/water mixtures) into the coexistence of an ordered (lamella) and a disordered (sponge) phase upon heating. 
When phase separation cannot follow the heating rate, usual viscoelastic phase separation is observed. 
The slow lamella phase, which has internal smectic order and anisotropic elasticity, cannot catch up with the fast domain deformation, 
and thus it transiently behaves like an elastic body and supports most of the mechanical stress.  
On the other hand, the less viscous sponge phase, which is an  isotropic Newtonian liquid, cannot support any stress. 
This dynamic asymmetry leads to formation of a well-developed network structure of the lamella phase (see Fig. \ref{fig:univ}(d)). 

\begin{figure}[!h]
\begin{center}
\includegraphics[width=6cm]{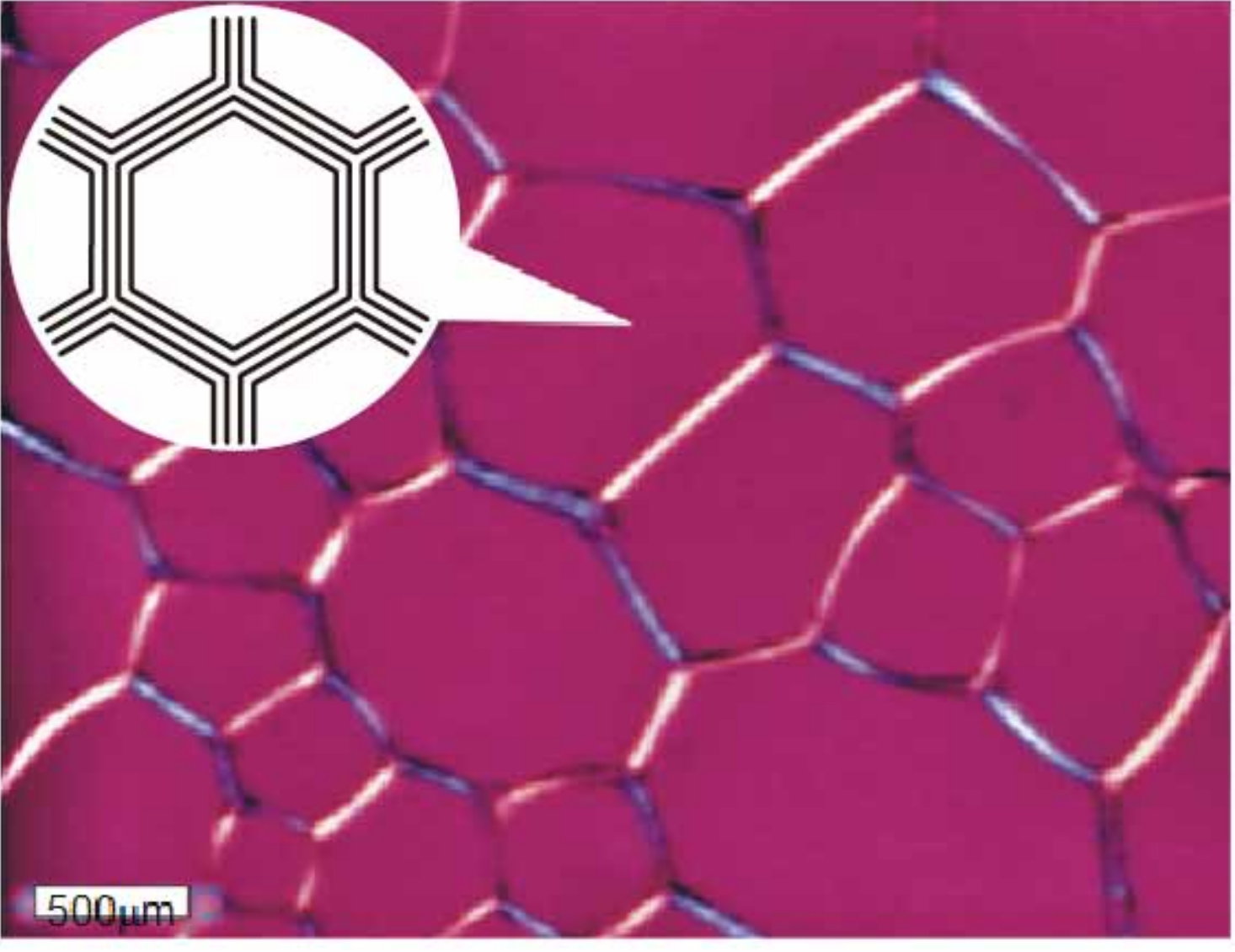}
\end{center}
\caption{Cellular pattern formation in a lyotropic liquid crystal (a C$_{10}$E$_3$/water mixture of 19.9 wt\% 
C$_{10}$E$_3$) observed at 42.15 $^\circ$C with polarizing microscopy. 
The inset schematically shows how membranes are organized in a cellular 
structure. In a border region, there is a disclination line of topological charge -1/2. 
This figure is reproduced from Fig. 8 of Ref. \protect\cite{tanaka2012viscoelastic}. 
} 
\label{fig:smectic}
\end{figure}

If phase separation is slow enough to satisfy the quasi-equilibrium condition, then membranes can homeotropically align along 
the interface between the two phases while keeping their connectivity, to lower the elastic energy. This leads to the formation of 
a cellular structure (see Figs. \ref{fig:univ}(e) and \ref{fig:smectic}). 
The lamellar films forming closed polyhedra cannot exchange material with layers in neighbouring polyhedra, except by permeation (i.e., diffusion of material normal to the layers). For slow permeation, which should be the case in our system, the lamellar films can 
exhibit dilational elasticity, which leads to a (quasi-)stable cellular structure. This is markedly different from a soap froth, 
where a stretching fluid film merely pulls in material from the others without any elastic cost. 
Thus, a high degree of smectic order in the cell walls and borders stabilises the cellular structure: 
Any deformation of the smectic order increases elastic energy and thus the structure is selected by 
the elastic force balance condition.  

We also note that this is an interesting example showing that the heating rate can be used to control the type of phase separation covering  
from droplet phase separation, network phase separation, and to foam-like phase separation \cite{iwashita2006self}. 
This may be relevant in pattern formation in various soft materials, where the change in a physical variable, such as temperature, is not instantaneous.

This phenomenon may be applied to phase separation of systems with smectic order, such as lyotropic and thermotropic 
smectic liquid crystals and block copolymers~\cite{yan2011foam}. The basic physical strategy may also be used for various types of soft matter 
and foods, which have other internal order that can support elastic stress. 
We emphasize that the kinetics of phase separation, more specifically, the characteristic domain deformation rate of phase separation, 
is a key factor for attaining lamellar order in the cell wall. It can in some cases be controlled by a quench rate, as demonstrated above. 

\section{Coexistence of mechanical and hydrodynamic nature in viscoelastic phase separation}

\subsection{Switching of the order parameter during viscoelastic phase separation} \label{sec:switching}

\subsubsection{Concept of order-parameter switching}

Here we describe that the dynamic behaviour of viscoelastic phase separation 
can be explained by the concept of ``order-parameter switching''.  
Phase separation is usually driven by the thermodynamic force 
and the resulting ordering process can be described by the temporal 
evolution of the relevant order parameter associated with 
the thermodynamic driving force. 
The primary order parameter describing phase separation 
of a binary mixture of isotropic components is only the composition difference 
between the two phases in both the solid and fluid model of phase separation.  
In the viscoelastic model, on the other hand, 
the phase-separation mode can be switched 
between ``fluid mode'' and ``elastic gel mode''.  This switching is 
caused by a change in the coupling between the stress and the velocity fields, 
which is described by Eq.~(\ref{sigma}): 
Equation~(\ref{sigma}) tells us that 
these two ultimate cases, namely, (i) fluid model 
($\kappa_{ij}^{p} \sim const$) and (ii) elastic gel model 
($G_S(t)$ and $G_B(t)$ $\sim const$), 
correspond to the situation of $\tau_{ts} \gg \tau_d$ 
and $\tau_{ts} \ll \tau_d$, respectively. 
Here $\tau_{ts}$ is the rheological relaxation time of the slow-component-rich phase and 
$\tau_d$ is the characteristic time of deformation. 
For $\tau_d \gg \tau_{ts}$ the primary order parameter is the composition as 
in usual classical fluids, 
whereas for $\tau_d \leq \tau_{ts}$ it is the deformation tensor as 
in elastic gels. 
The deformation tensor $u_{pij}$ is defined as 
\begin{equation}
u_{pij}=\frac{1}{2}(\frac{\partial u_{pi}}{\partial x_j}+
\frac{\partial u_{pj}}{\partial x_i}). 
\end{equation}
It is well known~\cite{sekimoto1989sponge,onuki} that the free energy of gel can be expressed only by 
the local deformation tensor as $f(u_{pij})$. 
Thus, we can say that the order-parameter switching 
is a result of the competition between the two time scales characterizing 
the domain deformation, $\tau_d$, and the rheological properties of domains, $\tau_{ts}$. 
As mentioned above, thus, this can be regarded as viscoelastic relaxation in pattern evolution. 
Here it should be noted that the above two order parameters are 
related with each other in a gel state as~\cite{onuki} 
\begin{equation}
\frac{\phi_0}{\phi}={\rm Det} \left[ \frac{\partial u_{pi}}{\partial x_j} \right],  
\end{equation}
where $\phi_0$ is the volume fraction in the relaxed state. 

\subsubsection{Crossover between the characteristic timescales}
\label{sec:timescale}

Next we consider how  $\tau_{ts}$ and $\tau_d$ change with time during 
phase separation. 
To simplify the problem, we estimate the temporal change of $\tau_d$ and 
$\tau_{ts}$, provided that they are independent with each other. 
Under this crude assumption, we can estimate the velocity field 
determining the deformation rate, neglecting 
the contribution of $\vec{\nabla}\cdot \mbox{\boldmath$\sigma$}$, 
from the relation 
\begin{equation}
\vec{v}=-\int d\vec{r'} \mbox{\boldmath$T$}(\vec{r}-\vec{r'}) \cdot 
\vec{\nabla}\cdot \left( C \nabla^2 \phi \nabla \phi \right), \label{eq:vel}
\end{equation}
where $\mbox{\boldmath$T$}(\vec{r})$ is the so-called Osceen tensor 
given by 
\begin{equation}
 \mbox{\boldmath$T$}(\vec{r})=\frac{1}{8 \pi \eta_s r}
 \left( \mbox{\boldmath$I$}+\frac{\vec{r}\vec{r}}{r^2} \right).
\end{equation}
According to the above equation (\ref{eq:vel}), the velocity fields should grow in the initial stage as 
$|\vec{v}| \sim (k_BT C/3 \eta \xi)\Delta \phi^2$ \cite{tanaka_araki_PS}, where $\Delta \phi$ 
is the composition difference between the two phases, and $\xi$ 
is the correlation length, or the interface thickness. 
Since $\Delta \phi$ approaches to $2 \phi_e$ 
($\phi_{e}$: the equilibrium composition) with time, 
this expression of $|\vec{v}|$ reduces to the well-known relation 
$|\vec{v}| \sim \gamma/\eta$ ($\gamma$: interface tension) 
in the late stage (note that $\gamma 
\sim k_{\rm B}T C (2\phi_e)^2/3 \xi$).   
Thus, the characteristic deformation time $\tau_d$
changes with time as 
$\tau_d \sim R(t)/v(t) \sim R(t)/\Delta \phi(t)^2$. 
In the initial stage, the domain size does not grow so much with time 
whereas $\Delta \phi$ rapidly increases with time; 
and, accordingly, $\tau_d$ decreases rapidly. 
On the other hand, $\tau_{ts}$ increases steeply 
with an increase in $\Delta \phi$, reflecting the 
increase in $\phi$ in the slow-component-rich domains. 
Thus, $\tau_{ts}$ becomes comparable to $\tau_d$ in this intermediate stage 
of phase separation. 
Once $\tau_d$ exceeds $\tau_{ts}$, the slower phase cannot follow a deformation 
speed and behaves as an elastic body: The mechanical force balance dominates 
a coarsening process in the intermediate stage. 
Next we consider what happens in the late stage. 
Since $\Delta \phi$ approaches to $2 \phi_e$ and becomes almost constant with time 
in the late stage, $\tau_d$ ($ \sim R \eta/\gamma$) 
increases with an increase in $R$ whereas $\tau_{ts}$ becomes almost constant. 
Thus, $\tau_d$ becomes longer than $\tau_{ts}$ again. 
This results in the fluid-like behaviour in the final stage of phase separation. 
We may regard $Wi=\tau_{ts}/\tau_d$ as the Weissenberg number for the self-generated 
deformation rate. Viscoelastic effects become significant when this $Wi$ significantly 
exceeds 1. 

\begin{figure}[!h]
\begin{center}
\includegraphics[width=8cm]{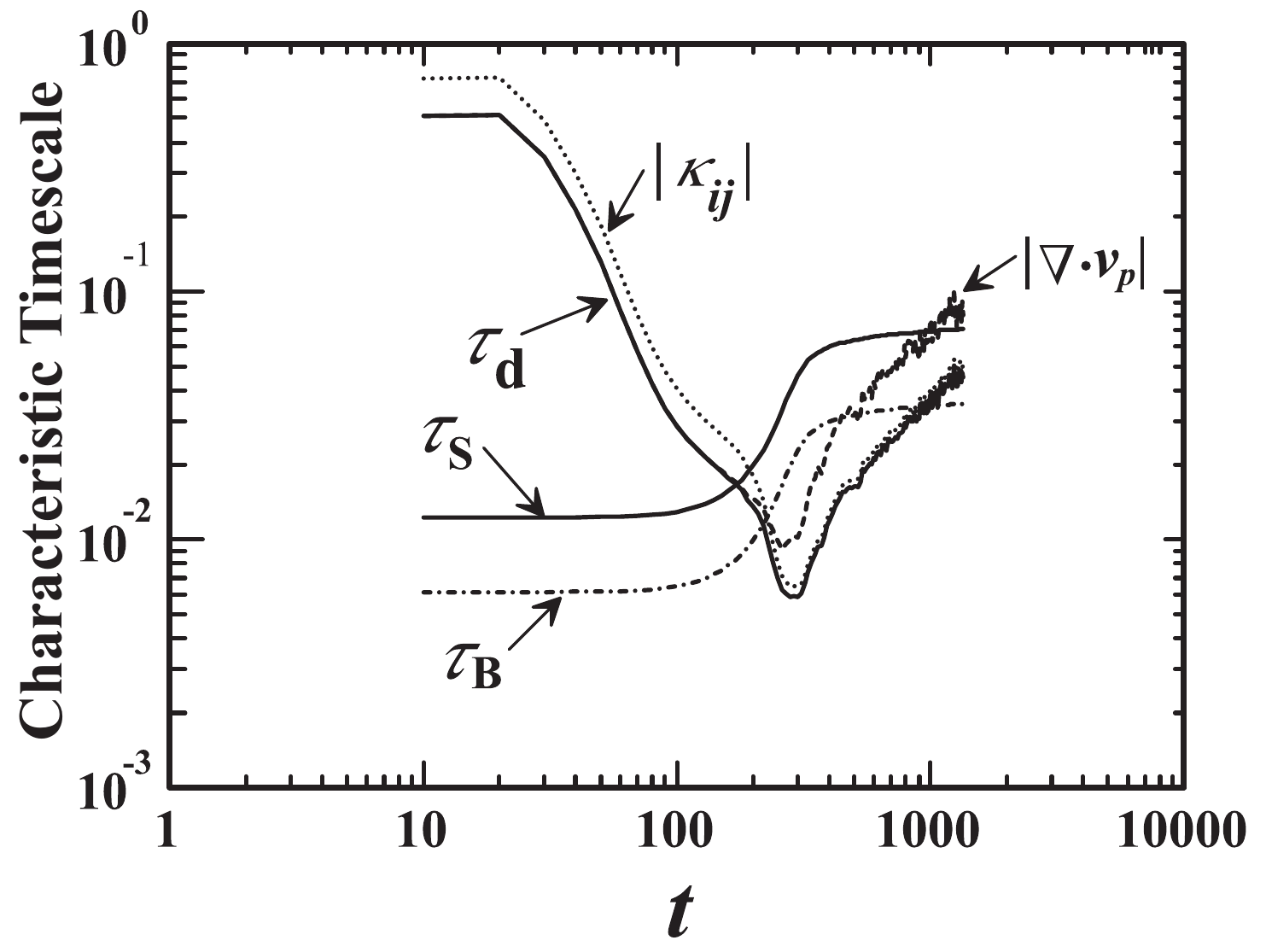}
\end{center}
\caption{Temporal change in the characteristic time scales, 
$\tau_{\rm S}$, $\tau_{\rm B}$, and $\tau_{\rm d}$, and the characteristic 
deformation rate, $|\kappa_{ij}|$ and $|\mbox{\boldmath$\nabla$} 
\cdot \mbox{\boldmath$v$}_{\rm p}|$, which is obtained by numerical simulation 
of our model. 
This figure is reproduced from Fig. 23 of Ref. \protect\cite{tanaka2000viscoelastic}. 
} 
\label{fig:simutau}
\end{figure}

In short, $\tau_d \gg \tau_{ts}$ in the initial stage, 
$\tau_d \leq \tau_{ts}$ in the intermediate stage, and 
$\tau_d \gg \tau_{ts}$ in the late stage again. 
Accordingly, the order parameter switches from the composition to 
the deformation tensor, and then switches back to the composition again. 
Such behaviour is indeed observed in our numerical simulations, as shown in Fig. \ref{fig:simutau}, which  
shows the temporal changes in the key characteristic timescales obtained by our numerical simulation. 
Here $\tau_{ts}$ for the shear deformation mode is represented by $\tau_S$, whereas 
that for the bulk volume deformation mode by $\tau_B$. 
When phase separation accompanies an ergodic-to-nonergodic transition such as glass transition, 
phase separation ends up with a dynamically arrested state, i.e., network-like and 
sponge-like structures can be frozen: gelation (see Sec. \ref{sec:gelation}). 

Next we consider possible effects of a difference in the two types of origins of dynamic asymmetry on pattern evolution: disparity 
in the size or $T_{\rm g}$ between the two components. 
In the above, the domain deformation rate is related to the interfacial tension $\gamma$, or the coefficient $C$. 
It is known that $\gamma$ is inversely proportional to $\xi^2$, $\gamma \sim 0.1 k_{\rm B}T/\xi^2$, 
according to the two scale-factor universality \cite{onuki}. 
Since the interfacial thickness, or the correlation length, $\xi$, is proportional to the size of a component, 
the interface tension $\gamma$ is known to be extremely small for systems of macromolecules, emulsions, 
and colloids simply because of their large sizes \cite{norton2001microstructure,aarts2004direct,royall2007bridging}. 
This leads to the large difference in $\gamma$ between the two types of systems. 
However, the large size of the slow component also results in the slow relaxation in proportional to 
$\xi^3$. Thus, the above-defined Weissenberg number $Wi$ can become very large even for a system of large size disparity.  

\subsection{Viscoelastic selection of phase-separation morphology}

\subsubsection{What physical factors determine the phase-separation morphology?}

Since the deformation tensor $u_{pij}$ has an intrinsic coupling to the mechanical stress, 
a pattern in the elastic regime 
is essentially different from 
that of usual phase separation in fluid mixtures, which is dominated by the balance between the thermodynamic and the viscous force. 
The domain shape during viscoelastic phase separation 
is determined by which of the mechanical and interface force is more dominant in the momentum 
conservation equation. 
Roughly, the elastic energy is scaled as $G_S e^2 R^d$ ($e$: strain and 
$d$: spatial dimensionality) for a domain of size $R$, 
since it is the bulk energy. 
On the other hand, the interface energy is estimated 
as $\gamma R^{d-1}$. 
For macroscopic domains, thus, the elastic energy 
is much more important than the interface energy 
in the intermediate stage where $\tau_d \leq \tau_{ts}$. 

The momentum conservation tells us that the domain shape is generally determined by the mechanical shear force balance 
condition~\cite{tanaka2000viscoelastic}:
\begin{eqnarray}
\partial_i \left[ C(\phi) \{ \partial_i \phi \partial_j \phi-\frac{1}{d} (\partial_i \phi)(\partial_j \phi) \delta_{ij} \}
-\sigma_{ij} \right]=0, \label{eq:balance}
\end{eqnarray}
which leads to network-like or spongelike morphology.  
In two dimensions, this force balance condition 
favours a three-armed treelike structure where the angle between 
the arms are about 120$^\circ$, whereas in three dimensions a four-arm (tetrapod-like) structure around its junction point is favoured. 
This is consistent with what is observed in Figs. \ref{fig:NPS}, \ref{fig:visco}, \ref{fig:3DVPS}, and \ref{fig:colloidsimu}. 
In the late stage of phase separation where $\tau_d \gg \tau_{ts}$, on the other hand, the interface energy dominates a domain shape 
since the mechanical stress becomes very weak. 

Here we note a possible difference between a system of size disparity and a system of disparity in $T_{\rm g}$. 
As mentioned above, a system of large size disparity is characterized by ultralow interface tension $\gamma$. 
For such a system, the above force balance condition can approximately be given by $\partial _i \sigma_{ij}=0$. 
For a system of disparity in $T_{\rm g}$, on the other hand, the interface tension plays a more important role 
when the mechanical stress is about the same. In the final stage of viscoelastic phase separation, where 
the Weissenberg number $Wi$ decreases and becomes smaller than 1, the interface tension leads to the breakage 
of a network structure, which transforms the morphology from network-like to droplet-like. 
This process may take place more slowly for a system of size disparity than for a system of disparity in $T_{\rm g}$ since the relative importance 
of the interfacial tension is less for the former than the latter. 

\subsubsection{Crucial roles of the boundary condition for a system in viscoelastic and fracture phase separation}

\begin{figure}[!h]
\begin{center}
\includegraphics[width=8cm]{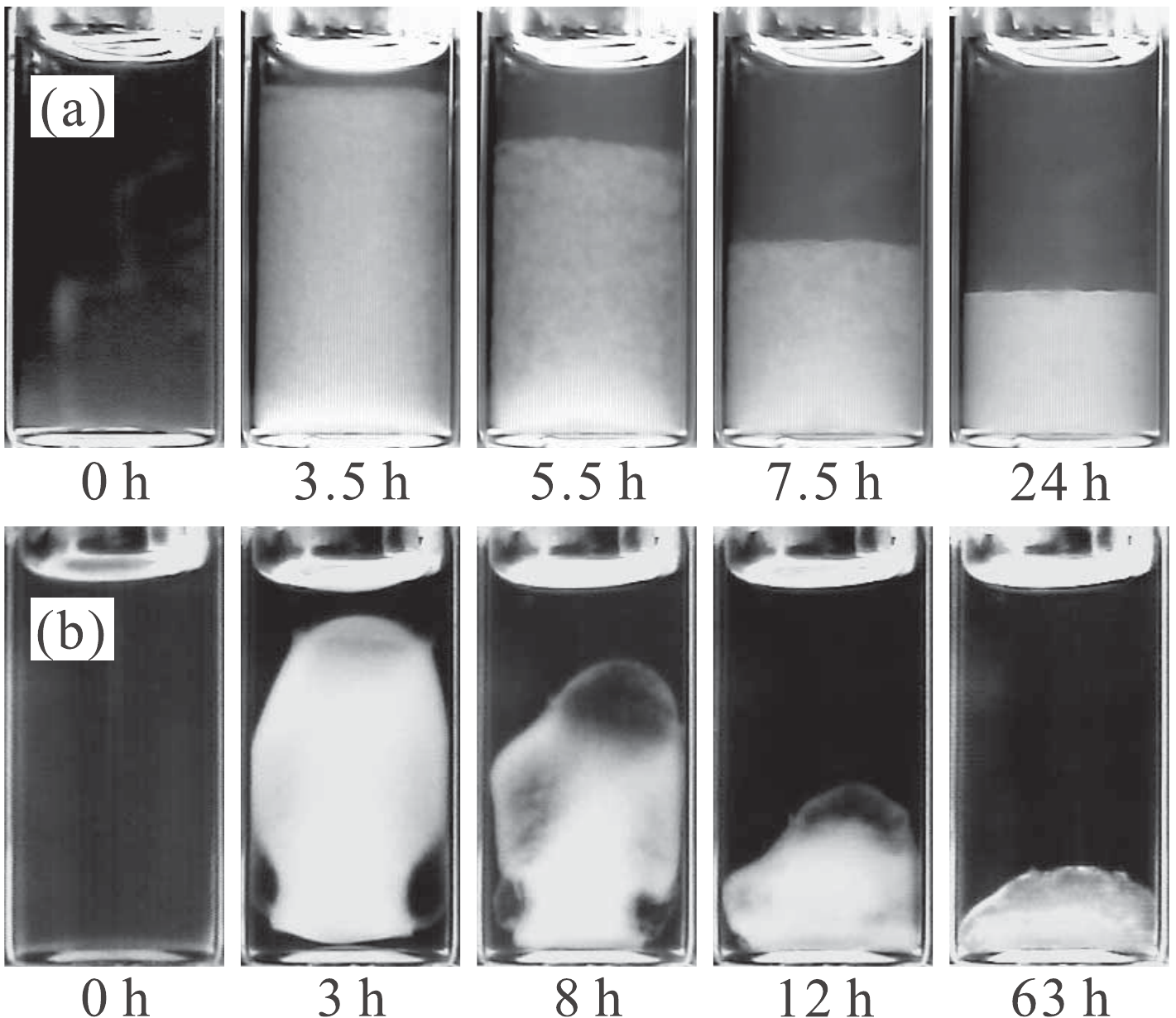}
\end{center}
\vspace{-0.5cm}
\caption{
Macroscopic phase separation observed in the critical solution of 
PS/DEM mixture ($M_w=3.84 \times 10^6$, $N=3.69 \times 10^{4}$, $\phi_c=2.14$ \%, $T_c=27.2$ $^\circ$ C).
The brighter phase is PS-rich, whereas the darker phase is DEM-rich. 
Note that the PS-rich phase is heavier than the DEM-rich phase. 
The inner diameter of these vials are 1 cm.
(a): $\Delta T=0.4$ K, which corresponds to the NPS regime. 
(b): $\Delta T=$ 8.0 K, which corresponds to the VPS regime.  
This figure is reproduced from Fig. 2 of Ref. \protect\cite{koyama2007generic}. 
}
\label{fig:macro_sed}
\end{figure}

As described above, the mechanical force balance plays a crucial role in pattern selection in viscoelastic phase separation. 
A transient gel always tends to shrink to reduce the elastic energy as a gel undergoing volume-shrinking transition does 
\cite{matsuo1988kinetics,matsuo1992patterns}. 
This means that the entire network tends to shrink its volume. This stress leading to volume shrinking of a whole sample must 
be supported by the boundary to have only internal mechanical instability inducing solvent-hole formation and crack formation. 
In simulations, the employment of a periodic boundary condition automatically allows us to avoid long-wavelength 
instabilities. 
The rate of volume shrinking is controlled by the rate of the transport of the fluid component under the stress fields. 
In many experimental situations, this elastic stress is supported by the boundary which prevents the shrinking of 
the overall volume of a transient gel. 
This can be realized by wetting or adsorption of the slow-component-rich phase to walls confining a sample.  
In our experiments using a quasi two-dimensional sample for optical microscopy observation, 
i.e., for an anisotropic confinement of a sample, on the other hand, 
volume shrinking in the lateral direction is strongly suppressed by a large friction of the sample to the walls. 
This boundary effect is the very origin of the mechanical stress acting against concentration 
diffusion ($\vec{\nabla} \cdot \vec{v}_p$). Even if there is no fixed boundary condition, for a very large sample  
there is a clear separation between the time scale of the volume shrinking of the entire sample and that of the 
local development of the mechanical instability. For a sample of a finite volume, however, the volume shrinking takes place 
and thus affects or interferes the internal mechanical instability. Such volume shrinking behaviour of the 
whole transient gel accompanying mechanical instability  was observed in a macroscopic sample 
undergoing viscoelastic phase separation~\cite{koyama2007generic}, 
as shown in Fig. \ref{fig:macro_sed}. 

\begin{figure}
\begin{center}
\includegraphics[width=8cm]{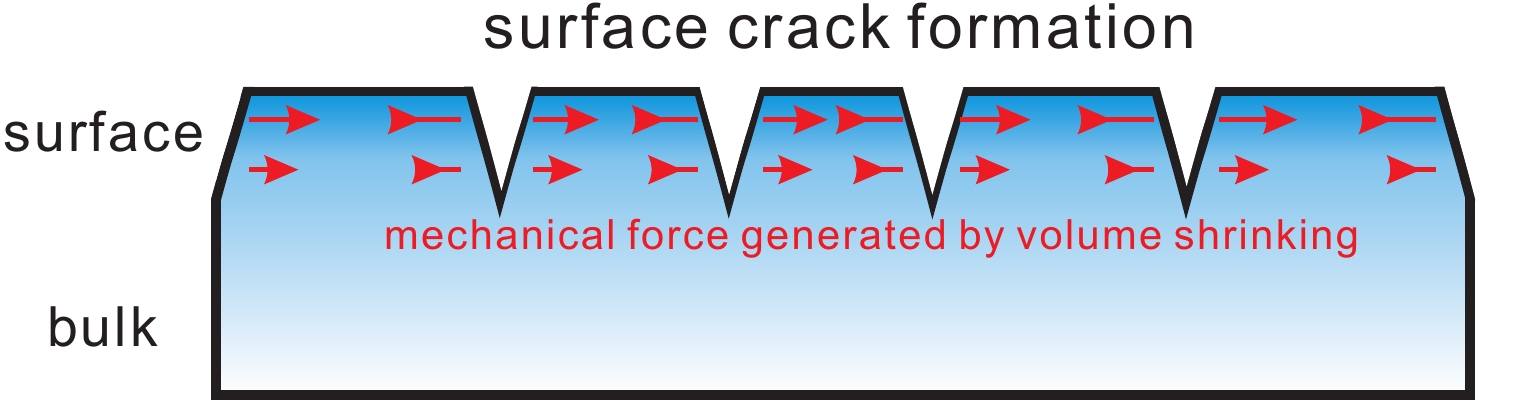}
\end{center}
\caption{Schematic figure showing the localization of mechanical stress near the surface of drying soft matter. 
The mechanical stress is a consequence of volume shrinking induced by solvent evaporation. 
This figure is reproduced from Fig.6 of Ref. \protect\cite{tanaka2012viscoelastic}. 
} 
\label{fig:crack}
\end{figure}

This special role of the boundary condition in phase separation is a manifestation of the mechanical nature of 
phase separation, which is common to both viscoelastic and fracture phase separation. 
We note that surface crack formation is also affected by such a boundary condition (see Fig. \ref{fig:crack}). 
During evaporation of the liquid component, the volume shrinking of the surface part takes place much faster than the bulk part 
far from the surface. 
Thus, the bulk part plays the same role as a fixed boundary condition and supports the mechanical stress, which leads to the formation of surface crack patterns. 
Surface crack patterns can also be induced by bulk expansion: The slow (or solid-like) surface layer 
cannot catch up with the expansion of the bulk. 
We also note that surface crack formation can also be caused by cooling of a glassy material from its surface. 
This is because surface cooling leads to larger volume shrinking near the surface. This causes the extensional mechanical stress on the surface, 
which may induce mechanical fracture of the surface region that becomes solid-like near and below the glass transition upon cooling. 
This may be the case for formation of grazes on ceramic or glass mugs. 

\section{General mechanism of the formation of cellular, foam-like, or sponge-like structures in materials}

\subsection{Viscoelastic phase separation as a general model of mechanical pattern formation}

Next we discuss the universal nature of the physical mechanism producing a spongelike morphology. 
It is known that gel undergoing volume-shrinking 
phase transition forms a foam-like structure~\cite{matsuo1988kinetics,matsuo1992patterns,sekimoto1989sponge}. 
We argue that the physical origin of the appearance of a honeycomb structure in plastic 
foams (e.g., polystyrene and urethane foams) and breads 
is also similar to that of a network structure in viscoelastic phase separation. 
For the formation of network patterns in ordinary viscoelastic 
phase separation, the pressure, $p$, plays only a
minor role in the force balance (or, momentum conservation) equation: $p$ is determined to satisfy the incompressibility
condition. However, the formation of foam structures, which is usually
induced by the liquid-to-gas transformation of one of the
components of a mixture, accompanies its large volume expansion. Thus the pressure $p$ plays a primary role in the pattern selection. 

As such an example, we briefly describe a typical formation process of plastic foams (see, e.g., Ref.~\cite{Lemstra}).  
First, a polymer matrix absorbing a low-boiling-point solvent is prepared.
Then, its temperature is raised above the boiling point of the
solvent, which induces bubble formation in the polymer matrix.
These bubbles nucleate and grow as the result of evaporation of the solvent from the
polymer matrix. The total volume of the system expands as a
result of the liquid-gas transformation of the solvent. In this
process, a pattern is dominated by the mechanical force balance
condition with the contribution of the gas pressure $p$. This is caused by the
strongly asymmetric stress division: gas bubbles cannot support
any mechanical stress besides the hydrostatic pressure and only the polymeric phase can support it. 
In this way, a cellular pattern is formed. 
As mentioned above, it was pointed out that strain hardening plays an important role in the formation of well-developed cellular patterns \cite{spitael2004strain}. 
This may be because strain hardening prevents the liquid-type rapture of cell walls. Thus, it can be viewed as the enhancement of 
the importance of mechanical stress over interfacial tension in the force balance condition (see, e.g., Eq. (\ref{eq:balance})). 
Finally, foam structures are stabilized by glass transition or crystallization, when the stress generated by high internal gas pressure in cells 
becomes lower than the yield stress of the matrix phase.

To describe this phenomenon we need to use the dynamic equations for compressible liquids. 
The force balance can be satisfied
only when a gas bubble is surrounded by the matrix phase: the
internal gas pressure is balanced with the mechanical stress created
by the stretched matrix phase surrounding the gas bubble. It is
this feature that leads to the formation of cellular foam structures.
As in the case of network formation in viscoelastic phase separation, we can say that the foam
structure formation is a mechanically selected pattern formation, and thus can be regarded as 
a special case of viscoelastic phase separation.

Besides the above-explained difference in the morphological selection, the processes of network and cellular pattern formation 
have an important common feature: Holes of a less 
viscoelastic fluid phase (gas in plastic foams, water in gels, solvent 
in polymer solutions, and so on) are nucleated in a phase-separation 
process. Then the structure develops while keeping a mechanical force balance condition and 
a network or form-like structure is transiently formed, depending upon the type of the mechanical force generated.  
In this process, the relative volume fraction of the more viscoelastic phase in the whole system  decreases with time.  
Furthermore, the limiting process of material transport between the two phases is 
that in the slower phase.  
This suggests that a network or sponge-like structure is the 
morphology {\it universal} to phase separation in dynamically asymmetric mixtures. 

\subsection{Selection principles of patterns for viscoelastic and elastic phase separation}

It is worth noting that the pattern selection in viscoelastic phase separation essentially differs from that in phase separation 
of elastic solid mixtures (e.g., metal alloys). 
We note that elasticity, which is static, does not involve any time scales (or, velocity fields). 
For elastic phase separation, thus, the momentum conservation, or force balance condition, is irrelevant for the selection 
of morphology and a pattern evolves solely to lower the elastic energy while obeying the diffusion equation alone. 
We emphasize that the momentum conservation is relevant only when a mixture contains a fluid or viscoelastic material as its component. 

\begin{figure}[!h]
\begin{center}
\includegraphics[width=9cm]{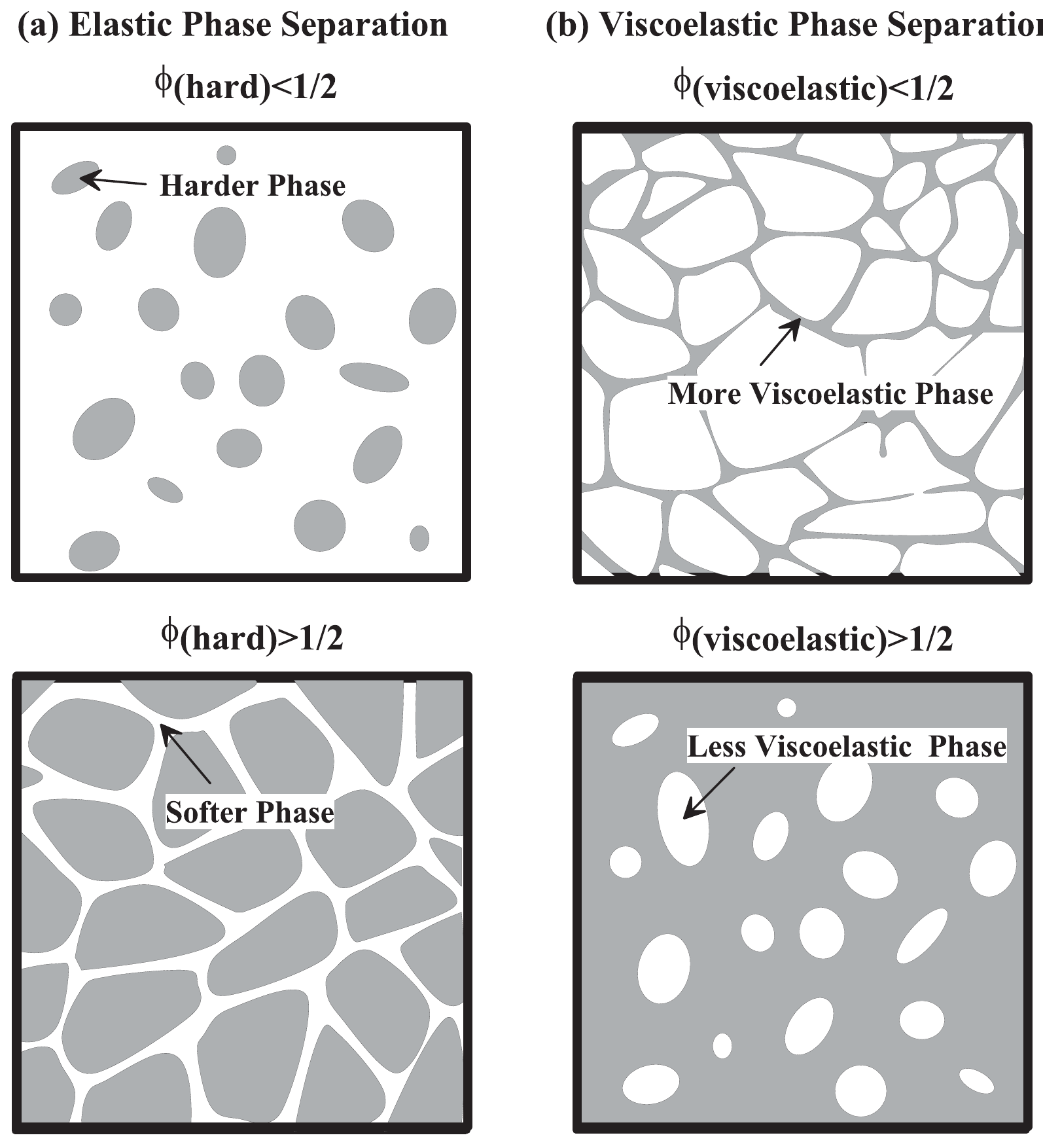}
\end{center}
\caption{Difference in pattern evolution 
between (a) phase separation in a solid mixture with elastic deformation energy 
and (b) viscoelastic phase separation. 
This figure is reproduced from Fig. 25 of Ref. \protect\cite{tanaka2000viscoelastic}. 
} 
\label{fig:vis-ela}
\end{figure}

Elastic effects often originate from a lattice mismatch between the two atomic components in solid alloys. 
First of all, solid phase separation accompanies little volume change of each phase.  
Furthermore, the softer phase always forms a network-like continuous phase 
to minimize the total elastic energy~\cite{onuki}, 
in contrast to the case of viscoelastic phase separation (see Fig. \ref{fig:vis-ela}).  
For example, the elastic energy is expressed in terms of the displacement vector $\vec{u}$ as follows for an isotropic elastic body: 
\begin{eqnarray}
 \mathscr{H}_{el}=\int d\vec{r} \ \left[ \frac{1}{2} G_B (\vec{\nabla} \cdot \vec{u})^2+G_S \sum_{i,j} \left( \frac{\partial u_j}{\partial x_i} 
+\frac{\partial u_i}{\partial x_j} -\frac{2}{d} \delta_{ij} \vec{\nabla} \cdot \vec{u} \right)^2 \right], 
\end{eqnarray}
where $G_B$ is the bulk modulus, $G_S$ is the shear modulus, and $d$ is the spatial dimension. 
The leading-order coupling between $\phi$ and $\vec{u}$ is then given by 
\begin{eqnarray}
 \mathscr{H}_{int}=\int d\vec{r} \ \alpha \phi  \vec{\nabla} \cdot \vec{u},  
\end{eqnarray}
where $\alpha$ is the coupling constant. 
If pattern evolution is slow enough, $\vec{u}$ is determined by the local equilibrium condition 
$\delta ( \mathscr{H}_{el}+ \mathscr{H}_{int})/\delta \vec{u}=0$. 
Here we note that $G_S$ generally depends on $\phi$. 
Thus, the softer phase with smaller $G_S$ is deformed to form a network structure since the deformation 
of the softer phase costs less elastic energy than that of the harder phase. 
The phase-separation morphology is determined to minimize the elastic energy in solid mixtures, whereas 
to satisfy the momentum conservation (or, the force balance) in liquid mixtures.  

Concerning the momentum conservation, we note that hydrodynamic degrees of freedom play a significant role 
in the initial and final stage of viscoelastic phase separation. For example, network formation in colloid phase separation 
is significantly influenced by hydrodynamic interactions between colloids \cite{tanaka2000simulation,tanaka2006simulation,furukawa2010key}. 
In the final stage, hydrodynamic effects are important to describe Rayleigh instability of tubes (or networks) and the resulting 
breakup of tube-like structures which are component structures of a network. 
In the intermediate stage, on the other hand, hydrodynamic effects are not so significant and only the force balance 
plays an important role 
in pattern evolution. To describe this regime, thus, we may use Langevin (Brownian) dynamics \cite{araki2005simple}, 
which neglects hydrodynamic interactions.

\section{Viscoelastic phase separation and shear-induced mechanical instability: Roles of dynamic asymmetry} 

\subsection{Shear-induced composition fluctuations and demixing}

\subsubsection{Basic mechanism of shear-induced instability}

Here we consider shear-induced composition fluctuations 
and demixing (or flocculations) in polymer solutions ~\cite{wolf1980phase,wolf1984,wu1991enhanced,hashimoto1991shear,moses1994shear,onuki1997phase,onuki1989elastic,helfand1989,doionuki,ji1995,milner1993} 
as well as in colloidal suspensions and emulsions~\cite{tanaka1999colloid}.
Shear-induced composition fluctuations are induced by a steep increase of $\eta$ with $\phi$. 
An intuitive explanation was given as follows~\cite{wolf1980phase,wolf1984}. 
Shear-induced demixing is caused by a certain mechanism to store elastic energy under shear. 
This elastic energy effectively leads to a change in the free energy 
functional, which results in an effective shift of the phase diagram and destabilizes a thermodynamically stable system. 
However, this picture was turned out to be too simplistic. 
It was shown by intensive theoretical studies 
\cite{onuki1997phase,onuki1989elastic,helfand1989,doionuki,ji1995,milner1993} 
that we need to treat dynamical couplings between the composition and stress fields properly for explaining shear-induced demixing of polymer solutions.  
This phenomenon is now widely known as ``shear-induced demixing'' 
in polymer solutions under shear~\cite{onuki1997phase}. 

Some time ago we considered whether similar phenomena can be observed in colloidal suspensions, emulsions, and protein solutions or not~\cite{tanaka1999colloid}. 
In polymers, the conformational degrees of freedom of chains and entanglement effects play a crucial role in shear-induced instability. 
Since such internal degrees of freedom are absent in suspensions of particle-like objects,  
the mechanism to store elastic energy under shear in colloidal suspensions might be 
essentially different from that in polymer solutions~\cite{tanaka1999colloid}. 
At first sight, thus, shear effects seem less pronounced for colloidal suspensions than for polymer solutions. 
Accordingly, this problem looks far from being obvious. However, it was shown that the basic mechanism 
of shear-induced instability is the same between the two cases~\cite{tanaka1999colloid}. 

In the following, we briefly discuss shear effects on colloidal suspensions. 
Under thermal fluctuations, local shear stress is stored 
inhomogeneously due to a strong nonlinear and asymmetric 
dependence of $G_S(\phi)$ and $\tau_S(\phi)$ on $\phi$~\cite{tanaka1999colloid}. 
Note that the mechanical stress relevant to a shear problem 
is the ``shear'' stress, $\mbox{\boldmath$\sigma$}_c^S$. 
The linear stability analysis tells us that this enhances composition 
fluctuations along the compressional axis of the flow, 
since this stress moves colloidal particles towards  
a more concentrated region. 
This positive feedback process results in shear-induced instability in a self-catalytic manner. 

In the linear Newtonian regime under the condition 
$\dot{\gamma}\tau_S \ll 1$, where $\dot{\gamma}$ is the shear rate, $\mbox{\boldmath$\sigma$}_c$ 
is given as 
\begin{eqnarray}
\mbox{\boldmath$\sigma$}_c 
\sim \eta(\phi)(\vec{\nabla}\vec{v}+(\vec{\nabla}\vec{v})^t) \sim \eta(\phi) 
\dot{\gamma}. 
\end{eqnarray}
Then, one can straightforwardly obtain the following 
expression for the relaxation rate of the composition 
fluctuations convected by shear flow on the basis of the two-fluid model of colloidal suspensions \cite{tanaka1999colloid}: 
\begin{eqnarray}
\Gamma_{eff}=L \left[ k^2(r_0+C k^2)-2 (\frac{\partial \eta}{\partial \phi})_T 
\phi^{-1} \dot{\gamma} k_x k_y \right]/(1+\xi_{ve} k^2). 
\end{eqnarray}
It is important to note that if $(\partial \eta/\partial \phi)_T>0$, 
$\Gamma_{eff}$ can be negative 
even for positive $r_0$ for $\dot{\gamma}>\dot{\gamma}_c$, 
indicating the (initially) exponential growth of composition fluctuations even in a thermodynamically 
stable region. Compare this equation for shear-induced instability with 
that for thermodynamic instability, Eq. (\ref{Aq}). 
The critical shear rate $\dot{\gamma}_c$ is thus obtained, using $r_\phi$ defined in Sec. \ref{sec:early}, as 
\begin{eqnarray}
\dot{\gamma}_c \sim r_\phi \phi/(\partial \eta/\partial \phi)_T. 
\end{eqnarray}

Recently it was demonstrated by Furukawa and Tanaka~\cite{furukawa2006violation} that this condition can be rewritten by using the 
osmotic pressure $\Pi$ as follows: 
\begin{eqnarray}
\dot{\gamma_c} \sim (\partial \eta/\partial \Pi)_T. \label{eq:pi}
\end{eqnarray}
We briefly discuss a general implication of this relation and its relevance to single-component glassy systems in Sec. \ref{sec:link}. 
For the details, please refer Refs. \cite{furukawa2006violation,furukawa2009inhomogeneous}.

\subsubsection{Shear-induced instability and pattern evolution}

Because of intrinsic dynamic asymmetry between the components of a mixture, shear 
flow enhances concentration fluctuations or induces phase separation \cite{onuki1997phase}. At the same time, shear gradient deforms and breaks up domains formed by phase separation 
\cite{onuki,imaeda2004viscoelastic}. 
Furthermore, flow can generate anisotropic structures such as layered structures and fibrous structures. 
In polymer mixtures \cite{hashimoto1995string,onuki1997phase} and colloidal suspensions \cite{derks2008phase}, for example, 
string-like phase separated structures are formed for a high shear rate. 
We note that string-like phase separation is observed for a system with rather weak dynamic asymmetry between the two phases. 
For strongly dynamically asymmetric cases, more chaotic and disordered structures are formed \cite{imaeda2004viscoelastic}. 
This indicates that string-like domain formation is of hydrodynamic origin and the interplay between shear deformation and 
interface tension may play a primary role in the selection of the string structure. 
We also note that string-like morphology, more precisely, leek-like structures, can also be formed along the flow direction 
by shear flow in a lyotropic lamellar phase \cite{miyazawa2007nucleation}. 

Lamella-like layered structures are often ascribed to so-called shear banding, which is a consequence of nonlinear rheology 
accompanying non-monotonic stress-strain rate relation \cite{cates2006rheology,olmsted2008perspectives}. Such non-linearity may come from a coupling between shear flow and internal degrees of freedom 
of slow components, e.g., orientation of polymer chains \cite{tanaka2000inhomogeneous}.  
A constitutive relation such as the nonlocal Johnson-Segalman (JS) model can describe rheological instability \cite{cates2006rheology,olmsted2008perspectives},  
which is very similar to the upper-convective Maxwell relation besides additional inclusion of the slippage effects and the so-called  
stress diffusion term in the nonlocal JS model. Thus, the viscoelastic model may describe rich pattern evolution in a nonlinear 
flow regime at least on a phenomenological level. 
Unlike a single-component description, the two-fluid model provides a coupling between shear, stress, and concentration fields, 
which plays a crucial role in the dynamical behaviour of multi-component systems.  
In relation to this, it is worth noting that in the two-fluid model, the nonlocal constitutive relation may not 
be required to have stable shear banding since similar 
nonlocal effects are expected to be produced by the concentration gradient and their couplings to stress and strain fields. 
\cite{onuki1997phase,jupp2003modeling,jupp2004dynamic,imaeda2004viscoelastic}. For theoretical analysis, we need to treat 
nonlinear effects properly, including the spatial variation not only the concentration field but also 
the stress and strain fields, and their couplings. 
This is a difficult theoretical task. 
Whether we fix the total stress or strain rate applied is also crucial for 
the selection of nonequilibrium steady states, e.g., gradient and vorticity banding, if they exist \cite{cates2006rheology,olmsted2008perspectives,dhont2008gradient}. 

In previous studies of shear instability, the steep dependence of the transport coefficient, the structural relaxation time, 
and the elastic modulus on the order parameter such as the composition $\phi$, has not been considered carefully. 
However, as emphasized above, it may induce instability of a different mechanism and thus play a crucial 
role in shear-induced phenomena  \cite{wolf1980phase,wolf1984,wu1991enhanced,hashimoto1991shear,moses1994shear,onuki1997phase,onuki1989elastic,helfand1989,doionuki,ji1995,milner1993,tanaka1999colloid,furukawa2006violation,furukawa2009inhomogeneous,nozieres1986,lequeux,besseling2010shear}. 
This problem needs further study in the future.  

\subsection{A link to mechanical fracture of gassy materials} \label{sec:link}

Recently Furukawa and Tanaka proposed that the mesoscopic dynamics of glassy materials 
may generally be described by the following set of 
equations for compressible viscoelastic materials.   
The mass density $\rho(\mbox{\boldmath$r$},t)$ obeys the continuity equation: 
\begin{eqnarray}
\dfrac{\partial}{\partial t}\rho= -\nabla\cdot(\rho\mbox{\boldmath$v$}),   
\label{continuity}
\end{eqnarray} 
where $\mbox{\boldmath$v$}(\mbox{\boldmath$r$},t)$ is the velocity field 
that obeys the generalized Navier-Stokes equation: 
\begin{eqnarray}
\rho\biggl(\dfrac{\partial}{\partial t}
+\mbox{\boldmath$v$}\cdot\nabla\biggr)\mbox{\boldmath$v$}= 
-\nabla \cdot \mbox{\boldmath$\Pi$}+\nabla \cdot \mbox{\boldmath$\sigma$}, 
\label{momentum}
\end{eqnarray}
where $\mbox{\boldmath$\Pi$}(\mbox{\boldmath$r$},t)$ 
is the pressure tensor due to density fluctuations. 
For small density fluctuations $\delta \rho$ from its average value $\rho_0$, the pressure tensor $\mbox{\boldmath$\Pi$}$ can be expressed, up to linear order in
$\delta \rho$, as $\Pi_{ij}=(p_0+K_T^{-1} \delta \rho/\rho_0)\delta_{ij}$, where $p_0$ is the average pressure
and $K_T=(\partial \rho/\partial P)_T/\rho$ is the isothermal compressibility. 
Then $\mbox{\boldmath$\sigma$}(\mbox{\boldmath$r$},t)$ is 
the viscoelastic stress tensor 
arising from the slow structural relaxation,  
whose time evolution is assumed to be governed by 
the following constitutive equation \cite{larson}:
\begin{eqnarray}
(\dfrac{\partial}{\partial t}+\mbox{\boldmath$v$}\cdot\nabla)
\mbox{\boldmath$\sigma$}-(\nabla\mbox{\boldmath$v$}^\dagger
\cdot \mbox{\boldmath$\sigma$}+\mbox{\boldmath$\sigma$} 
\cdot \nabla \mbox{\boldmath$v$})
=G_S(\rho)\bigl(\nabla\mbox{\boldmath$v$}^\dagger
+\nabla\mbox{\boldmath$v$}\bigr)
-\dfrac{1}{\tau_S(\rho)}{\mbox{\boldmath$\sigma$}}, 
\label{eq:viscoelastic}
\end{eqnarray} 
where the right hand side represents the upper convective time derivative, 
which ensures the frame invariance 
of the tensor properties of $\mbox{\boldmath$\sigma$}$ \cite{larson}. 
In Eq. (\ref{eq:viscoelastic}), 
$\tau_S(\rho)$ and $G_S(\rho)$ are the rheological relaxation time and shear elastic modulus, 
respectively, which characterize the material properties 
and in general depend upon the density $\rho$. 
In most materials, the higher the density 
(i.e., the smaller the free volume), the more solid-like behaviour 
(longer $\tau$ and larger $G_S$). 
This dynamic asymmetry with respective to the change in the density is particularly pronounced near the glass transition. 
It is these intrinsic density dependencies 
of $\tau_S$ and $G_S$ that lead to a coupling between density fluctuations and shear deformation, 
which is the key to shear-induced instability. 

On the basis of the above set of equations, it was shown by Furukawa and Tanaka that the above mechanism of shear-induced instability is much more general and 
can be relevant for not only a critical mixture but also a single component liquid~\cite{furukawa2006violation,furukawa2009inhomogeneous}. 
The criterion (\ref{eq:pi}) on the shear-induced demixing is further linked to the condition for the mechanical instability of a simple fluid:
\begin{eqnarray}
\dot{\gamma_c} \sim (\partial \eta/\partial P)_T, \label{eq:criterion}
\end{eqnarray}
where $P$ is the pressure.
In a single-component system, the dynamical asymmetry arises from the asymmetry 
of the viscosity, the elastic modulus, and/or the shear relaxation time against 
the change in the density. Near the glass transition temperature, these quantities 
quite asymmetrically depend on the density increase and decrease. 
This provides a general scenario of shear-induced instability for a system whose transport coefficient (viscosity)  
depends upon the conserved order parameter, such as density and composition~\cite{furukawa2006violation}. 
This scenario can also explain a crossover between liquid, ductile and brittle fracture on the basis of the concept of 
viscoelastic relaxation~\cite{furukawa2009inhomogeneous}. 
The validity of the prediction (\ref{eq:criterion}) was confirmed by comparing it with experimental results 
\cite{furukawa2006violation,furukawa2009inhomogeneous}. 
This liquid-ductile-brittle transition has a striking similarity to the crossover between fluid, viscoelastic and fracture phase separation. 

Finally, we stress that the above $\tau_\gamma=1/\dot{\gamma}_c$ should be regarded as a new intrinsic rheological time scale of a system. 
This time scale is much shorter than the structural relaxation time $\tau_\alpha$ and is not relevant in ordinary situations.   
However, near the glass transition, $(\partial \eta/\partial \Pi)_T$ or $(\partial \eta/\partial P)_T$ becomes very large 
and thus this time scale can become much slower than $\tau_\alpha$.  
This is why the instability can take place even in the linear Newtonian regime. 
For a single component liquid, $\dot{\gamma}_c$ is the critical shear rate above which the incompressibility 
condition is violated. Thus, this instability provides a novel mechanism for the violation of the incompressible condition of fluids.

If we compare the above set of equations and those for the viscoelastic model, 
we can see that there is a tight analogy between viscoelastic phase separation and 
shear-induced instability (mechanical fracture) of single-component liquids and glasses \cite{furukawa2009inhomogeneous}. 
To describe the unique feature of transient gel formation in soft matter, we need the two-fluid viscoelastic model. 
However, to describe dynamic asymmetry due to the difference in $T_g$ between the two components of a mixture, 
the composition dependence of the diffusion constant $D(\phi)$ and the structural relaxation time $\tau_\alpha$ and/or 
the shear modulus $G_S$ may be enough to describe the viscoelastic phase separation. 
Then the analogy between viscoelastic phase separation and mechanical fracture becomes even more transparent. 
The only difference is whether the stress is self-generated by the thermodynamic driving force of phase separation 
or externally applied.  

\section{Analogy between classifications  
of rheological behaviour of materials, phase separation, and mechanical fracture}
\label{sec:analogy}

First we consider the general nature of the basic equations describing viscoelastic phase separation~\cite{tanaka1997general}: 
(i) If we set $G_S(t-t')=G_S(\phi(\vec{r},t))$, 
$G_B(t-t')=G_B(\phi(\vec{r},t))$, and the absence of the velocity fields ($\vec{v}=0$), 
a viscoelastic model reduces to the elastic solid model~\cite{nishimori1990pattern}. 
(ii) If we further assume that $G_S$ and $G_B$ do not depend on the composition $\phi$, 
it reduces to the solid model (model B~\cite{hohenberg}).  
(iii) If we assume dynamic symmetry between the two components of 
a mixture in the viscoelastic model, it reduces to a new model of symmetric viscoelastic model 
\cite{tanaka1997general}. If we further assume slow enough deformation, then, it reduces to 
the fluid model (model H~\cite{hohenberg}). 
(iv) If we assume only $G_S(t)=G_S$ and $G_B(t)=G_B$, the viscoelastic model reduces  
to the ``elastic gel model''~\cite{sekimoto1989sponge,onuki} that describes 
phase separation in elastic gels.  
Note that the time integration of the velocity 
becomes the deformation $u_p$, and 
$\sigma_{ij}=G_S[\frac{\partial u_{pj}}{\partial x_i} +
\frac{\partial u_{pi}}{\partial x_j} -\frac{2}{3} 
(\nabla \cdot u_p)\delta_{ij}]+G_B(\nabla \cdot u_p)\delta_{ij}$.  
Here we note that the above mappings of the viscoelastic model to the various models support 
that the introduction of the bulk relaxation modulus is essential for the description of viscoelastic phase separation, 
or the general model of phase separation.  
These mappings clearly tell us that the viscoelastic model 
is the general model of phase separation that can describe any types of 
phase separation in mixtures of isotropic condensed matter, 
as its special cases~\cite{tanaka1997general}.

\begin{figure}[!h]
\begin{center}
\includegraphics[width=10cm]{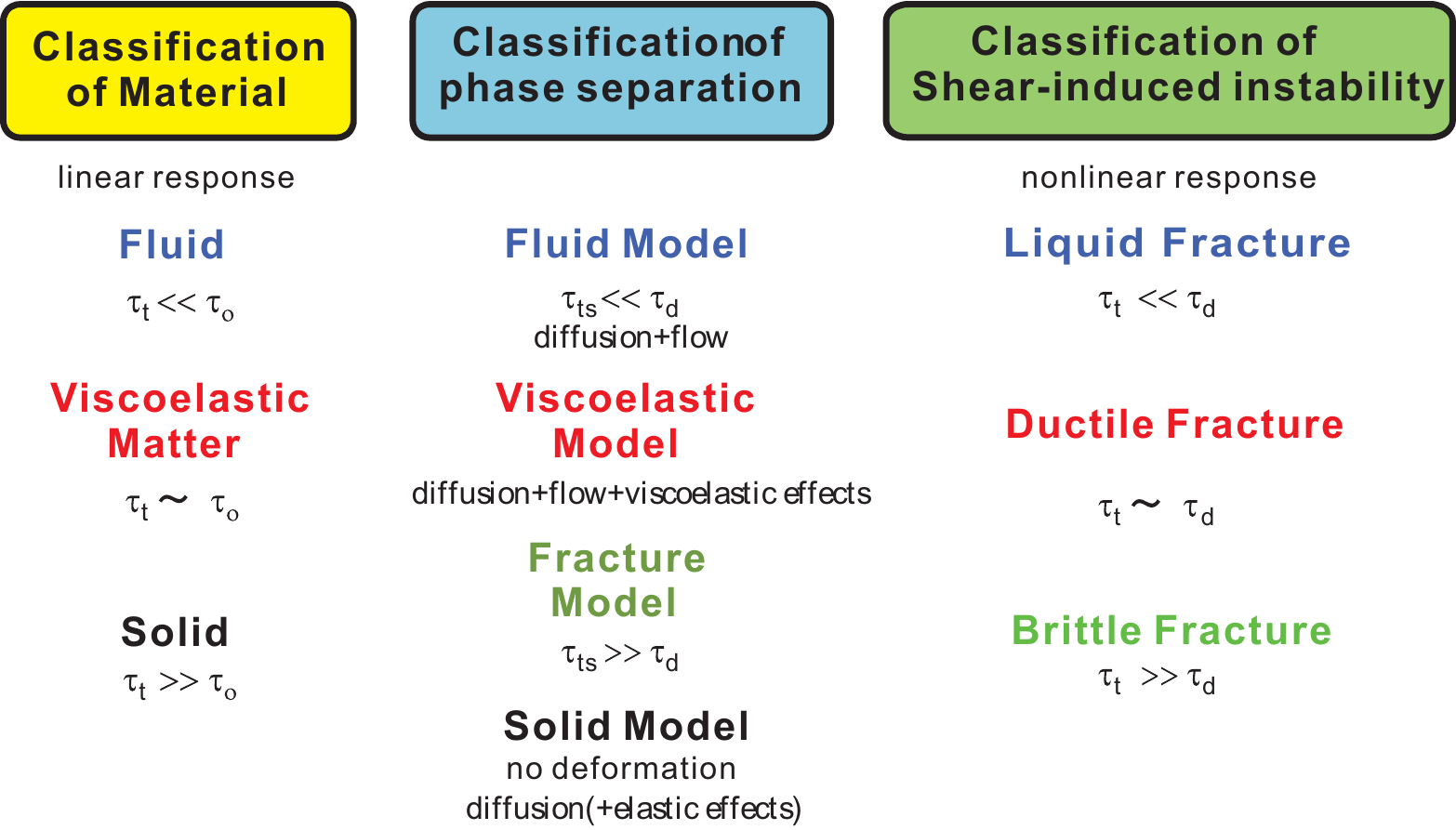}
\end{center}
\caption{
Schematic figure explaining the classification of phase separation of isotropic matter 
and its relation to the classification of materials (linear response) and mechanical fracture (nonlinear response). 
In the classification of materials (left), the ratio of the structural relaxation time $\tau_t$ 
to the observation time $\tau_o$, which is known as the Deborah number, is a key number. 
In the classification of phase separation (middle), the ratio of $\tau_{ts}$ to $\tau_d$, plays a crucial role, as discussed in Sec. \ref{sec:timescale}.  
This ratio can be regarded as the Weissenberg number for deformation self-induced by phase separation. 
On the classification of mechanical fracture (right),  liquid-type fracture should be observed if 
fracture takes place in the time regime $t \ll \tau_t$,  
solid-type (brittle) fracture if fracture takes place in the time regime 
$0<t < \tau_t ((\partial \ln G_S/\partial \rho)/(\partial \ln \tau_t/\partial \rho))$, and  
viscoelastic (ductile) fracture in the intermediate time regime. Please refer Ref. \protect\cite{furukawa2009inhomogeneous} for the details. 
This figure is reproduced from Fig. 10 of Ref. \protect\cite{tanaka2012viscoelastic}. 
}
\label{fig:general}
\end{figure}

The viscoelastic model in the classification of isotropic phase separation 
corresponds to viscoelastic matter in the classification 
of isotropic condensed matter. 
Viscoelastic matter includes any condensed matter ranged from solid to fluid. 
The key factor for the classification of materials  
is the relation between the characteristic internal 
rheological time, $\tau_t$, and the characteristic observation time, $\tau_o$. 
Corresponding to this, the key physical factor 
for the classification of isotropic phase separation 
is the relation between the characteristic time of phase separation 
(domain deformation), $\tau_d$, 
and the characteristic rheological time of the slower phase, $\tau_{ts}$. 
The above analogy is schematically summarized in Fig.~\ref{fig:general}. 

Furthermore, this classification may also be common to that of mechanical fracture 
\cite{furukawa2009inhomogeneous}, which is determined by the relation between the time when the mechanical instability sets in, 
the mechanical relaxation rate, and the deformation rate (see Fig.~\ref{fig:general}). 
The only difference between the two is just whether the deformation is induced by 
phase separation or externally imposed. 
See Ref. \cite{furukawa2009inhomogeneous} on the details of the physical mechanism of mechanical fracture and 
the theoretical prediction for the occurrence and types of fracture.

\section{Summary}
In summary, we show that viscoelastic phase separation and the concept of 
dynamic asymmetry are very useful for understanding not only phase separation and gelation 
in soft matter but also the mechanical instability of soft matter and glassy matter under shear deformation. 
We argue that ``dynamic asymmetry'' is one of the key physical concepts in soft matter since it is a consequence 
of the spatio-temporal hierarchical nature of a system. 
We demonstrated that the viscoelastic model including both bulk and shear stress 
contributions is a very general model that can universally describe phase separation of any isotropic condensed matter. 
However, our understanding is still a level of phenomenology and far from the full understanding 
of the phenomena. In particular, how to describe the constitutive equations of various soft matter 
and how to generalize them to unstable states where phase separation occurs are of fundamental importance.  

We also demonstrate that the formation of heterogeneous network or cellular structures in various materials 
may be regarded as mechanically driven pattern evolution and understood in the framework of viscoelastic phase separation. 
Dynamic asymmetry may be a key to the physical understanding of not only phase separation 
but also mechanical instability of materials under deformation. 
The latter may be relevant not only to materials science, but also in engineering (shrinkage crack pattern formation of concretes) 
and geology (crack in dry mud or rocks).   
These phenomena of mechanically driven inhomogeneization can be understood in a unified manner on the basis 
of the concept of dynamic asymmetry with respect to the change in a conserved order parameter.  

Finally we mention some fundamental remaining problems of the current viscoelastic model. 
(1) The dissipation in a dynamically asymmetric mixture may not be given by a simple sum of friction due to the relative motion 
of the components and hydrodynamic dissipation. Here the nonlocal nature of the transport, which is characterized by the viscoelastic length 
$\xi_{\rm ve}$, should also be considered properly in the process of the coarse-graining. 
(2) The phenomenological constitutive relation crucially depends on the composition dependence 
of the elastic moduli and the mechanical relaxation times. However, there is no firm basis for the physical description 
of these quantities primarily because there is no theoretical basis for the description of the constitutive relation in an 
unstable state. For more quantitative understanding of viscoelastic phase separation it is crucial to overcome 
these difficult problems. 

We hope that this article would contribute to better understanding of pattern formation and 
mechanical instability of various condensed matter systems, particularly, soft matter, and encourage young researchers to get interested in these 
intriguing problems.

\section*{Acknowledgements}
The author is grateful to T. Araki, Y. Iwashita, T. Koyama, and Y. Nishikawa for their collaboration on 
viscoelastic phase separation and A. Furukawa for his collaboration on mechanical instability and fracture of materials. 
He also thanks T. Araki and A. Furukawa for fruitful discussions over many years. 
This work was partly supported by Grants-in-Aid for Scientific Research (S) and Specially Promoted Research from JSPS 
and Aihara Project, the FIRST program from JSPS, initiated by CSTP.

\end{document}